\documentclass[fleqn,usenatbib]{mnras}

\usepackage{newtxtext,newtxmath}

\usepackage[T1]{fontenc}

\DeclareRobustCommand{\VAN}[3]{#2}
\let\VANthebibliography\thebibliography
\def\thebibliography{\DeclareRobustCommand{\VAN}[3]{##3}\VANthebibliography}

\usepackage{graphicx}	
\usepackage{amsmath}	
\usepackage[dvipsnames]{xcolor}		
\usepackage{multirow}	
\usepackage{hyperref}	
\usepackage{threeparttable} 

\newcommand{\Ha}{H$\alpha$}
\newcommand{\Hb}{H$\beta$}
\newcommand{\Lya}{Ly$\alpha$}
\newcommand{\OIII}{[O{\sc{iii}}]}
\newcommand{\NII}{[N{\sc{ii}}]}
\newcommand{\CII}{[C{\sc{ii}}]}
\newcommand{\CIII}{C{\sc{iii}}]}
\newcommand{\SII}{[S{\sc{ii}}]}
\newcommand{\BrG}{Br$\gamma$}
\newcommand{\Ks}{$K_{s}$}

\newcommand{\um}{$\mu$m}
\newcommand{\comment}[1]{}

\newcommand{\eazypy}{{\sc{eazy-py}}}
\newcommand{\magphys}{{\sc{magphys}}}

\newcommand{\Msun}{${\rm M}_{\odot}$}

\newcommand{\ujy}{$\mu$Jy}

\title[A narrowband study of SMG environments at $z \sim 2\text{--}3$]{An ALMA survey of submillimetre galaxies in the Extended \textit{Chandra} Deep Field South: an unbiased study of SMG environments measured with narrowband imaging}

\author[T.\ M.\ Cornish et al.]{Thomas M.\ Cornish$^{1,2}$\thanks{E-mail: thomas.cornish@physics.ox.ac.uk},
Julie Wardlow$^{1}$\thanks{E-mail: j.wardlow@lancaster.ac.uk},
Heather Wade$^{1}$,
David Sobral$^{1,3}$,
W.\ N.\ Brandt$^{4,5,6}$,
\newauthor
Pierre Cox$^{7}$,
Helmut Dannerbauer$^{8,9}$,
Roberto Decarli$^{10}$,
Bitten Gullberg$^{11,12}$,
Kirsten Knudsen$^{13}$,
\newauthor
John Stott$^{1}$,
Mark Swinbank$^{14}$,
Fabian Walter$^{15}$,
Paul van der Werf$^{16}$
\\
$^{1}$Department of Physics, Lancaster University, Lancaster, LA1 4YB, UK\\
$^{2}$Department of Physics, University of Oxford, Denys Wilkinson Building, Keble Road, Oxford, OX1 3RH, UK\\
$^{3}$BNP Paribas Corporate \& Institutional Banking, Lisbon, Portugal\\
$^{4}$Department of Astronomy \& Astrophysics, The Pennsylvania State University, 525 Davey Lab, University Park, PA 16802, USA\\
$^{5}$Institute for Gravitation and the Cosmos, The Pennsylvania State University, University Park, PA 16802, USA\\
$^{6}$Department of Physics, The Pennsylvania State University, University Park, PA 16802, USA\\
$^{7}$Sorbonne Universit\'{e}, CNRS UMR 7095, Institut d’Astrophysique de Paris, 98bis bvd Arago, 75014 Paris, France\\
$^{8}$Instituto de Astrofísica de Canarias (IAC), E-38205 La Laguna, Tenerife, Spain\\
$^{9}$Universidad de La Laguna, Dpto. Astrofísica, E-38206 La Laguna, Tenerife, Spain\\
$^{10}$INAF –- Osservatorio di Astrofisica e Scienza dello Spazio di Bologna, Via Gobetti 93/3, 40129 Bologna, Italy\\
$^{11}$DTU Space, Technical University of Denmark, Elektrovej 327, 2800 Kgs. Lyngby, Denmark\\
$^{12}$Cosmic Dawn Centre (DAWN), Copenhagen, Denmark\\
$^{13}$Department of Space, Earth and Environment, Chalmers University of Technology, SE-412 96 Gothenburg, Sweden\\
$^{14}$Centre for Extragalactic Astronomy, Department of Physics, Durham University, South Road, Durham DH1 3LE, UK\\
$^{15}$Max-Planck-Institut f\"{u}r Astronomy, K\"{o}nigstuhl 17, D-69117, Heidelberg, Germany\\
$^{16}$Leiden Observatory, Leiden University, P.O. Box 9513, 2300 RA Leiden, The Netherlands
}

\date{Accepted XXX. Received YYY; in original form ZZZ}

\pubyear{2024}

\begin{document}
\label{firstpage}
\pagerange{\pageref{firstpage}--\pageref{lastpage}}
\maketitle

\begin{abstract}
Submillimetre galaxies (SMGs) are some of the most extreme star-forming systems in the Universe, whose place in the framework of galaxy evolution is as yet uncertain. It has been hypothesised that SMGs are progenitors of local early-type galaxies, requiring that SMGs generally reside in galaxy cluster progenitors at high redshift. 
We test this hypothesis and explore SMG environments using a narrowband VLT/HAWK-I+GRAAL study of \Ha\ and \OIII\ emitters around an unbiased sample of three ALMA-identified and spectroscopically-confirmed SMGs at $z \sim 2.3$ and $z \sim 3.3$, where these SMGs were selected solely on spectroscopic redshift.
Comparing with blank-field observations at similar epochs, we find that one of the three SMGs lies in an overdensity of emission-line sources on the $\sim4$~Mpc scale of the HAWK-I field of view, with overdensity parameter $\delta_{g} = 2.6^{+1.4}_{-1.2}$. A second SMG is significantly overdense only on $\lesssim 1.6$~Mpc scales and the final SMG is consistent with residing in a blank field environment. 
The total masses of the two overdensities are estimated to be 
$\log(M_{h}/{\rm M}_{\odot}) =$~12.1--14.4, 
leading to present-day masses of 
$\log(M_{h,z=0}/{\rm M}_{\odot}) =$~12.9--15.9. 
These results imply that SMGs occupy a range of environments, from overdense protoclusters or protogroups to the blank field, suggesting that while some SMGs are strong candidates for the progenitors of massive elliptical galaxies in clusters, this may not be their only possible evolutionary pathway.
\end{abstract}




\section{Introduction}
\label{S:intro}

Since their discovery more than two decades ago, submillimetre galaxies \citep[SMGs; e.g.][]{Smail+1997,Barger+1998,Hughes+1998,Eales+1999,Blain+2002, Coppin+2006,Casey2014,Hodge&daCunha2020} have proven to be important laboratories for exploring galaxy formation and evolution. 
These galaxies are identified in (sub)millimetre surveys and have typical infrared (IR) luminosities of $L_{\rm IR} \sim 10^{12-13}\,L_{\odot}$ corresponding to star formation rates of ${\sim}10^{2-3}$~\Msun\,yr$^{-1}$ \citep[e.g.][]{Chapman+2005,Pope+2006,Wardlow+2011,Magnelli+2012,Swinbank+2014,Mackenzie+2017,Michalowski+2017,RowanRobinson+2018,Cheng+2019,Greenslade+2020}. 
They are massive, with stellar masses of $M_{\ast} \sim 10^{10-11}$~\Msun\  \citep[e.g.][]{Hainline+2011,Michalowski+2012,Gruppioni+2013,Simpson+2014,daCunha+2015,Dudzeviciute+2020a,Pantoni+2021}, 
dust masses of ${\gtrsim}10^8$~\Msun\  \citep[e.g.][]{Clements+2010,Miettinen+2017,Dudzeviciute+2020a,Pantoni+2021}, 
and cold gas masses of ${\sim}10^{11}$~\Msun\ \citep[e.g.][]{Greve+2005,Tacconi+2006,Bothwell+2013, Birkin+2021}, 
but with gas depletion times of just a few hundred Myr \citep[e.g.][]{Tacconi+2006,Birkin+2021}. 
The SMG redshift distribution peaks at $z \sim  2.5$ \citep[e.g.][]{Chapman+2005,Wardlow+2011,Koprowski+2014,Danielson+2017,Smith+2017,Stach+2019,Dudzeviciute+2020a,daCunha+2021}, making these massive dusty galaxies the most intense star-forming systems in the Universe during its peak epoch of star formation \citep{Madau+Dickinson2014}. 
SMGs contribute up to ${\sim}20$ percent of the cosmic star-formation rate density at $z \sim 2$ \citep[e.g.][]{Coppin+2006,Barger+2012,Swinbank+2014}.

The extreme properties of SMGs has long made them a good test of galaxy evolution models \citep[e.g.][]{Baugh+2005,Lacey+2008,Lacey+2010,Dave+2010,Narayanan+2010,Narayanan+2015,Bethermin+2011,Niemi+2012,Hayward+2021,Lovell+2021}, yet questions about their evolution and role in the evolution of other galaxies remain. 
SMGs have similar properties to those expected of the progenitors of local massive elliptical galaxies, which formed most of their stars in short bursts at $z \gtrsim 2$ \citep{Ellis+1997,Blakeslee+2003}. Indeed, the dust emission from SMGs is typically compact \citep[e.g.][]{Hodge+2016,Gullberg+2019}, which is consistent with a scenario in which a gas-rich $z\gtrsim2$ galaxy undergoes a compact starburst, leading to a compact quiescent galaxy, which eventually evolves into a local elliptical galaxy \citep{Toft+2014,Simpson+2014,Ikarashi+2015}. 
Since local ellipticals are predominantly found in galaxy clusters \citep[e.g.][]{Dressler1980} then if SMGs are indeed a progenitor phase in their formation, then it is expected that SMGs should reside in early galaxy clusters, or `protoclusters', at  $z \gtrsim 2$. 

Galaxy protoclusters \citep[for a review, see][]{Overzier2016} are typically defined as structures that will collapse and virialise to form a galaxy cluster by $z = 0$. Simulations have shown that in a $\Lambda{\rm CDM}$ universe, protoclusters form hierarchically at the highest density regions of the matter distribution in the universe \citep[the `cosmic web';][]{Bond+1996} at $z \sim$ 4--6 \citep[e.g.][]{Baugh+1998,DeLucia+2007b}. As such, protoclusters are characterised by overdensities of galaxies relative to the average galaxy density in the coeval blank field. Conversely to their present-day descendants, galaxies in a protocluster are generally not bound to a single halo; they instead occupy large structures extended over megaparsec (Mpc) scales, with the main halo containing as little as 20 percent of the member galaxies \citep[e.g.][]{Chiang+2013,Muldrew+2015}.

Unfortunately, observationally identifying protoclusters is challenging. Methods of detecting galaxy clusters from their X-ray emission \citep[e.g.][]{Truemper1993,Bohringer+2001,Henry+2006,Pacaud+2016} or by searching for their imprint on the cosmic microwave background at millimetre wavelengths \citep[e.g.][]{Staniszewski+2009,Williamson+2011,Hasselfield+2013,Bleem+2015,Planck2016} via the Sunyaev-Zel'dovich effect \citep{SunyaevZeldovich1972} are rendered impractical due to the lack of a hot intracluster medium (ICM). Similarly, searches for high concentrations of passive early-type galaxies occupying a tight `red sequence' in colour-magnitude space \citep[e.g.][]{GladdersYee2000,GladdersYee2005,Muzzin+2009,Wilson+2009,Gilbank+2011} become ineffective since the stellar populations of galaxies in protoclusters typically have not evolved sufficiently for a significant 4000~\AA\ break to be detected. 
Consequently, the majority of protocluster surveys resort to searching for overdensities of galaxies at high redshift. Such searches depend on the existence of accurate redshift information across large cosmological volumes, and several protoclusters have been discovered serendipitously through large spectroscopic surveys \citep[e.g.][]{Steidel+1998,Steidel+2000,Steidel+2005,Cucchiati+2014,Lemaux+2014}. In lieu of expensive large-scale spectroscopic observations, an alternative method is to use wide-field narrowband photometric surveys to search for overdensities of galaxies with strong emission lines at a particular observed-frame wavelength \citep[e.g. \Lya\ or \Ha\ emitters;][]{Venemans+2002, Shimasaku+2003, Matsuda+2004, Palunas+2004, Venemans+2005, Hatch+2011, Kuiper+2011a, Matsuda+2011, Tanaka+2011, Hayashi+2012, Koyama+2013b, Zheng+2021}. 

Whether SMGs commonly inhabit protoclusters or protocluster-like environments is as yet uncertain. Several examples of SMGs residing in protoclusters have been documented \citep[e.g.][]{Ivison+2000, Smail+2003a, Geach+2005, Daddi+2009, Matsuda+2011, Ivison+2013, Casey+2015}, but these systems were selected for detailed follow-up because of prior evidence of high galaxy densities. I.e.\ they comprise a highly biased subset and therefore cannot be used to make inferences about the general SMG population. 

Clustering studies have been used to obtain statistical measurements indicative of the whole SMG population. Results from single-dish clustering measurements suggest that on average SMGs at $z \sim 2.5$ reside in dark matter halos of mass ${\sim}10^{13}$~\Msun\ \citep[e.g.][]{Hickox+2012,Wilkinson+2017}. 
This is marginally lower than expected for the progenitors of massive ellipticals, and implies that SMGs are instead more likely to evolve into 2--3$L^{\ast}$ galaxies in groups and small clusters. 
However, these halo mass measurements have typical uncertainties of ${\sim}0.5$~dex due to the difficulties associated with obtaining accurate photometric redshifts for SMGs. Furthermore, these clustering measurements relied on the statistical identification of optical/near-IR counterparts to submillimetre sources detected in low-resolution single-dish surveys, which are incorrect in ${\sim}30$~percent of cases and incomplete in a further ${\sim}30$~percent \citep[e.g.][]{Hodge+2013a}. 
More recently, \citet{Garcia-Vergara+2020} and \citet{Stach+2021} measured the clustering of SMGs which had been followed up interferometrically with the Atacama Large Millimetre/submillimetre Array (ALMA). 
Using a small sample of 50 ALMA-identified SMGs with spectroscopic redshifts and employing forward modelling to correct for incompleteness, \citet{Garcia-Vergara+2020} estimated halo masses that are a factor of $\sim 3.8$ lower than other studies of SMGs.  
From a significantly larger parent sample, \citet{Stach+2021} selected a complete sample of $\sim350$ ALMA-identified SMGs with photometric redshifts to measure halo masses consistent with the results from the single-dish studies, particularly at $z>2$ \citep{Hickox+2012,Wilkinson+2017}. 
Overall, the picture from clustering measurements is complex, and differing results from different studies may be methodological, due to sample selection or cosmic variance. Other ways of measuring the environments of SMGs are required.

Statistical photometric redshifts have identified galaxy overdensities around ${\sim}$5--60~percent of SMGs \citep[e.g.][]{Davies+2014,Simpson+2014,Smolcic+2017a}, but these are subject to significant selection biases \citep[e.g.\ see Section~6 in][]{Smolcic+2017a}, and few overdensities have been spectroscopically confirmed.
Instead, in order to determine the nature of a `typical' SMG environment, and thus confirm whether SMGs really are the progenitors of massive elliptical galaxies in local clusters, we need targeted observations of individual SMGs, but with no prior knowledge of their environments in order to avoid biases. To this end, we have conducted a wide-field narrowband survey of the environments of three SMGs identified as part of the ALESS project \citep{Hodge+2013a},
in which follow-up observations of submillimetre sources detected in the LABOCA ECDFS Submillimetre Survey \citep[LESS;][]{Weiss+2009} were conducted using ALMA. We search for overdensities of \Ha\ or \OIII\ emitters around these three SMGs to assess whether they reside in protocluster-like environments. The target SMGs were selected on the basis of redshift only, and with no prior information about their environments. 
Our method is similar to that employed by \citet{Matsuda+2011}, which combined narrowband photometry with submillimetre observations to identify an overdensity of \Ha\ emitters around two SMGs in SSA\,13. However, the SMGs targeted by \citet{Matsuda+2011} were already known to be closely grouped with each other and three optically-faint radio-galaxies. Our study is the first to perform such an analysis around a sample of  SMGs that are selected without prior knowledge of their environment.

The structure of this paper is as follows: in \S2 we describe the SMG sample selection, our observations and data reduction, and the identification of candidate companion galaxies for each target SMG; \S3 includes our main results, analysis and discussion; our conclusions are presented in \S4. Throughout this paper we use a $\Lambda$CDM cosmology with $\Omega_{\textrm{M}} = 0.315$, $\Omega_{\Lambda} = 0.685$ and $H_{0} = 67.4$ km s$^{-1}$ Mpc$^{-1}$ \citep{Planck2018}. 
All magnitudes are presented in the AB system, where a $1~\mu$Jy source has a magnitude of $23.9$ \citep{Oke+Gunn1983}.

\section{Observations and galaxy identification}

In this study we use the High Acuity Wide-field K-band Imager \citep[HAWK-I;][]{HAWKI2004,HAWKI2006,HAWKI2008,HAWKI2011} on the Very Large Telescope (VLT) to investigate the environments of three ALMA-identified SMGs from ALESS. 
As part of ALESS, extensive follow-up was undertaken to obtain spectroscopic redshifts of the SMGs \citep[][]{Danielson+2017, Birkin+2021}, 
which enables a search for galaxies that share environments with these submillimetre sources. 
The wide-field imaging capabilities and narrowband photometric filters of HAWK-I allow for an efficient emission-line survey of their environments, which are expected to span physical scales on the order of several Mpc if consistent with being protoclusters \citep[e.g.][]{Chiang+2013,Muldrew+2015,Yajima+2022}.

\subsection{Sample selection}\label{SS:samp_select}

The blank-field LABOCA ECDFS Submillimetre Survey (LESS) observed  $0.5\times0.5$~degrees in ECDFS with  APEX/LABOCA  and detected 126  sources at ${>}3.7\sigma$ at 870-$\micron$ \citep{Weiss+2009}. Each of these sources was followed-up with ALMA to yield the 131 ALESS sources described in \citet{Hodge+2013a}, divided into a main catalogue of 99 SMGs and a supplementary catalogue of 32 SMGs. The SMGs in the main catalogue all lay within the ALMA primary beam full width at half-maximum (FWHM) of the highest-quality maps, while those in the supplementary catalogue were either extracted from outside the primary beam or from lower-quality maps \citep{Hodge+2013a}. 

Spectroscopic redshifts were obtained for 52 of the 131 ALESS SMGs by \citet{Danielson+2017},  
and targets for our study are selected from these 52 ALMA-identified and spectroscopically-confirmed SMGs. 
We require SMGs with spectroscopic redshifts that shift the \Ha\ or \OIII$\lambda5007$ emission lines into the wavelength coverage of the HAWK-I \BrG\ filter; this requires that the SMGs are located at $z = 2.299\pm0.023$ or $z=3.324\pm0.060$. Of the 52 ALMA-identified SMGs with spectroscopic redshifts from \citet{Danielson+2017}, five (ALESS\,6.1, 75.2, 87.1, 102.1, and 112.1) have spectroscopic redshifts within the desired range for H$\alpha$. A sixth SMG (ALESS\,5.1) has a CO-derived spectroscopic redshift of  $z = 3.303$ \citep{Birkin+2021}, which places \OIII\ in the \BrG\ coverage.

These six SMGs were the proposed targets for observations in four HAWK-I pointings (PID: 0103.A-0668). The six SMGs were selected purely based on their spectroscopic redshifts, with no consideration of their environments. Of the four pointings, only two were  observed during the service-mode observations and the choice of pointings was random. The two observed pointings contain three of the six proposed targets: ALESS\,5.1, ALESS\,75.2, and ALESS\,102.1, whose spectroscopic redshifts are $z = 3.303$, 2.294, and 2.296, respectively. 
Details of these three targeted SMGs are provided in Table \ref{tab:SMG_info}. A total of 16 other ALESS SMGs lie within the two HAWK-I pointings, but these are not considered in this study as their redshifts are such that no bright emission lines are expected in the narrowband filter. 
Indeed, those that are detected in our HAWK-I observations fail to meet our criteria for being emission-line galaxies (see \S\ref{SS:NB_selection} and Figure \ref{fig:NB_selection}).

\begin{table}
	\centering
	\caption[]{Details of each of the three SMGs included in our sample.}
	\label{tab:SMG_info}
	\begin{tabular}{ccccc}
		\hline
		SMG 	& $z_{\textrm{spec}}$ & $S_{870}$ [mJy]$^{a}$ & mag \Ks$^{b}$ & Target line$^{c}$\\
		\hline
		\hline
		ALESS 5.1	&	$3.303^{d}$ & $7.8 \pm 0.7$ & $19.79 \pm 0.01$	&	\OIII$\lambda5007$ \\
		ALESS 75.2	&	$2.294^{e}$ & $5.0 \pm 1.2$ & $20.67 \pm 0.01$	&	\Ha \\
		ALESS 102.1	&	$2.296^{f}$ & $3.1 \pm 0.5$ & $21.07 \pm 0.08$	&	\Ha \\
		\hline
	\end{tabular}
	\begin{flushleft}
        $^{a}$ Primary-beam-corrected ALMA 870 \um\ flux densities from \citet{Hodge+2013a}.\\
		$^{b}$ From \citet{Simpson+2014}.\\
		$^{c}$ The emission line used in this study to identify companion galaxies for each SMG, exploiting the fact that these lines shift into the wavelength coverage of the HAWK-I \BrG\ filter at the redshifts of the SMGs (see \S\ref{SS:samp_select}).\\
		$^{d}$ Obtained via detection of the CO(4--3) and \CII\ emission lines \citep{Birkin+2021}.\\
		$^{e}$ Based on \Ha+\NII\ and \SII\ detections \citep{Danielson+2017}.\\
		$^{f}$ Determined using a combination of \Lya, \CIII\ and continuum measurements \citep{Danielson+2017}.
	\end{flushleft}
\end{table}

\subsection{HAWK-I data}\label{SS:HAWKI_data}

Each pointing was imaged using the HAWK-I instrument \citep{HAWKI2004,HAWKI2006,HAWKI2008,HAWKI2011} on the VLT in the \Ks\ (central wavelength: $\lambda_{c} = 2.146$~\um; FWHM: $\Delta\lambda = 0.324$~\um) and \BrG\ ($\lambda_{c} = 2.165$~\um; $\Delta\lambda = 0.030$~\um) filters \citep{HAWKI2008}. The FWHM of the \BrG\ filter is equivalent to $\Delta z = 0.046$ (0.060) at $z \sim 2.3$ (3.3), corresponding to a velocity width of $\Delta v = 4200$~km~s$^{-1}$. All three observing blocks (OBs) for the field containing ALESS~5.1 and 75.2 (hereafter Pointing~5+75) were executed on 2019 August 21, while the OBs for the field containing ALESS~102.1 (hereafter Pointing~102)  were split among three separate nights from 2019 August 21 to 2020 January 01. The total exposure times for Pointing~5+75 (Pointing~102) were 7.2~ks (6.6~ks) and 900~s (660~s) in the \BrG\ and \Ks\ filters, respectively. 
Individual exposures of 120~s (\BrG) and 30~s (\Ks) were 
taken using the ``HAWKI img obs AutoJitter'' template, with five random 
dither positions within a 20\arcsec\ box for each filter in each OB. 
Each pointing covers a $7\farcm5 \times 7\farcm5$ area, except for a cross-shaped gap of width 15{\arcsec} between the detector's four 2k $\times$ 2k Hawaii 2RG arrays. Using HAWK-I's GRAAL system \citep[GRound layer Adaptive optics system Assisted by Lasers;][]{HAWKI_GRAAL2008,HAWKI_GRAAL2010}, we achieved point spread functions (PSFs) of ${\sim}0\farcs4$ in \BrG\ and ${\sim}0\farcs3$ in \Ks\ (see Table~\ref{tab:lim_mags}). 

The data were reduced using a custom Python-based pipeline, with each of the four detector chips treated separately. Briefly, the pipeline begins by dark-subtracting the data and subsequently using twilight flats to perform flat-fielding. We then use {\sc{SExtractor}} \citep{SExtractor1996} to detect sources in each of the flattened frames and produce individual masks. 
A final flat field is produced for each frame by median combining all masked frames from the same OB except the frame being flattened; the frames are then flattened using their unique final flat fields. 
The astrometry of each flattened frame is then calibrated by using {\sc{scamp}} \citep{SCAMP2006} to match our detected sources with sources detected in a reference $K$-band image, correcting for any distortions across the field of view by fitting a third-order polynomial. The reference images used for Pointing~5+75 and Pointing~102 were taken from the Taiwan ECDFS NIR Survey \citep[TENIS;][]{TENIS2012} and the MUltiwavelength Survey by Yale-Chile \citep[MUSYC;][]{MUSYC2006b,MUSYC2009}, respectively. Different reference images were required for each pointing because while TENIS is deeper and has higher resolution, roughly a quarter of the Pointing~102 field of view lies outside of the TENIS coverage. Finally, the astrometrically-corrected frames were median combined using {\sc{SWarp}} \citep{SWarp2010}. The resultant stacks in both bands were photometrically calibrated using MUSYC \Ks\ data \citep{MUSYC2009,Simpson+2014} such that they all had a zeropoint magnitude of 30.0 mag and ensure a median (\Ks\ -- \BrG) colour of 0.

Source detection and photometry were conducted using {\sc{SExtractor}} \citep{SExtractor1996} operating in dual-image mode; the stacked \BrG\ images were used to identify the positions of sources, and then photometry was extracted at these positions in both the \BrG\ and \Ks\ images to ensure that any difference between the measured \BrG\ and \Ks\ photometry is a purely intrinsic property of the sources and not caused by positional offsets. After masking noisy regions near the edges of the stacked images, we detected a total of 2175 sources in Pointing~5+75 and 1754 in Pointing~102. Apertures with a diameter of $1\farcs25$ were used, as this is large enough to contain the majority of the flux for all detected sources while minimising the amount of additional background noise captured. The $3\sigma$ limiting magnitudes measured in these $1\farcs25$ diameter apertures are provided for each filter and each HAWK-I detector chip in Table \ref{tab:lim_mags}. To account for the variation in size of the detected sources, we then estimated total magnitudes in each filter by selecting all bright ($m_{K_{s}} < 19.5$) sources and (for each pointing separately) calculating the median difference between their fixed-aperture magnitudes and their magnitudes measured by {\sc{SExtractor}} in adaptively-scaled \citep{Kron1980} apertures ({\sc mag\_auto}; \citealt{SExtractor1996}); this difference was then added to the fixed-aperture magnitudes of all sources in the pointing to obtain their total magnitudes. 
The choice to use the same aperture size for all sources and apply a correction (as opposed to simply using the {\sc mag\_auto} values) ensures that estimates of the total magnitudes are self-consistent whilst also closely matching existing photometry in the same band.

\begin{table}
	\centering
	\caption[]{Limiting $3\sigma$ AB magnitudes and resolution for each stacked image. Limiting magnitudes are measured using randomly-placed $1\farcs25$ diameter apertures. Due to variation in the four HAWK-I detector chips, each quadrant is considered separately.}
	\label{tab:lim_mags}
	\begin{tabular}{cccccc}
		\hline
		Pointing 			& Quadrant$^{a}$ & \multicolumn{2}{c}{$m_{\textrm{lim}}^{3\sigma}$} & \multicolumn{2}{c}{PSF ($\arcsec$)} \\
				 			& 		   & \BrG 		 & \Ks 		& \BrG	& \Ks   \\
		\hline
		\hline
		\multirow{4}{*}{Pointing~5+75} & Q1	   & 24.29		 & 24.01	& 0.28	& 0.27  \\
							& Q2	   & 24.19		 & 24.27		& 0.27	& 0.27	\\
							& Q3	   & 24.30		 & 24.19		& 0.35	& 0.26	\\
							& Q4	   & 24.22		 & 24.27		& 0.32	& 0.26	\\
		\hline
		\multirow{4}{*}{Pointing~102} & Q1	   & 24.09		 & 23.85	& 0.38	& 0.30     \\
							& Q2	   & 24.08		 & 24.00	& 0.37	& 0.30	\\
							& Q3	   & 24.19		 & 24.01	& 0.44	& 0.30	\\
							& Q4	   & 24.11		 & 24.09	& 0.41	& 0.29	\\
		
		\hline
	\end{tabular}
  \begin{flushleft}
    $^{a}$ Quadrants are assigned the same labels as in \citet{HAWKI2008}.
  \end{flushleft}
\end{table}

\subsection{Ancillary data}\label{SS:ancillary}

There exists a wealth of photometric data in the ECDFS, which  supplements our HAWK-I photometry. Archival TENIS \citep{TENIS2012}, MUSYC \citep{MUSYC2009} and HAWK-I (Zibetti, priv. comm.) \Ks\ data were collated by \citet{Simpson+2014} (hereafter S14) and then used to calibrate our astrometry and photometry (see \S\ref{SS:HAWKI_data}). 

In \S\ref{SS:contaminants} and \S\ref{SSS:MS} we fit spectral energy distributions to galaxies in our sample in order to first derive properties such as photometric redshift, stellar mass and star formation rate. For the photometric redshifts we use \eazypy\footnotemark \footnotetext{\url{https://github.com/gbrammer/eazy-py}} -- an updated version of the photometric redshift code {\sc{eazy}} \citep{Brammer+2008} written in {\sc Python} (see \S\ref{SS:contaminants}) -- while for the other galaxy properties we use \magphys\ \citep[][see \S\ref{SSS:MS}]{daCunhaMAGPHYS+2008}.
To this end, we also make use of existing ECDFS images spanning the ultraviolet (UV) to mid-infrared (MIR; see \S\ref{SS:contaminants}). 
These images were sourced either from the MUSYC 2010 Public Data Release \citep{Cardamone+2010} or from TENIS \citep{TENIS2012}. 
The MUSYC dataset consists of $UU_{38}BVRI$ broadband images from the Wide Field Imager (WFI) on the MPG/ESO 2.2m telescope \citep{Hildebrandt+2006}; 5000 \AA\ narrowband and $z^{\prime}$ broadband imaging from the Mosaic-II camera on the CTIO Blanco 4m telescope \citep{MUSYC2006b,MUSYC2006a}, with $JK_{s}$ broadband imaging from the Infrared Sideport Imager on the same telescope \citep{MUSYC2009}; 18 medium-band (IA427, IA445, IA464, IA484, IA505, IA527, IA550, IA574, IA598, IA624, IA651, IA679, IA709, IA738, IA767, IA797, IA827, IA856) images taken with the Subaru telescope's Suprime-Cam \citep{Cardamone+2010}; \textit{Spitzer}/IRAC images at 3.6, 4.5, 5.8 and 8.0 \um\ \citep{Cardamone+2010,SIMPLE2011}\footnotemark\footnotetext{$H$-band data are also available but our pointings are not covered.}.

We also make use of spectroscopic redshifts in the ECDFS from studies whose areas overlap with our pointings, obtained from publicly-available composite catalogues\footnotemark\footnotetext{\url{https://www.eso.org/sci/activities/garching/projects/goods/MasterSpectroscopy.html}}$^{,}$\footnotemark\footnotetext{\url{http://member.ipmu.jp/john.silverman/CDFS_vlt.html}} \citep{Silverman+2010}. The spectroscopic redshifts used are from the VIMOS VLT Deep Survey \citep[VVDS;][]{LeFevre+2005}; the GOODS/VIMOS survey \citep{Popesso+2009,Balestra+2010}; the Extended \textit{Chandra} Deep Field-South Survey \citep{Silverman+2010}; and \citet{Treister+2009}. 
Additionally we utilise the results of the spectroscopic study conducted as part of ALESS by \citet{Danielson+2017}.

Finally, we make use of the \citet{Lehmer+2005} {\it Chandra} point source catalogue for the identification of AGN in our final sample (see \S\ref{SS:contaminants}).

\subsection{Emission line galaxy selection}\label{SS:NB_selection}

Star-forming galaxies at the same redshifts as our target SMGs ($z \sim 2.295 \pm 0.023$ for ALESS~75.2 and ALESS~102.1; $z \sim 3.324 \pm 0.030$ for ALESS~5.1) have emission lines that are redshifted into the narrow wavelength coverage of the \BrG\ filter. Since the \BrG\ filter is near the centre of the \Ks\ transmission a galaxy without line emission at these wavelengths will have a (\Ks--\BrG) colour of zero. However, due to the narrow width of the \BrG\ filter relative to the \Ks\ filter, galaxies with redshifts that place an emission line in the narrow \BrG\ filter will  have a (\Ks--\BrG) colour that is significantly greater than zero. 
We employ the same methodology as previous narrowband surveys \citep[e.g.][]{Moorwood+2000,Geach+2008,Sobral+2013} to identify line-emitting galaxies. This methodology uses two parameters to select sources with a significant, physically-driven narrowband excess, as opposed to an excess due to random noise. 

\begin{figure*}
  \includegraphics[width=0.95\columnwidth]{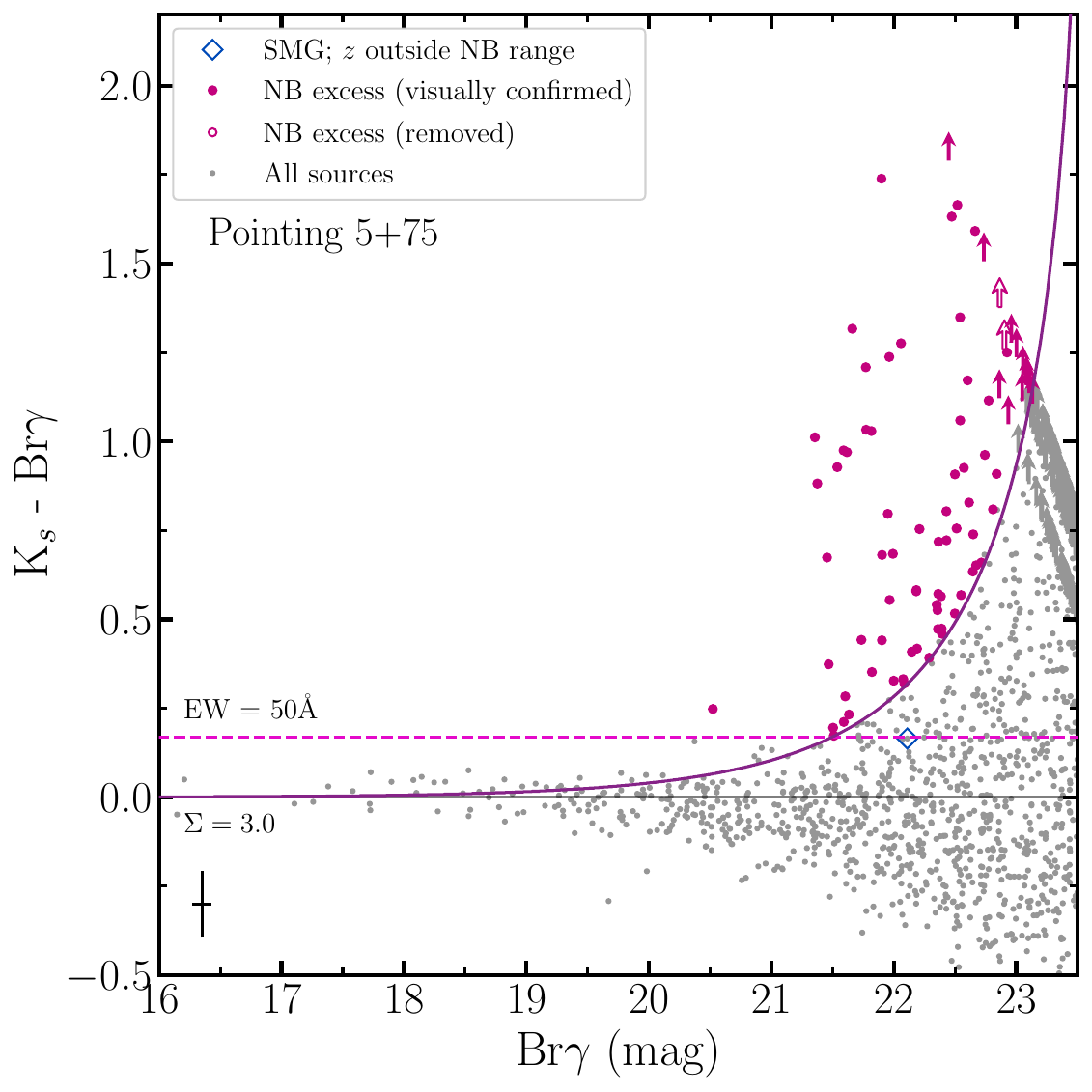}
  \includegraphics[width=0.95\columnwidth]{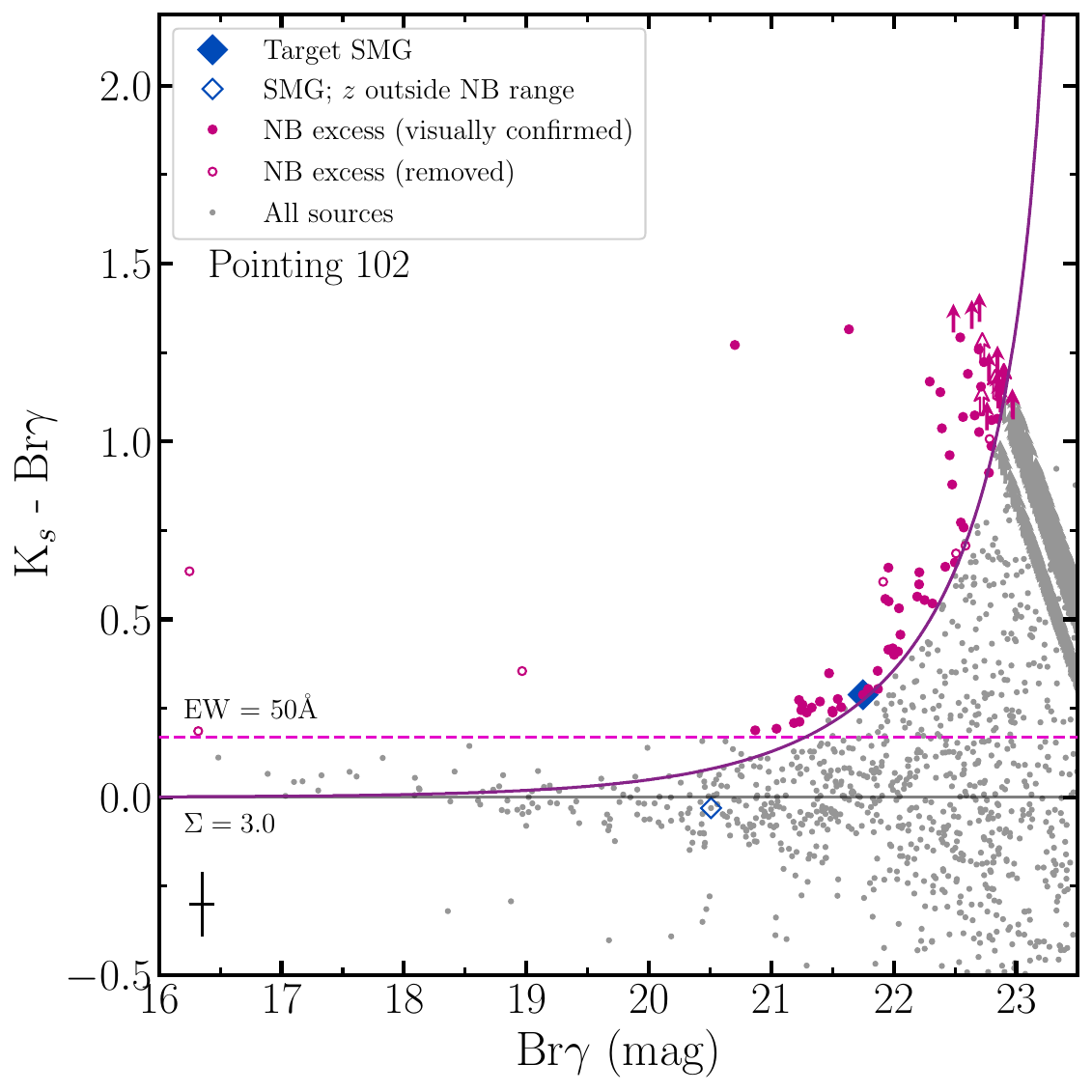}
  \caption{Colour-magnitude diagrams demonstrating the criteria described in \S\ref{SS:NB_selection}  for the selection of candidate narrowband emitters in the HAWK-I pointings containing ALESS~5.1 \& 75.2 (left), and ALESS~102.1 (right). All sources detected in the \BrG\ observations are shown and sources identified as narrowband emitters are highlighted. Open symbols represent candidate narrowband emitters which were removed from the sample following visual checks. Sources that are detected in \BrG\ but are undetected in our \Ks\ data and have no \Ks\ photometry in S14, are shown as lower limits. The $\Sigma = 3$ curve for the average properties and the observed-frame equivalent width cut for each field are shown. The solid horizontal line marks a \Ks$-$\BrG\ colour of zero. The target SMGs and other ALESS SMGs with counterparts in our HAWK-I data are highlighted. Two of the target SMGs (ALESS~5.1 and 75.2) and several other ALESS SMGs in these fields are not detected and are therefore not shown. The black cross in the bottom left corner of each panel shows the mean uncertainties in colour and \BrG\ magnitude. } 
  \label{fig:NB_selection}
\end{figure*}

The first of these parameters, $\Sigma$, quantifies the significance of the narrowband excess compared to the expected random scatter for a source with zero (\Ks--\BrG) colour \citep{Bunker+1995}. $\Sigma$ is given by:
\begin{equation}\label{eq:Sigma}
  \Sigma = \frac{1 - 10^{-0.4(\textrm{BB} - \textrm{NB})}}{10^{-0.4(\textrm{ZP} - \textrm{NB})}\sqrt{\textrm{rms}_{\textrm{NB}}^{2} + \textrm{rms}_{\textrm{BB}}^{2}}}\,,
\end{equation}
\noindent where NB and BB are the apparent magnitudes in the narrowband (\BrG) and broadband (\Ks) filters, respectively; ZP is the zeropoint magnitude of the narrowband images; $\textrm{rms}_{\textrm{NB}}$ and $\textrm{rms}_{\textrm{BB}}$ are the rms counts in $1\farcs25$ apertures for the individual narrowband and broadband quadrants, respectively. 
We require candidate line-emitters  have $\Sigma > 3$, which is consistent with previous narrowband studies \citep[e.g.][]{Bunker+1995,Sobral+2013}; see Figure \ref{fig:NB_selection}. Note that this $\Sigma$ does not correspond directly to signal-to-noise (SNR) in the \BrG\ filter, but is a separate quantity based on counts; $\Sigma > 3$ implicitly excludes sources with SNR $\lesssim 8$ in \BrG\ \citep[for details see e.g.][]{Sobral+2009}. 

In addition to having $\Sigma > 3$ line emitters are required to have an observed equivalent width (EW) $> 50$\AA. The EW is calculated for each source using:
\begin{equation}
  \textrm{EW} = \Delta\lambda_{\textrm{Br}\gamma} \frac{f_{\textrm{Br}\gamma} - f_{K_s}}{f_{K_s} - f_{\textrm{Br}\gamma} (\Delta\lambda_{\textrm{Br}\gamma}/\Delta\lambda_{K_s})}\,,
\end{equation}
\noindent where $\Delta\lambda_{\textrm{Br}\gamma}$ and $\Delta\lambda_{K_s}$ are the widths of the two filters and $f_{\textrm{Br}\gamma}$ and $f_{K_s}$ are the flux densities of the source in each filter. The 50 \AA\ lower limit on EW for a source to be selected as a line emitter was chosen to lie above the $3\sigma$ scatter in (\Ks--\BrG) colours for bright (\BrG\ $> 19.5$ mag) sources in both pointings (Figure \ref{fig:NB_selection}).

Before applying the selection criteria, we first account for sources that are detected with ${\geq}3\sigma$ significance in the \BrG\ filter but ${<}3\sigma$ in \Ks. We classify these sources as non-detections in \Ks, and replace their aperture magnitudes with the relevant $3\sigma$ limiting magnitude (see Table \ref{tab:lim_mags}). However, several of these non-detections have counterparts in the S14 catalogue (within a $1\arcsec$ matching radius) and thus have \Ks\ magnitudes from either TENIS \citep{TENIS2012}, archival HAWK-I observations (Zibetti, priv. comm.), or MUSYC \citep{MUSYC2009}. For these sources, we replace our HAWK-I \Ks\ photometry with values from one of these surveys, preferentially using TENIS photometry as it is the deepest of the three (with a limiting $3\sigma$ magnitude of $m_{\rm lim}^{3\sigma} = 24.45$~mag); if no TENIS photometry is available then we opt for the archival HAWK-I values ($m_{\rm lim}^{3\sigma} = 24.36$~mag), using MUSYC ($m_{\rm lim}^{3\sigma} = 22.55$~mag) only when no photometry exists for either of the other two. Note that while MUSYC \Ks\ observations are the shallowest of all the data considered here (including our own), there are 11 sources for which only MUSYC photometry is available. However, all of these sources reside in regions of Pointing~102 that are (a) outside of the coverage of the TENIS and archival HAWK-I observations, and (b) close to the quadrant edges in our HAWK-I observations where the noise is at its greatest.

Using the $\Sigma > 3$ and EW $> 50$ \AA\ selection criteria, 81 and 80 candidate line emitters are identified in Pointing~5+75 and Pointing~102, respectively (Figure \ref{fig:NB_selection}). Of these candidates, 30 are \Ks\ non-detections with no \Ks\ photometry in the S14 catalogue, and thus $\Sigma$ and EW are calculated by assuming that their \Ks\ magnitudes are equal to the $3\sigma$ limiting magnitudes of our data. Since this can only provide a lower limit for the (\Ks--\BrG) colour and thereby underestimate $\Sigma$ for these sources, we do include these sources in our sample of candidate line emitters. 

Finally, we visually inspect all 161 candidate line emitters, removing stars/quasi-stellar objects and image artefacts. 
The final sample consists of 79 and 68 candidate line emitters in Pointing~5+75 and Pointing~102, respectively (147 sources in total).

\subsection{Identifying line emitters associated with the SMGs}\label{SS:contaminants}

Narrowband excess alone is not sufficient to identify star-forming galaxies in the same environments as our target SMGs; such an excess could be caused by a number of possible emission lines at different redshifts (see Figure~\ref{fig:NB_photozs}). We therefore use the available multi-band photometric data covering our pointings to estimate photometric redshifts for the narrowband emitters in our sample. The S14 catalogue contains photometric redshift estimates for sources across the ECDFS, however after cross-matching with our data (using a matching radius of $1\arcsec$), a significant fraction (${>}30$ percent) of the line emitters identified in \S\ref{SS:NB_selection} do not have broadband counterparts in this catalogue and thus lack any redshift information. 
We therefore perform our own spectral energy distribution (SED) fitting using \eazypy.

To maximise the number of sources for which we can derive photometric redshifts, we extract fixed-aperture photometry at their HAWK-I \BrG\ positions in the UV-to-MIR images described in \S\ref{SS:ancillary}. 
Each image is astrometrically calibrated using {\sc{scamp}} \citep{SCAMP2006} and {\sc{SWarp}} \citep{SWarp2010} to match the astrometry of our HAWK-I images, and then photometrically recalibrated so that all images have a zeropoint magnitude of 30.0 mag. Photometry is extracted in fixed apertures using the {\sc{photutils}} {\sc Python} package \citep{Bradley_photutils+2022}; apertures of diameter $2\farcs0$ are used for all images except those from \textit{Spitzer}/IRAC, for which we use apertures of diameter $3\farcs8$ due to the larger point spread function (PSF). 
Aperture corrections are determined for each filter by measuring the median difference between the magnitudes measured in these apertures and those measured in adaptively scaled apertures with {\sc{SExtractor}} for bright point sources. 
Final corrections are applied to each filter to account for Galactic attenuation, using values from \citet{Cardamone+2010} and \citet{TENIS2012}.

\eazypy\ operates using a $\chi^{2}$-minimisation procedure in which linear superpositions of template SEDs are tested at different redshifts to find an optimal fit to the observed fluxes \citep{Brammer+2008}. In keeping with other recent studies which implement \eazypy\ \citep[e.g.][]{Stevans+2021,Finkelstein+2022}, we use the ``tweak\_fsps\_QSF\_12\_v3'' set of 12 template SEDs, which cover a wide range of galaxy types and utilise a \citet{Chabrier2003} initial mass function (IMF) and a \citet{KriekConroy2013} dust attenuation law while assuming solar metallicity. An advantage of these templates is that they include emission lines, such that a narrowband excess can provide a relatively tight constraint on the redshift. 

\begin{figure}
  \includegraphics[width=0.95\columnwidth]{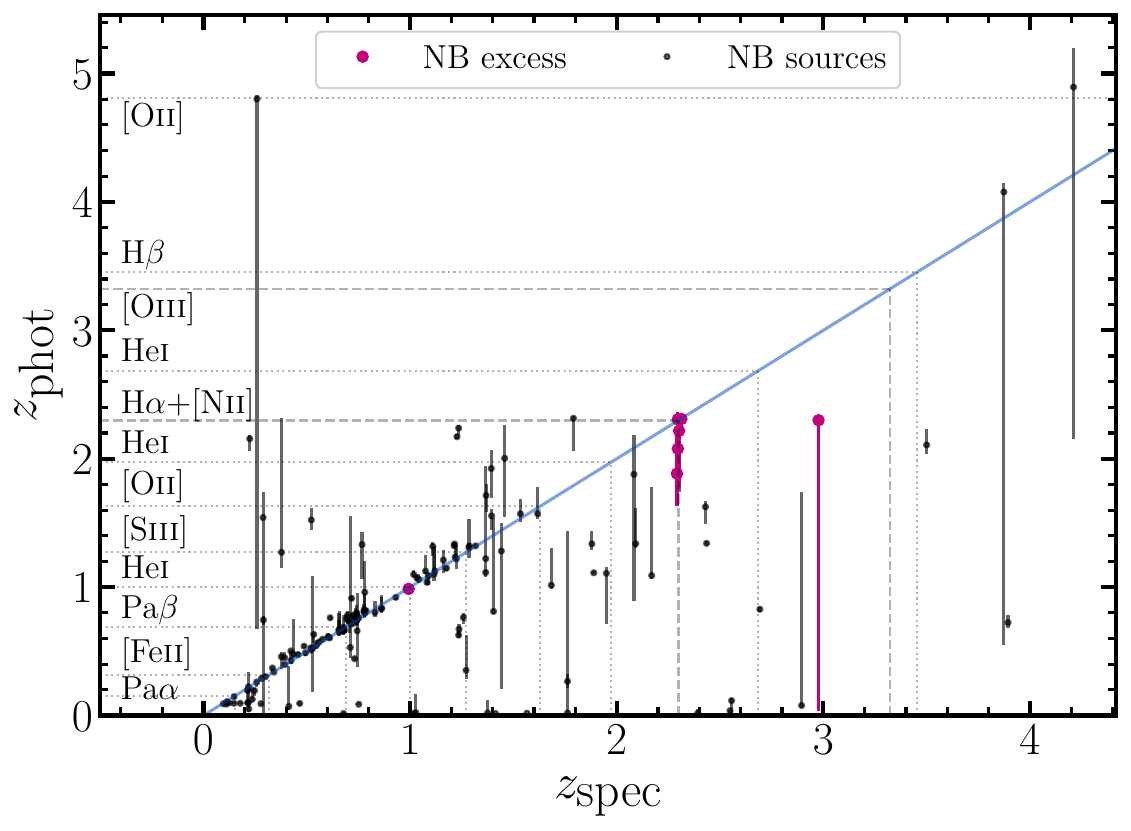}
  \caption{Photometric redshifts derived using {\sc eazy-py} compared to spectroscopic redshifts  for  all sources detected in HAWK-I \BrG\ with archival spectroscopic redshifts \citep[from ][]{LeFevre+2005,Popesso+2009,Treister+2009,Balestra+2010,Silverman+2010,Danielson+2017}. 
    Galaxies included in our final sample of candidate line emitters (see \ref{SS:NB_selection}) are highlighted. The redshifts at which common extragalactic emission lines enter the \BrG\ filter are shown using horizontal and vertical lines; dashed lines highlight \Ha\ and \OIII, which the emission lines of interest in this study. The diagonal line shows a one-to-one correspondence; the scatter is low and majority of sources have photometric redshifts that are consistent with their spectroscopic redshifts. 
    }
  \label{fig:zphot_zspec}
\end{figure}

\begin{figure*}
  \includegraphics[width=0.95\columnwidth]{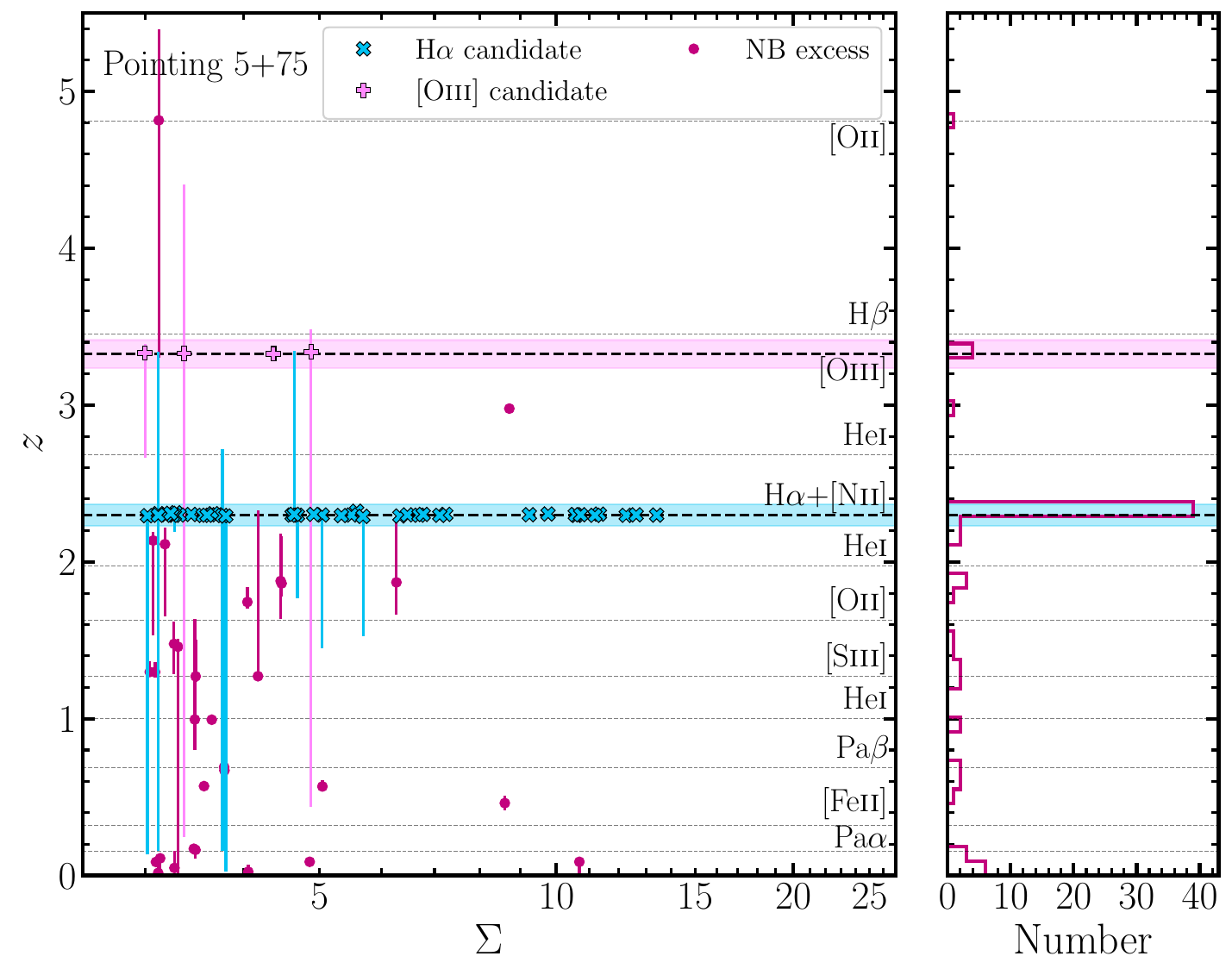}
  \includegraphics[width=0.95\columnwidth]{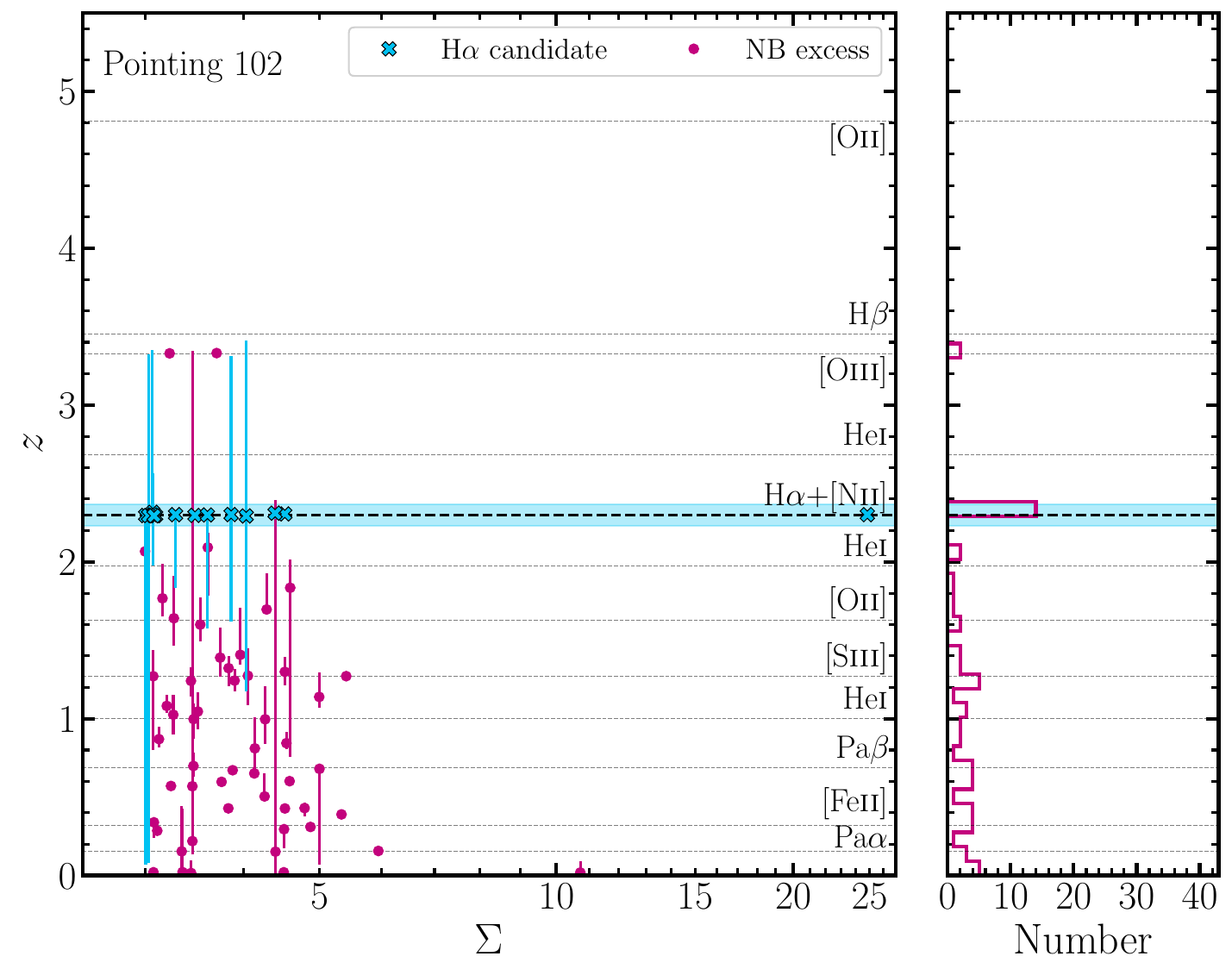}
  \caption{The distributions of redshifts for the emission line galaxies in Pointing~5+75 (left) and Pointing~102 (right) compared with their emission-line significance, $\Sigma$. Photometric redshifts are computed using {\sc{eazy-py}}, with archival spectroscopic redshifts included where available (\S\ref{SS:contaminants}). \Ha\ and \OIII\ emitters are highlighted, and shaded regions show the redshift ranges used to select them. Peaks in the redshift distributions at these redshifts may be driven by overdensities of these line emitters. Horizontal dashed lines show the redshifts at which other common extragalactic emission lines enter the \BrG\ filter. 
  }
  \label{fig:NB_photozs}
\end{figure*}

As discussed in \S\ref{SS:ancillary}, there have been several spectroscopic studies in the ECDFS, from which spectroscopic redshifts have been obtained for a number of galaxies across the field. Using a matching radius of $1\farcs5$, we cross-match our data with catalogues from VVDS \citep{LeFevre+2005}, the GOODS/VIMOS survey \citep{Popesso+2009}, the ECDFS spectroscopic survey \citep{Silverman+2010}, and the spectroscopic studies conducted by \citet{Treister+2009} and \citet{Danielson+2017}. This gives spectroscopic redshifts for 163 (${\sim}4.1$ percent) of the 3929 sources detected in our HAWK-I imaging, including nine for which photometric redshifts could not be estimated due to insufficient photometry. Seven of the 163 sources with spectroscopic redshifts are emission line galaxies selected in \S\ref{SS:NB_selection}; the spectroscopic redshifts for these sources are used for our analyses. We compare the photometric and spectroscopic redshifts for our HAWK-I sources in Figure \ref{fig:zphot_zspec}. There is strong agreement between the photometric and spectroscopic redshifts, which is quantified using the normalised median absolute deviation (NMAD) of $\Delta z$:
\begin{equation}
\label{eq:nmad}
  \sigma_{\textrm{NMAD}} = 1.48 \times \textrm{median}\left(\left|\frac{\Delta z - \textrm{median}(\Delta z)}{1 + z_{\textrm{spec}}}\right|\right)\,,
\end{equation}
\noindent where $z_{\textrm{spec}}$ is the spectroscopic redshift and $\Delta z = z_{\textrm{spec}} - z_{\textrm{phot}}$. We obtain $\sigma_{\textrm{NMAD}} = 0.062$ when considering all 152 HAWK-I detections with photometric and spectroscopic redshifts.

Only five (${\sim}3.4$ percent) of our 147 emission line galaxies have neither spectroscopic nor photometric redshifts, the latter being due to a lack of photometry with sufficient depth. Figure~\ref{fig:NB_photozs} shows the redshifts of the remaining 142 emission line galaxies, compared with the significance of their narrowband excess ($\Sigma$; Equation~\ref{eq:Sigma}). 
Peaks in the redshift distribution are visible at $z \sim 2.3$ (both pointings) and $z \sim 3.3$ (Pointing~5+75 only), as expected of \Ha\ and \OIII\ in the environments of the target SMGs. We select as \Ha\ (\OIII) emitters any galaxies for which $2.23 < z < 2.37$ ($3.23 < z < 3.41$), where these redshift ranges correspond to $3\times$ the FWHM of the \BrG\ filter when \Ha\ (\OIII) has redshifted to the centre. We represent these selection criteria with shaded regions in Figure \ref{fig:NB_photozs}; the highlighted galaxies are henceforth assumed to be \Ha\  and \OIII\ emitters at similar redshifts as the target SMGs. 
We identify 44 \Ha\ emitters and 4 \OIII\ emitters in Pointing~5+75, and in Pointing~102 there are  11 \Ha\ emitters (\OIII\ emitters in Pointing~102 are not further considered because there is no SMG at $z \sim 3.3$ in this pointing). 
Table \ref{tab:sample_summary} summarises the results of each step in the sample selection. We note that all of these \Ha\ and \OIII\ candidates have an SNR $> 8.5$ in the \BrG\ filter as a natural consequence of our selection process (see also \S\ref{SS:NB_selection}). We therefore do not expect the sizes of these samples to be significantly affected by Eddington bias \citep{Eddington1913}.

To identify any AGN in the sample we use a 1\arcsec\ matching radius to locate counterparts in the \citet{Lehmer+2005}  {\it Chandra} point source catalogue. None of the \OIII\ emitters and only one  of the \Ha\ emitters (2.3\%) is an X-ray luminous AGN, which is consistent with the rate of X-ray AGN in blank-field surveys of \Ha\ emitters at the same redshift \citep[e.g.\ $1.8\pm1.3\%$ in][]{Calhau+2017}. Since the AGN fraction is the same as in field surveys then this galaxy is kept in our sample to enable a fair like-for-like comparison between the SMG fields and blank field \Ha\ emitters.

\begin{figure*}
  \includegraphics[width=0.95\columnwidth]{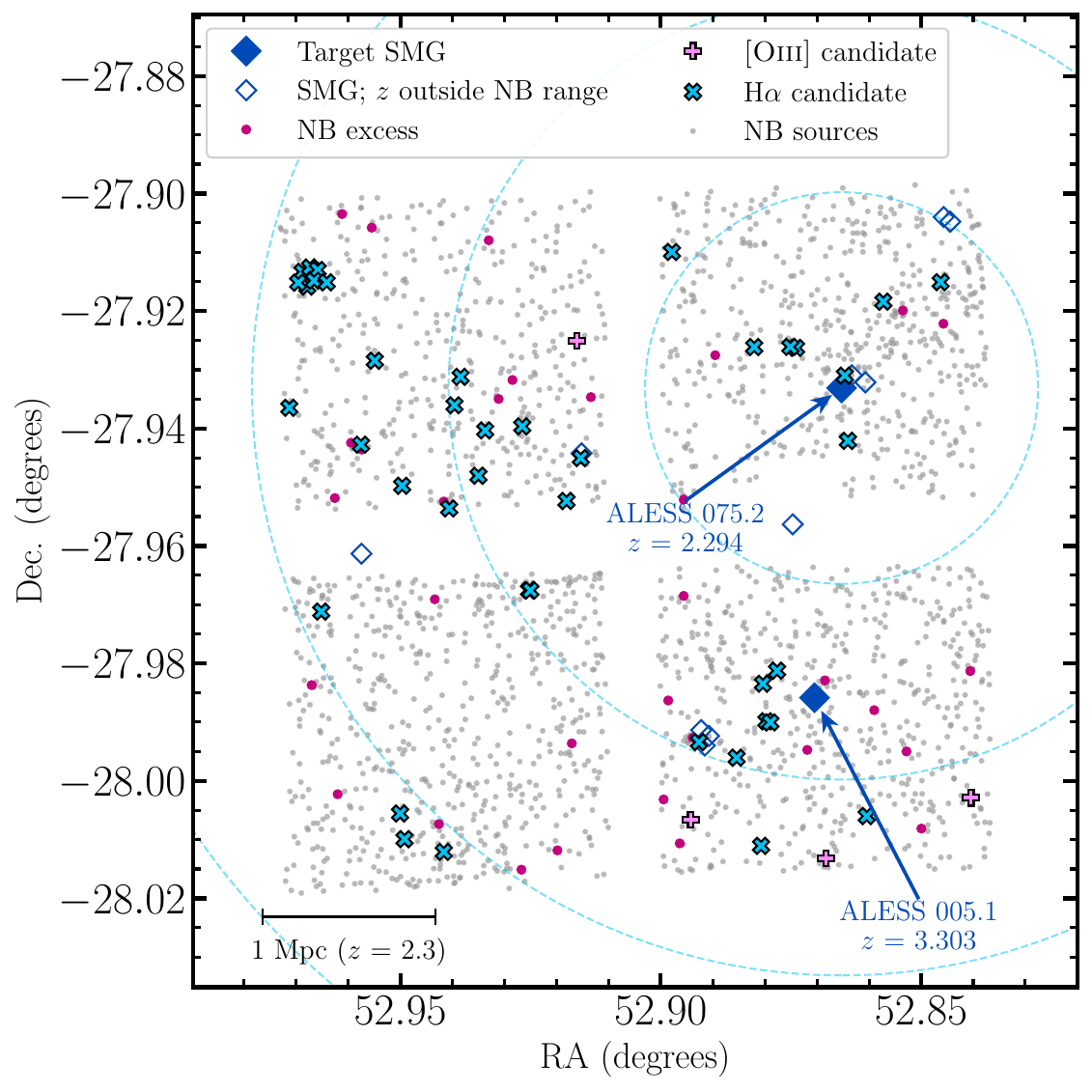}
  \includegraphics[width=0.95\columnwidth]{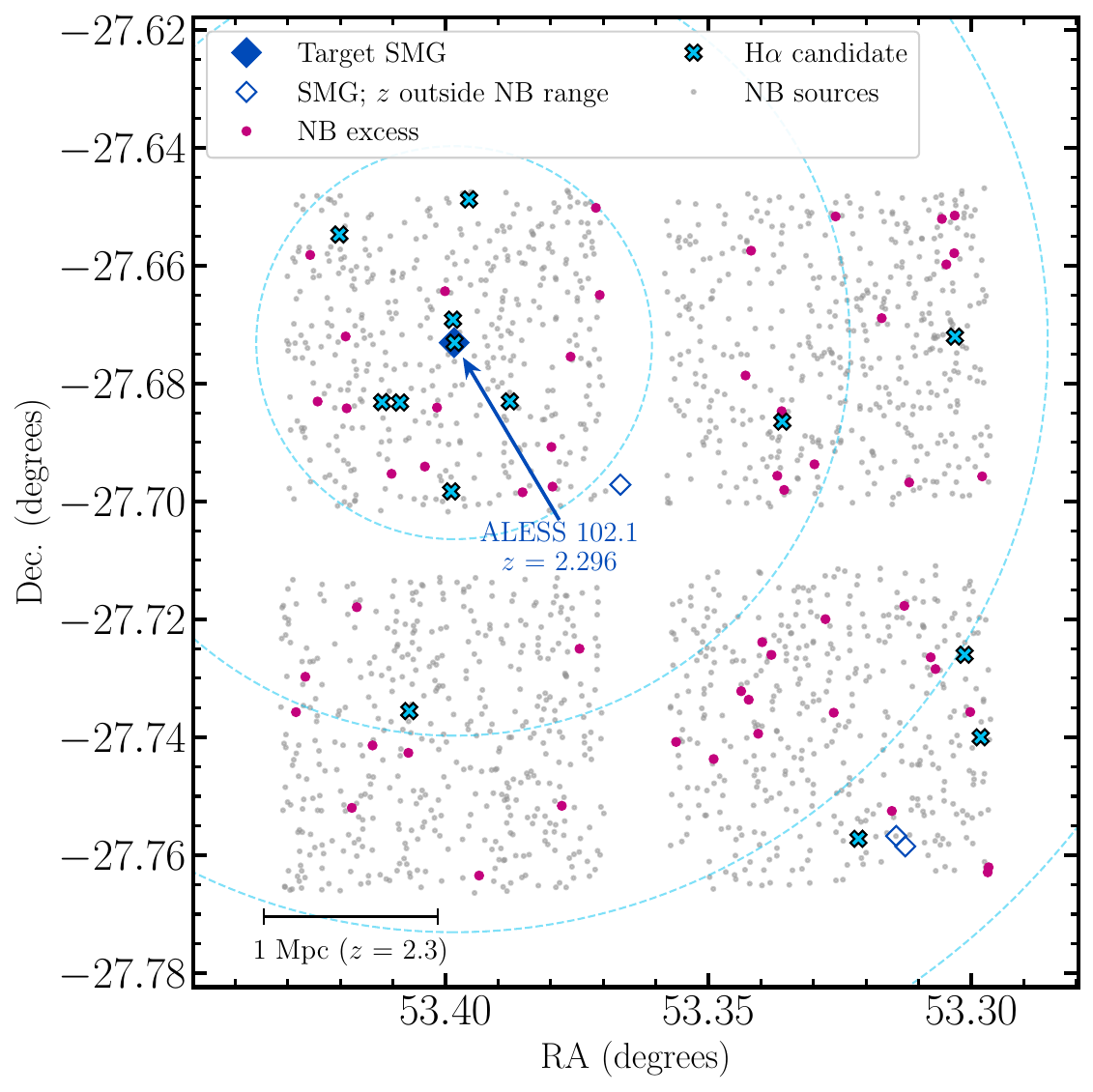}
  \caption{Spatial distribution of the emission line galaxies and the other \BrG\ detections in Pointing~5+75 (left) and Pointing~102 (right). Also shown are the positions of the target SMGs and other ALESS SMGs in these areas, although the redshifts of the non-target SMGs are either unknown or outside the ranges that would place the \Ha\ or \OIII\ emission lines in the \BrG\ filter \citep{Danielson+2017,Birkin+2021}. \Ha\ and \OIII\ candidates are indicated. While Pointing~102 does contain \OIII\ candidates, they are not shown here because there are no ALESS SMGs at $z \sim 3.3$ in this pointing. For all three SMGs, the candidate companion galaxies are distributed across the entire HAWK-I field of view, corresponding to physical spans of a few Mpc, as expected from protocluster simulations \citep[e.g.][]{Chiang+2013,Muldrew+2015,Yajima+2022}. Dashed circles show the boundaries of annuli used to measure radial trends in  the density of companion galaxies (\S\ref{SS:annuli}).}
  \label{fig:NB_positions}
\end{figure*}

Figure~\ref{fig:NB_positions} shows the distributions of emission line galaxies across the HAWK-I pointings, with \Ha\ and \OIII\ emitters highlighted. For all three target SMGs, the companion galaxies are spread across the entire field of view and therefore span several physical Mpc. This is consistent with expectations from simulations \citep[e.g.][see also \S\ref{SS:annuli}]{Chiang+2013,Muldrew+2015,Yajima+2022}, in which protoclusters are seen to extend over several Mpc, such that the entire structure is unlikely to be captured by a single HAWK-I pointing. We also note the presence of a dense clump of seven \Ha\ emitters (three of which are spectroscopically confirmed at $z \sim 2.3$) in the northeast of Pointing~5+75, which coincides with a photometrically-identified \Lya\ blob at $z \sim 2.3$ \citep[CDFS-LAB03;][]{Yang+2010}. This system will be discussed further in \S\ref{SS:annuli}. 

Of the three SMGs targeted, only ALESS\,102.1  is identified as an \Ha\ (or \OIII) emitter in our data. \citet{Danielson+2017} did not identify  ALESS\,102.1 as an \Ha\ emitter in their spectroscopy, because the wavelength coverage with VLT/FORS2 and VLT/VIMOS does not cover \Ha\ at $z\sim2.3$. 
The original spectroscopic redshift for ALESS\,5.1 is from CO(4-3) \citep{Birkin+2021} and no emission lines were observed in Keck/DEIMOS, Keck/MOSFIRE or VLT/XSHOOTER observations \citep{Danielson+2017}; this is likely because the redshifted \OIII\ line clashes with a bright OH$^-$ sky line for this source \citep{Ramsay+1992}. For ALESS\,75.2 the original spectroscopic redshift was measured, in part, thanks to a faint \Ha\ line detected in Keck/MOSFIRE observations \citep{Danielson+2017}, which is below the detection limit of our data.

\begingroup
\setlength{\tabcolsep}{4pt}
\begin{table}
  \centering
  \caption{Summary of the sample at each stage of the selection process described in \S\ref{SS:NB_selection} and \S\ref{SS:contaminants}.}
  \label{tab:sample_summary}
  \begin{tabular}{cccc}
    \hline
    & \multicolumn{3}{c}{Number per pointing}\\
    & Pointing~5+75 & Pointing~102 & Total\\
    \hline
    \hline
    \BrG\ detections & 2175 & 1754  & 3929\\
    Line emitter candidates (initial) & 81 & 80 & 161\\
    Line emitter candidates (confirmed) & 79 & 68 & 147\\
    \hline
    \Ha\ candidates & 44 & 11 & 55\\
    \OIII\ candidates & 4 & 2 & 6\\
    \hline

  \end{tabular}
\end{table}
\endgroup

\section{Results, Analysis \& Discussion}

\subsection{Measuring luminosity functions}\label{SS:LFs}

In order to quantify whether the SMGs reside in significant overdensities of \Ha\ or \OIII\ emitters, a comparison to the blank field needs to be drawn. The High Redshift ($z$) Emission Line Survey \citep[HiZELS;][]{Geach+2008} is a large narrowband survey of emission line galaxies, including \Ha\ emitters at $z = 2.23$ \citep{Sobral+2013} and \OIII\ emitters at $z = 3.24$ \citep{Khostovan+2015} in COSMOS. The HiZELS results are therefore representative of \Ha\ and \OIII\ emitters in regions of average density at redshifts similar to those of our target SMGs, and we use their luminosity functions as a blank-field sample for comparison with our results.

To construct luminosity functions for our \Ha\ and \OIII\ emitters we bin them according to line luminosities, making corrections to the observed number counts in each bin to account for completeness, contamination from other emission lines, dust attenuation, and the shape of the narrowband filter profile. Each of these steps is described in more detail below.

\subsubsection{Survey volumes}\label{SSS:volumes}

Approximating the \BrG\ filter profile as a top-hat function with width equal to the FWHM of the filter ($\Delta\lambda = 0.030$ \um), the \Ha\ emission line should be detectable from $z_{\textrm{min}} = 2.276$ to $z_{\textrm{max}} = 2.322$, corresponding to a comoving volume per square degree of $5.86\times10^{5}$ cMpc$^{3}$ deg$^{-2}$. The redshift range within which \OIII\ can be detected extends from $z_{\textrm{min}} = 3.294$ to $z_{\textrm{max}} = 3.354$, which gives $1.04\times10^{6}$ cMpc$^{3}$ deg$^{-2}$. Accounting for the high-noise regions that were masked prior to source detection (see \S\ref{SS:HAWKI_data}) and the gaps between the HAWK-I detector chips, the surveyed areas in Pointing~5+75 and Pointing~102 are 0.0117 and 0.0118 deg$^{2}$, respectively. The volumes probed are therefore: 6859 cMpc$^{3}$ for \Ha\ in Pointing~5+75; 6891 cMpc$^{3}$ for \Ha\ in Pointing~102; 12180 cMpc$^{3}$ for \OIII\ in Pointing~5+75; 12230 cMpc$^{3}$ for \OIII\ in Pointing~102. 
In \S\ref{SSS:vol_corr} we correct the derived luminosity functions to account for the fact that the \BrG\ filter is not a perfect top-hat function, which leads to the volume probed being slightly different for sources with different luminosities.

\subsubsection{Completeness correction}\label{SSS:completeness}

It is possible that real galaxies with weak emission lines were missed in our selection process (\S\ref{SS:NB_selection}) despite actually meeting the selection criteria: the sample is incomplete at low emission line fluxes. We correct for this using the method employed by \citet{Sobral+2013}, applying it separately for each quadrant of each pointing due to the variation in depth between detector chips (see Table \ref{tab:lim_mags}). For each emission line (\Ha\ and \OIII), we select sources that failed to meet the emission line galaxy selection criteria (i.e. sources for which EW $< 50$ \AA\ and/or $\Sigma < 3$) with redshifts within the range used to identify the targeted emission line (see \S\ref{SS:contaminants}). Due to the size of these samples, we generate ${\sim}$1000 mock galaxies by randomly varying the \Ks\ and \BrG\ magnitudes of the selected galaxies according to their uncertainties (assuming a Gaussian probability distribution for each magnitude) and placing them at random positions within their quadrant, removing any sources for which these changes result in them being classed as a line emitter. We then artificially inject line flux to each galaxy in this bolstered sample of non-line emitters, beginning with $10^{-22}$\,erg\,s$^{-1}$\,cm$^{-2}$ and incrementally increasing it by 0.05\,dex. Line fluxes are calculated as
\begin{equation}
F_{\textrm{line}} = \Delta\lambda_{\textrm{Br}\gamma} \frac{f_{\textrm{Br}\gamma} - f_{K_{s}}}{1 - (\Delta\lambda_{\textrm{Br}\gamma} / \Delta\lambda_{K_{s}})}\,,
\end{equation}
\noindent where $f_{\textrm{Br}\gamma}$ and $f_{K_{s}}$ are the \BrG\ and \Ks\ flux densities, respectively, in units of erg s$^{-1}$ cm$^{-2}$ \AA$^{-1}$. With each increment of injected line flux, we recalculate the EW and $\Sigma$ and reapply the line emitter selection criteria to determine the catalogue completeness as a function of line flux. This is used to estimate the completeness corrections for our luminosity functions. The uncertainty in the completeness at a given line flux is estimated by regenerating the mock galaxies 1000 times and measuring the standard deviation in the completeness across all iterations.

\subsubsection{Removing \NII\ contamination}\label{SSS:NII_correction}

The \Ha\ emission line lies in between a doublet of \NII\ lines at rest-frame wavelengths of 6548\,\AA\ and 6583\,\AA, which will contribute to the measured \BrG\ flux density and therefore affect the observed EW and emission line flux. Using spectroscopic data taken with Subaru/FMOS and Keck/MOSFIRE, \citet{Sobral+2015} observed an anti-correlation between the \NII$\lambda6583$-to-\Ha\ line flux ratio and the rest-frame EW (EW$_{\textrm{rest}} = \textrm{EW} / (1 + z)$) for the \Ha\ emitters in HiZELS, deriving the following empirical relation:
\begin{equation}
\frac{F_{[\textrm{N{\sc{ii}}}]}}{F_{\textrm{H}\alpha}} = -0.296 \times \log_{10}(\textrm{EW}_{\textrm{rest,H}\alpha+\textrm{N{\sc{ii}}}}) + 0.8\,.
\end{equation}
\noindent We adopt this relation to apply corrections to the line fluxes of all \Ha\ emitters in our sample, resulting in a median decrease of $11^{+9}_{-5}$\,percent in  emission line flux.

\subsubsection{Relative contributions from \OIII$\lambda5007$, \OIII$\lambda{4959}$ and \Hb}

Thus far only the \OIII$\lambda5007$ emisson line has been considered in the discussion of \OIII\ emitters at $z \sim 3.3$. However, this line is part of a doublet, with its counterpart residing at a rest-frame wavelength of 4959 \AA, and there is a narrow range of redshifts ($z = 3.336$--$3.344$) in which both lines can contribute to a galaxy's \BrG\ flux. Furthermore, while the \Hb\ emission line is sufficiently separated from the \OIII\ doublet to avoid contaminating the measured \OIII\ line flux, it is still close enough such that there is the potential for \Hb\ emitters to be misidentified as \OIII\ emitters (see Figure \ref{fig:NB_photozs}). Therefore, rather than try and separate our sample into \OIII$\lambda5007$, \OIII$\lambda4959$ and \Hb\ emitters, we present a combined \OIII+\Hb\ luminosity function; this also allows for a consistent comparison with the blank-field \OIII+\Hb\  luminosity function from \citet{Khostovan+2015}. 

We do however take into account the results of \citet{Sobral+2015} when estimating the total volumes probed by the \BrG\ filter in the search for \OIII\ emitters: using spectroscopy, \citet{Sobral+2015} find that for HiZELS ${\sim}50$ percent of photometrically-selected \OIII+\Hb\ emitters at $z \sim 1.4$ are \OIII$\lambda5007$, ${\sim}27$ percent are \OIII$\lambda4959$, ${\sim}16$ percent are \Hb, with the remaining ${\sim}7$ percent being simultaneous detections of \OIII$\lambda5007$ and \OIII$\lambda4959$. Based on these results, \citet{Sobral+2015} then add to the total volume probed (i.e. the volume probed if searching for \OIII$\lambda5007$ emitters) 16 percent of the volume that would be probed had their search been for \Hb, and 25 percent of the volume had they been searching for \OIII$\lambda4959$. We thus apply similar corrections to our total volume probed for \OIII$\lambda5007$ emitters.

\subsubsection{Corrections for dust attenuation}\label{SSS:attenuation}

Dust in star-forming galaxies reprocesses light emitted in the rest-frame UV and optical, and can therefore reduce the amount of \Ha\ and \OIII\ flux observed. In order to estimate the intrinsic brightness of the emission lines (i.e. their integrated luminosities), one has to correct for the effect of dust attenuation. For an attenuation of $A_{\textrm{line}}$ (mag) at the emission line wavelength, the conversion from line flux to intrinsic line luminosity is
\begin{equation}
L_{\textrm{line}} = 4 \pi D_{\textrm{L}}^{2}F_{\textrm{line}} \times 10^{0.4A_{\textrm{line}}}\,,
\end{equation}
\noindent where $D_{\textrm{L}}$ is the luminosity distance. We follow \citet{Sobral+2013} and assume an attenuation at the \Ha\ wavelength of $A_{\textrm{H}\alpha} = 1$ mag, which is based on previous HiZELS studies \citep{Garn+2010,Sobral+2012}. \citet{Khostovan+2015} do not correct for dust attenuation when plotting their luminosity functions, so we also leave our \OIII+\Hb\ luminosities uncorrected to ensure a consistent comparison. \citet{Khostovan+2015} later go on to calculate the dust-corrected star formation rates (SFRs) of their galaxies, where they then assume an attenuation of $A_{\textrm{\OIII+\Hb}} = 1.35$ mag, derived by assuming $A_{\textrm{H}\alpha} = 1$ mag and using a \citet{Calzetti+2000} dust attenuation curve. We thus adopt the same correction when calculating our own SFRs (see \S\ref{SSS:MS}).

\subsubsection{Filter profile volume corrections}\label{SSS:vol_corr}

The comoving volumes used for our luminosity functions (\S\ref{SSS:volumes}) are calculated by approximating the \BrG\ filter as a top-hat filter with width equal to the \BrG\ FWHM. Since the filter profile is not a top-hat in reality, this introduces two main effects which need to be accounted for when estimating the galaxy number densities. Firstly, bright emitters whose line falls near the edges of the \BrG\ filter will suffer a significant loss of line flux and thus appear to be fainter than they really are. This produces an overall bias towards faint sources in our sample. Secondly, any faint emitters close to the filter edges might be missed from our sample, and are therefore only detectable over a narrower redshift range (and thus a smaller volume) than their bright counterparts.

To correct for these effects, we follow the method used by \citet{Sobral+2013} and \citet{Khostovan+2015}, as first proposed in \citet{Sobral+2009}. An initial Schechter fit is performed to the uncorrected\footnotemark\footnotetext{`Uncorrected' only in terms of the filter profile correction; the results have already been corrected for line completeness and dust attenuation by this stage.} data. We then generate a mock sample of $10^{5}$ fake sources with a luminosity distribution that is weighted by the uncorrected Schechter function. These sources are randomly assigned redshifts with a uniform distribution across the whole possible \BrG\ coverage. They are then convolved through the \BrG\ filter profile such that their luminosities decrease according to their assigned redshift (i.e. according to the position of the redshifted emission line in the filter profile) and rebinned using the same luminosity bins as for the uncorrected data. Comparing the resultant distribution to the input distribution reveals that bright sources are underestimated relative to the fainter sources, as expected. The real data are corrected using the ratio of these distributions.

\subsubsection{Fitting Schechter functions}\label{SSS:Schechter}

Finally, we perform fits to the corrected data using a Schechter function \citep{Schechter1976}:
\begin{equation}
\Phi(L)dL = \ln(10) \Phi^{\ast} \left(\frac{L}{L^{\ast}}\right)^{\alpha + 1} e^{-(L/L^{\ast})} d\log L\,,
\end{equation}
\noindent where $\Phi(L)$ is the number density at luminosity $L$, $\Phi^{\ast}$ is the normalisation of the luminosity function, $L^{\ast}$ is the characteristic luminosity, and $\alpha$ is the slope at the faint end of the luminosity function, where the power-law component dominates. 

The faintest bins (open symbols in Figure \ref{fig:LFs}) are excluded from each fit due to their low completeness. For the \Ha\ emitters, we take `low completeness' to mean that the low-luminosity edge of the bin lies below the 30 percent completeness limit. 
The line luminosities of all \OIII+\Hb\ emitters lie above the 90 percent completeness threshold; we therefore do not exclude any from the fit.

Due to the small number of bins left available for fitting, it is impossible to reliably constrain all three free parameters of the Schechter function simultaneously. For the \Ha\ luminosity functions, we therefore fix the faint-end slope, $\alpha$, to the value of $-1.59$ obtained by \citet{Sobral+2013} at $z = 2.23$. For the \OIII+\Hb\ emitters, we only have one available bin and thus fix both $\alpha$ and $\log(L^{\ast}/\textrm{erg s}^{-1})$ to the values for the $z = 3.24$ sample of \OIII+\Hb\ from \citet{Khostovan+2015}, which are $-1.60$ and 42.83, respectively. Thus our \OIII+\Hb\ luminosity function is effectively a renormalised version of the \citet{Khostovan+2015} result, with $\Phi^{\ast}$ being the only free parameter. In addition to the \Ha\ luminosity functions of the individual pointings, we also construct fits to the combined sample of \Ha\ emitters from both SMG fields, as this provides a more general view of SMGs at $z \sim 2.3$ with improved statistics. 

The best-fit parameters for each luminosity function are summarised in Table \ref{tab:LF_summary}. Uncertainties are estimated for each free parameter by randomly perturbing the bin heights according to their uncertainties and then recalculating the fit, and repeating this process until $10^{5}$ fits have been made. The $1\sigma$ confidence interval for each parameter is then estimated using the 16th and 84th percentiles of the best-fit values.

\subsection{Analysing luminosity functions}\label{SS:BF_compare}

\begin{figure*}
	\includegraphics[width=0.95\columnwidth]{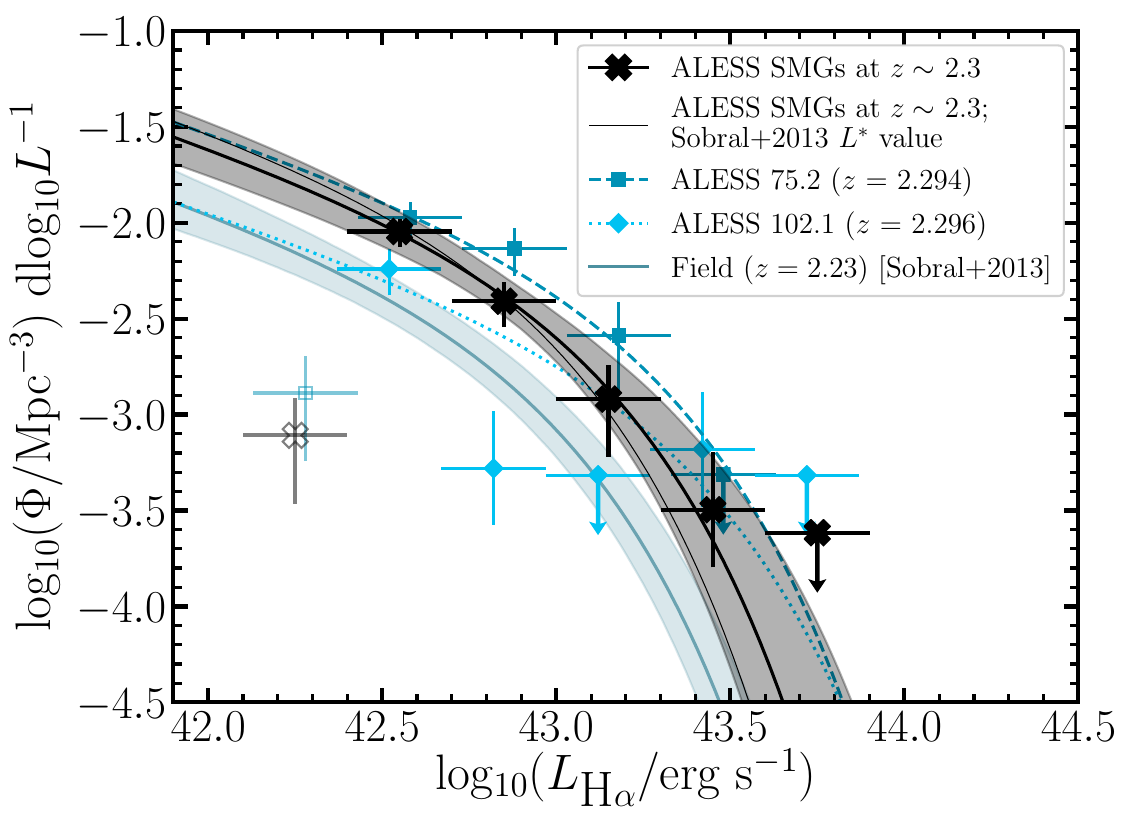}
	\includegraphics[width=0.95\columnwidth]{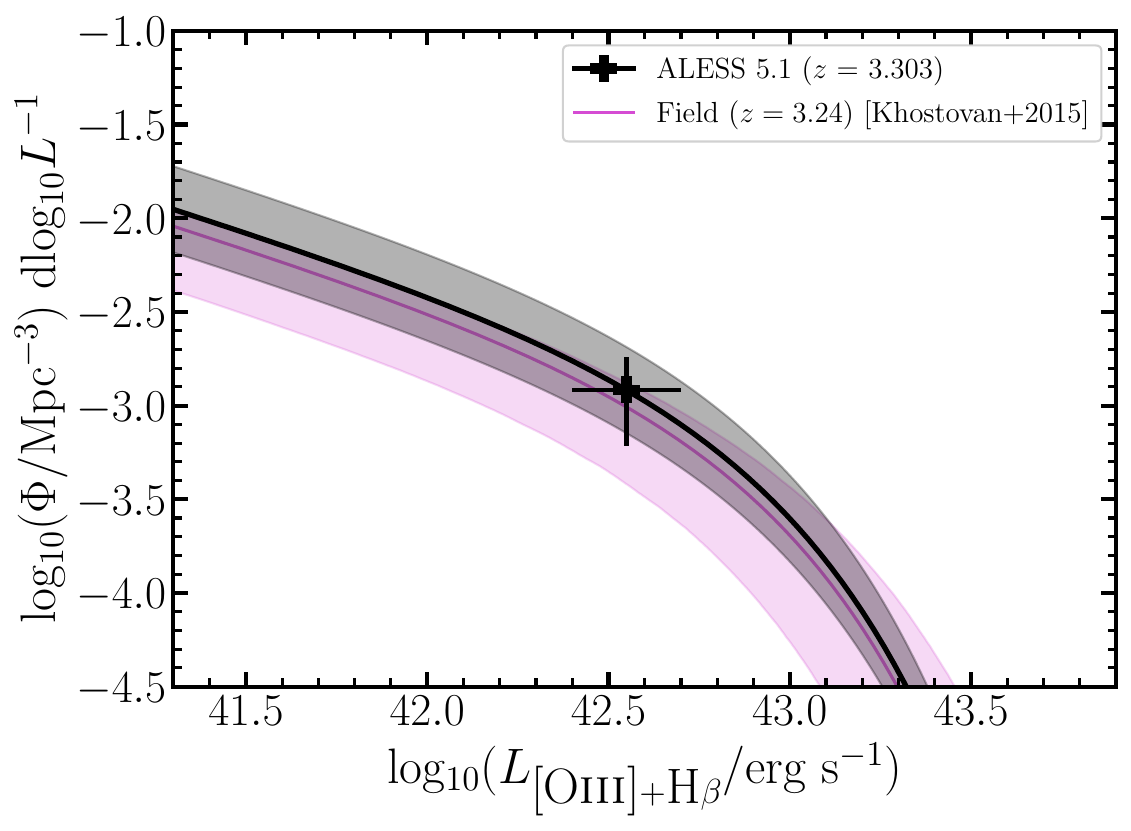}
	\caption{Luminosity functions of \Ha\ (left) and \OIII\ (right) emitters identified around SMGs at $z \sim 2.3$ and $z \sim 3.3$, respectively. Open symbols represent bins that are highly incomplete and are thus excluded from the fits (see \S\ref{SSS:Schechter}). The data are compared with luminosity functions from blank field studies at similar redshifts \citep{Sobral+2013,Khostovan+2015}, which are highlighted with coloured solid curves. Shaded regions represent the 1$\sigma$ uncertainties on each luminosity function. For our \Ha\ luminosity functions (left) the faint-end slope is fixed to the value from the corresponding blank-field luminosity function: $\alpha = {-}1.59$ \citep{Sobral+2013} and
 dashed and dotted curves show the fitted Schechter functions for the environments of ALESS~75.2 and ALESS~102.1, respectively. The thick, solid black curve shows the Schechter function obtained by fitting to the data for our combined sample of \Ha\ emitters at $z \sim 2.3$ (black data points); the grey shaded region shows the $1\sigma$ confidence region for this fit. The thin black line shows another Schechter function obtained by fitting to the black data points, but with $L^{\ast}$ fixed to the blank-field value of $\log(L^{\ast}/\textrm{erg s}^{-1}) = 42.87$ \citep{Sobral+2013}. 
  In the right panel, the thick solid black curve and grey shaded region shows the result of scaling up the blank-field luminosity function from \citet{Khostovan+2015} to fit to the single bin of \OIII\ emitters from the environment of ALESS~5.1. 
  Comparison with the blank-field luminosity functions reveals that ALESS~5.1, ALESS~75.2, and ALESS~102.1 reside in environments with overdensity parameters of $\delta_{g} = 0.2^{+2.5}_{-0.7}$, $2.6^{+1.4}_{-1.2}$, and $0.2^{+0.6}_{-0.5}$, respectively. On average, the SMGs at $z \sim 2.3$ reside in environments with galaxy overdensities of $\delta_{g} = 1.5^{+1.0}_{-0.8}$.
  }
	\label{fig:LFs}
\end{figure*}

We next use the luminosity functions to assess whether the targeted SMGs reside in overdensities of emission-line galaxies. Figure \ref{fig:LFs} compares the observed luminosity functions from the SMG fields to those from the blank field surveys of \citet{Sobral+2013} and \citet{Khostovan+2015}. 
The environment of ALESS~75.2 shows signs of being overdense relative to the field at $z \sim 2.3$, with most bins lying significantly above the blank field luminosity function. Conversely the environments of ALESS~5.1 and 102.1 are broadly consistent with the blank field luminosity functions at their respective epochs. 
An overdensity remains when the \Ha\ emitters from both fields at $z \sim 2.3$ are combined, implying that on average SMGs at this epoch reside in overdense, protocluster-like environments, which qualitatively consistent with clustering results \citep[e.g.][]{Hickox+2012, Wilkinson+2017, Stach+2021}. The contrast between the individual \Ha\ luminosity functions suggests that there is significant variation across SMG environments, although observations of additional SMGs are required to confirm and quantify the field-to-field variation. Furthermore, as explored later in this section and shown in Figure~\ref{fig:NB_positions}, the field around ALESS~102.1 is itself overdense on smaller scales.

\begin{figure}
	\includegraphics[width=0.95\columnwidth]{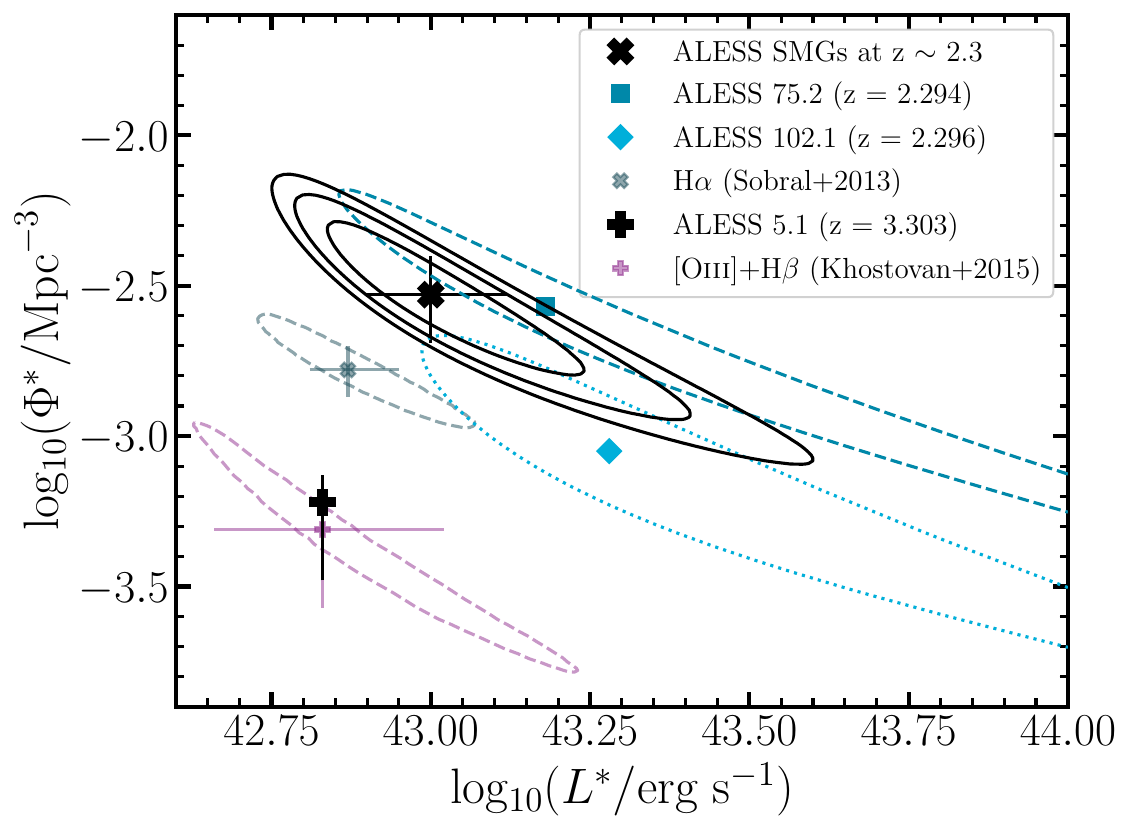}
	\caption{Contours showing the correlated uncertainties on the Schechter parameter fits to the luminosity functions shown in Figure \ref{fig:LFs}. For all the SMG fields, the faint-end slope, $\alpha$, is fixed to match to the blank-field luminosity functions from \citet{Sobral+2013} and \citet{Khostovan+2015}. Single contours are at the 1$\sigma$ level; the combined $z\sim2.3$ data has 1, 2 and 3$\sigma$ contours shown. For ALESS~5.1, only the 1$\sigma$ error bars are shown as $L^{\ast}$ is fixed. 
 Both of the luminosity functions for SMGs at $z \sim 2.3$ are separated from the corresponding blank-field luminosity function in $L^{\ast}$--$\Phi^{\ast}$ space, although for ALESS~102.1 this is only at the $\sim1\sigma$ level. Increases in $L^{\ast}$ relative to the blank-field, as seen for both SMGs at $z \sim 2.3$ (and for their combined luminosity function), imply that their environments may preferentially harbour brighter galaxies than those in the field. ALESS~5.1 exibits an offset of ${<}1\sigma$ relative to the blank-field luminosity function at $z \sim 3.3$.
 } 
	\label{fig:contours}
\end{figure}

To quantitatively compare the SMG field and blank field luminosity functions we consider the parameters of the Schechter function fits (\S~\ref{SSS:Schechter}). The parameters of Schechter function fitting are often correlated, so in Figure~\ref{fig:contours} we show the uncertainties of the luminosity function parameters in the $L^{\ast}$--$\Phi^{\ast}$ plane (as described in \S\ref{SSS:Schechter} the faint-end slope, $\alpha$, is fixed), comparing our SMG fields with the blank fields at similar redshifts. 
For the individual SMG environments at $z \sim 2.3$ the fit parameters are offset from those of the blank field, although for the ALESS~102.1 region the offset is only at the $\sim1\sigma$ level. 
These separations are driven by a higher $L^{\ast}$, and,  in the case of ALESS~75.2, by a larger $\Phi^{\ast}$, which implies that this environment is preferentially overdense in bright line emitters compared to the blank field. 
Meanwhile, the environment of ALESS~5.1 exhibits an offset of ${<}1\sigma$ relative to the blank-field value of $\Phi^{\ast}$ at $z \sim 3.3$, implying this SMG does not reside in an overdnensity.

We quantify the galaxy overdensity in each sample of \Ha\ and \OIII+\Hb\ emitters in two ways. Firstly, we calculate the ratio of the $\Phi^{\ast}$ from the best fit Schechter function to those from the relevant blank-field luminosity functions, $\Phi^{\ast}/\Phi^{\ast}_{\textrm{field}}$. This ratio tells us how much higher the `knee' of each SMG-field luminosity function is relative to the blank-field. These ratios are presented in Table \ref{tab:LF_summary}. The value for the \OIII+\Hb\ luminosity function suggests that the environment of ALESS~5.1 at $z\sim3.3$ is $1.2^{+0.6}_{-0.4}$ times as dense as the field, i.e.\ it is consistent with the blank field. For the \Ha\ emitters at $z\sim2.3$, $L^{\ast}$ is also a free parameter in the Schechter fits, and the $L^{\ast}$--$\Phi^{\ast}$ correlation means that we must first refit the data with $L^{\ast}$ fixed to the $z \sim 2.3$ blank-field value from \citet{Sobral+2013}, i.e.\ $\log(L^{\ast} / \textrm{erg s}^{-1}) = 42.83$. 
These fits give $\Phi^{\ast}/\Phi^{\ast}_{\textrm{field}}$ values of $3.6^{+0.6}_{-0.6}$ and $1.7^{+0.3}_{-0.3}$ for the ALESS~75.2 and ALESS~102.1 fields, respectively. The combined sample of \Ha\ emitters from both SMG environments suggests that the average SMG environment at $z \sim 2.3$ is $2.6^{+0.4}_{-0.4}$ times more dense than the blank field at this epoch. 

To derive a more representative estimate of the galaxy overdensity in each environment, we also estimate the number of \Ha\ or \OIII+\Hb\ emitters that one would expect to find in a blank field with the volumes probed by our observations. To do this, we integrate the field luminosity functions across the luminosity range covered by our data, excluding low completeness bins; i.e. we integrate across the ranges $42.5 \leq \log(L^{\ast}_{\textrm{\Ha}} / \textrm{erg s}^{-1}) < 43.8$ and $42.4\leq \log(L^{\ast}_{\textrm{\OIII+\Hb}} / \textrm{erg s}^{-1}) < 42.7$ and multiply by the volumes probed in each HAWK-I pointing to estimate the expected number counts in an equivalent blank field, $N_\textrm{field}$. Since these field galaxies would have contributed to the observed number counts, we quantify the galaxy overdensity in each environment using
\begin{equation}\label{eq:delta_g}
\delta_g = \frac{N_{\textrm{total}} - N_\textrm{field}}{N_\textrm{field}}
\end{equation}
\noindent where $N_{\textrm{total}}$ is the sum of the counts in our complete bins. Uncertainties in $N_{\textrm{total}}$ are determined by adding in quadrature the uncertainties in the bin counts. For $N_{\textrm{field}}$ the uncertainties are estimated by randomly permuting the blank-field Schechter parameters within their uncertainties prior to integrating, then repeating the process $10^{5}$ times and using the 16th and 84th percentiles of the resultant number counts to define the $1\sigma$ confidence interval.

The values of $\delta_{g}$ for each sample of \Ha\ and \OIII+\Hb\ emitters are summarised in Table \ref{tab:LF_summary} along with the significance of this overdensity, $\sigma_{\delta}$. Based on these values, the environments of ALESS~5.1, 75.2, and 102.1 are overdense by factors of $0.2^{+2.5}_{-0.7}\ (0.3\sigma_{\delta})$, $2.6^{+1.4}_{-1.2}\ (2.3\sigma_{\delta})$, and $0.2^{+0.6}_{-0.5}\ (0.5\sigma_{\delta})$, respectively. If the samples of \Ha\ emitters from both pointings are considered as one, then the SMG environments at $z \sim 2.3$ are overdense by a factor of $1.5^{+1.0}_{-0.8}\ (1.9\sigma_{\delta})$ on average. 
The above uncertainties do not account for cosmic variance, which could cause a factor $\sim2$ difference in number counts, as based on the \Ha\ emitters in two equal depth HAWK-I pointings in COSMOS and UDS \citep{Sobral+2013}. Including cosmic variance in our calculations does not change our overall conclusions that the ALESS~75.2 and the combined $z\sim2.3$ SMG fields are overdense, nor does it affect the conclusions that ALESS~5.1 and ALESS~102.1 reside in environments consistent with the blank field. 

While Pointing~102 as a whole is not overdense, the majority of the \Ha\ emitters in this pointing are contained within the same quadrant as the SMG, as can be seen in Figure \ref{fig:NB_positions} (see also \S\ref{SS:annuli}). We therefore recalculate $\delta_{g}$ for this SMG environment, this time considering only the volume probed within that quadrant (1722 cMpc$^{3}$), finding $\delta_{g} = 3.8^{+2.4}_{-1.8} (2.1\sigma)$ in this area, which suggests that ALESS~102.1 actually does reside in an overdense environment with a physical scale of ${\sim}1.6$~Mpc. This high concentration of galaxies surrounding the SMG could be indicative of structure formation on smaller scales than those of protoclusters and it is possible that ALESS~102.1 resides in a protogroup \citep[e.g.][]{Diener+2013}. We discuss the spatial distribution of companion galaxies in each SMG environment in more detail in \S\ref{SS:annuli}.

\begin{table*}
	\centering
	\caption{Summary of the best-fit Schechter parameters for companion galaxies in the environments of the target SMGs, along with comparisons to the blank field at similar redshifts. In all cases, the faint-end slope of the luminosity function, $\alpha$, is fixed to the value from the relevant blank-field luminosity function and, where indicated, the characteristic luminosity, $L^{\ast}$, is also fixed to the blank-field values. The values of $\Phi^{\ast}_{\textrm{field}}$ are taken from the relevant blank-field Schechter functions. All blank-field parameters are from \citep{Sobral+2013} and \citep{Khostovan+2015}.
 }
	\label{tab:LF_summary}
	\begin{tabular}{cccccccc}
		\hline
		SMG environment & $\log(L^{\ast} / \textrm{erg s}^{-1})$ & $\log(\Phi^{\ast} / \textrm{Mpc}^{-3})^a$ & $\log(\Phi^{\ast}_{\textrm{fixed}\,L^{\ast}} / \textrm{Mpc}^{-3})^b$ & ${\Phi^{\ast}/\Phi^{\ast}_{\textrm{field}}}^a$ & ${\Phi^{\ast}_{\textrm{fixed}\,L^{\ast}}/\Phi^{\ast}_{\textrm{field}}}^b$ & ${\delta_{g}}^c$ & ${\sigma_{\delta}}^d$  \\
		\hline
		\hline
		\vspace{0.5em}
		ALESS~75.2  & $43.18^{+0.42}_{-0.28}$ & $-2.57^{+0.29}_{-0.39}$ & $-2.22^{+0.05}_{-0.09}$ & $1.62^{+1.05}_{-0.80}$ & $3.63^{+0.58}_{-0.59}$ & $2.61^{+1.42}_{-1.15}$ & 2.3 \\
		\vspace{0.5em}
		ALESS~102.1  & $43.28^{+0.50}_{-0.28}$ & $-3.05^{+0.17}_{-0.42}$ & $-2.55^{+0.08}_{-0.12}$ & $0.54^{+0.22}_{-0.23}$ & $1.70^{+0.30}_{-0.32}$ & ${0.21^{+0.55}_{-0.45}}^e$ & $0.5^e$ \\
		\hline
		\vspace{0.5em}
		SMGs at $z \sim 2.3$  & $43.00^{+0.12}_{-0.10}$ & $-2.53^{+0.12}_{-0.17}$ & $-2.36^{+0.04}_{-0.08}$ & $1.78^{+0.41}_{-0.43}$ & $2.63^{+0.41}_{-0.42}$ & $1.51^{+0.98}_{-0.80}$ & 1.9 \\
		\hline
		ALESS~5.1 & $42.83$ (fixed) & -- & $-3.22^{+0.09}_{-0.26}$ & -- & $1.23^{+0.58}_{-0.40}$ & $0.22^{+2.50}_{-0.66}$ & 0.3 \\
		\hline
	\end{tabular}
	\begin{flushleft}
         $^a$ $\Phi^{\ast}$ as measured when both $\Phi^{\ast}$ and $L^{\ast}$ are free parameters ($\alpha$ is always fixed to the blank-field values).\\
		 $^b$ $\Phi^{\ast}_{\textrm{fixed}\,L^{\ast}}$ is obtained by fitting a Schechter function to the data with both $L^{\ast}$ and $\alpha$ fixed to the blank-field values. The values of $L^{\ast}$ are taken from the relevant blank-field Schechter functions.\\ 
		$^c$ Galaxy overdensity, $\delta_{g} = (N_\textrm{total}/N_{\textrm{field}}) - 1$; see \S\ref{SS:BF_compare}.\\
		$^d$ Significance of the galaxy overdensity, $\delta_{g}$.\\
        $^e$ ALESS~102.1 has $\delta_g=3.8^{+2.4}_{-1.8}$ (i.e.\ $\sigma_{\delta}=2.1$) when considering only the HAWK-I quadrant containing the SMG (\S\ref{SS:BF_compare}).
	\end{flushleft}
\end{table*}

The question remains as to whether the target SMGs reside in protoclusters, which will evolve into bound clusters by the present day. 
To learn more, we compare the overdensities in the SMG fields with previous studies of protoclusters. However, protoclusters exhibit a wide range of galaxy overdensities; a `typical' value of $\delta_{g}$ is not well-defined, though we highlight here structures  at similar redshifts to our target SMGs. For example, $\delta_{g} = 2.5$ for the $z = 1.99$ protocluster in the GOODS-N field \citep{Chapman+2009}. The protoclusters 4C 10.48 ($z = 2.35$) and 4C 23.56 ($z = 2.48$), which were both identified using narrowband selection of \Ha\ emitters around luminous radio galaxies, were found to have overdensities of $\delta_{g} = 11^{+2}_{-2}$ and $\delta_{g} = 4^{+5}_{-3}$, respectively \citep{Hatch+2011,Tanaka+2011}. 
Similarly, \citet{Matsuda+2011} used a narrowband search for \Ha\ emitters at $z = 2.23$ to identify overdensities of $\delta_{g} \sim 3, 2,$ and 2 around a quasi-stellar object overdensity, a high-redshift radio galaxy, and an overdensity of SMGs and optically faint radio galaxies, respectively. 
The protocluster Cl J0227-0421 at $z = 3.29$ is overdense by a factor of $10.5 \pm 2.8$ \citep{Lemaux+2014}. Two protoclusters in the COSMOS field at $z = 2.10$ and $z = 2.47$ were found to have overdensities of $\delta_{g} \sim 8$ and $\delta_{g} \sim 3.3$, respectively \citep{Yuan+2014,Chiang+2017}. \citet{Zheng+2021} confirm overdensities of \Ha\ emitters in two protocluster candidates, BOSS1244 and BOSS1542, with overdensity factors of $\delta_{g} = 5.6 \pm 0.3$ and $\delta_{g} = 4.9 \pm 0.3$, respectively. 
Comparing $\delta_{g}$ for these protoclusters with our values, we posit that the environments of ALESS~75.2 and ALESS~102.1 (and thus of SMGs on average at $z \sim 2.3$) are consistent with being protoclusters, albeit at the lower-density end. For the remainder of the analyses we assume that members of these overdensities will form larger structures by $z = 0$, although we caution that the significance of these overdensities is relatively low (1.3--2.3$\sigma_{\delta}$) and thus it is uncertain whether they will coalesce by $z\sim0$ \citep[e.g.][]{Chiang+2013, Overzier+2009, Angulo+2012}.

\subsection{Spatial distribution of line emitters}
\label{SS:annuli}

\begin{figure}
	\includegraphics[width=0.95\columnwidth]{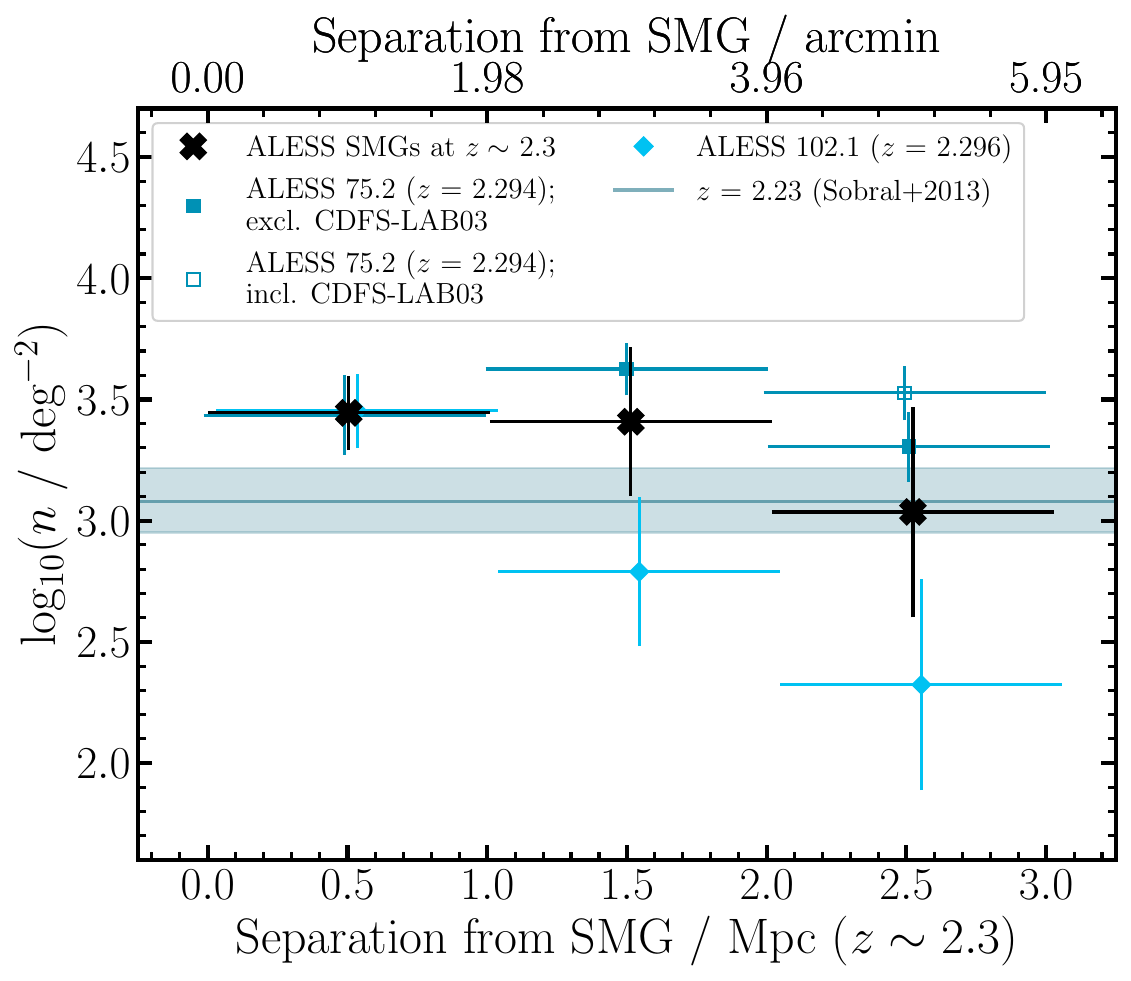}
	\caption{Surface density of \Ha\ emitters  measured in $2\farcm0$ annuli centred on the two target SMGs at $z \sim 2.3$ and compared with expected values  from the blank-field \Ha\ luminosity function \citep[horizontal line and shaded region;][]{Sobral+2013}.
  The shape of the field and positions of the SMGs means that coverage is incomplete with data for 74\% (81\%) of the inner, 36\% (31\%) of the middle, and 26\% (27\%) of the outer annuli for Pointing~5+75 (Pointing~102).
 Open symbols show the values calculated if the dense clump of \Ha\ emitters in the northeast of Pointing~5+75 (see \S\ref{SS:contaminants}) is included. Both $z \sim 2.3$ SMGs have high densities of \Ha\ emitters in the central $\sim1$~Mpc. For ALESS~102.1 the density falls with increasing separation from the SMG, though no significant trend exists for ALESS~75.2. The existence of a significant overdensity within ${\sim}2\arcmin$ of ALESS~102.1 with no evidence of further extension suggests it may reside in an early galaxy group, while ALESS~75.2 appears to reside in a larger structure that extends beyond the HAWK-I coverage.}
	\label{fig:annuli}
\end{figure}

\begin{figure*}
	\includegraphics[width=0.99\textwidth]{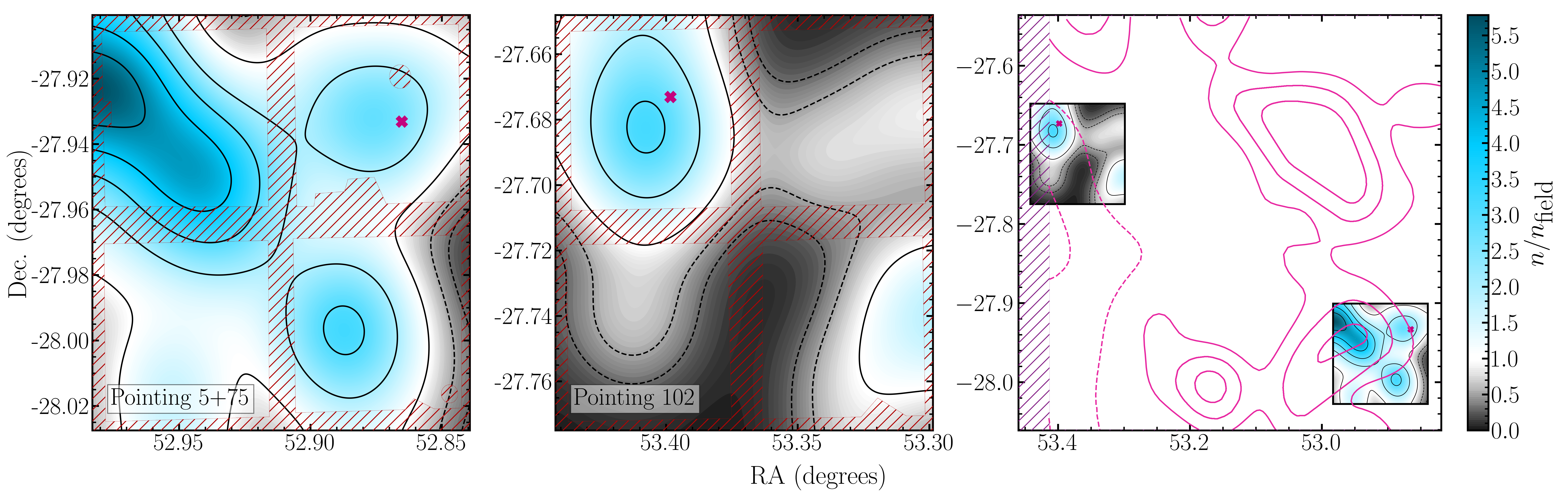}
	\caption{Maps showing the variation of surface overdensity, $n/n_{\textrm{field}}$, of \Ha\ emitters across Pointing~5+75 (left) and Pointing~102 (middle) fields, and in context of \Lya\ emitter (LAE) density at $z \sim 2.3$ in the wider ECDFS (right).
 The maps are smoothed using a 2D Gaussian kernel with a width of 0.5~Mpc. 
 Contour level are $n/n_{\textrm{field}} = 0.25, 0.5, 1$ and increasing in intervals of 1 thereafter; dashed lines represent underdensities. Crosses mark the positions of the target SMGs, ALESS~75.2 and ALESS~102.1 and hatching indicates regions outside out HAWK-I coverage (including chip-gaps) or that are masked (e.g.\ due to the presence of a bright star). Note that the smoothing implicitly assumes that no \Ha\ emitters reside in these regions, such that the densities here are conservative lower limits.
 Both SMGs are in/near \Ha\ density peaks, although ALESS~75.2 is not in the highest density region in P05\_75. 
  The rightmost panel shows the \Ha\ overdensities in the two SMG fields compared to the wider LAE density measured in the whole \citep{Yang+2010}; LAE contour levels are at $n/n_{\textrm{field}} = 0.3, 0.5, 1, 2\ \textrm{and}\ 3$. 
 The region of highest \Ha\ density in Pointing~5+75 corresponds to strong overdensity of LAEs,  which contains the \Lya\ blob CDFS-LAB03 \citep[][see \S\ref{SS:contaminants}]{Yang+2010} and there is an overall correlation between the \Ha\ and LAE overdensities in this region. Conversely, despite being in a small region of localised \Ha\ overdensity, ALESS~102.1 is in a region that is underdense in LAEs on the scales probed by \citet{Yang+2010}.} 
	\label{fig:density_maps}
\end{figure*}

In order to investigate the role of environment in shaping the evolution of SMGs, and to assess whether the target SMGs reside in special regions within any surrounding structures, we next explore the spatial distributions of coeval line emitters across our HAWK-I pointings. Due to the small size of the \OIII\ emitter sample, we limit this part of the analysis to the \Ha\ emitters around ALESS~75.2 and 102.1. 
Figure \ref{fig:annuli} compares the surface density of \Ha\ emitters as measured in annuli centred on each target SMG (where the density calculations account for masked and unobserved regions by assuming the density is the same as in the observed regions)  with the surface densities one would expect based on the blank-field luminosity function from \citet{Sobral+2013}. The annuli have inner and outer radii increasing in increments of $2\farcm0$, and are represented by dashed circles in Figure \ref{fig:NB_positions}. Note that these large annuli are necessary due to the sample sizes, but make it difficult to probe the protocluster structures in detail. 
We therefore also show in Figure \ref{fig:density_maps} the result of smoothing the distributions of \Ha\ emitters in Pointing~5+75 (left) and Pointing~102 (right) using a 2D Gaussian kernel with width corresponding to 0.5 Mpc at $z \sim 2.3$. This method of visualisation clarifies where the SMGs lie in relation to any density peaks and can highlight any substructures. For simplicity we assume that there are no \Ha\ emitters in the unobserved region between the detector chips, or in regions of the image that have been masked, and thus the densities shown in these regions are potentially underestimated. 

For ALESS~102.1, there is a noticeable decrease in the surface density of \Ha\ emitters as a function of radial distance from the SMG, with the innermost bin in Figure \ref{fig:annuli} being significantly overdense relative to the field despite the environment not being overdense as a whole (see also \S\ref{SS:BF_compare}). This is also clear from Figure \ref{fig:density_maps},  which shows that the SMG lies ${\sim}20\arcsec$ from a ${\sim}3\farcm25$ (${\sim}1.6$\,Mpc) density peak. Furthermore, Figure \ref{fig:NB_positions} demonstrates that the innermost $2\farcm0$ annulus contains more than half of the \Ha\ emitters detected across the Pointing~102 pointing.  
Figure~\ref{fig:density_maps} includes a panel showing the location of the SMGs and our \Ha\ emitter density maps in the context of the overdensity of \Lya\ emitters (LAEs) at $z \sim 2.3$ mapped by \citet{Yang+2010}. This shows that the small-scale overdensity around ALESS~102.1 is in a broader underdense region, and it is therefore unlikely to be a condensed infalling knot within a larger structure.

In the case of ALESS~75.2 we show two results in Figure \ref{fig:annuli}: one where we include all \Ha\ emitters in the pointing (open squares), and one where we exclude the dense clump of \Ha\ emitters in the north-east (filled squares; see \S\ref{SS:contaminants}). In both cases, there is no significant trend in the \Ha\ surface density as a function of separation from the SMG, although it does show signs of decreasing at the outermost radii if the dense north-easterly clump is excluded. This lack of trend implies that ALESS~75.2 does not reside in a particularly special region of the structure, and/or the structure extends beyond the HAWK-I field of view. 
The latter hypothesis is supported by the comparison of the \Ha\ emitter overdensity with that of the LAEs from \citet{Yang+2010} (Figure~\ref{fig:density_maps}, right), which shows that the whole structure around ALESS~75.2 is within a larger region of LAE overdensity. This suggests that the \Ha\ emitter structure likely spans a physical distance $\gtrsim 3.5$\,Mpc at $z \sim 2.3$, which is consistent with the simulations of e.g. \citet{Muldrew+2015}, in which protoclusters are expected to extend over $\gtrsim 10$\,Mpc at $z \sim 2$. The \citet{Yang+2010} structure in this region includes the \Lya\ blob CDFS-LAB03, which coincides with seven \Ha\ emitters (see also \S\ref{SS:contaminants}). The overall picture is consistent with previous findings, in which \Lya\ blobs are found to be associated with massive dark matter halos and filamentary large-scale structures \citep[e.g.][]{Geach+2016,Umehata+2019}.

\subsection{SMG companions: SFRs and stellar masses}\label{SSS:MS}

In this section we investigate the dust-corrected star formation rates (SFR) and stellar masses ($M_{\star}$) of the individual galaxies around each target SMG, to determine whether they lie on the main sequence of star formation at their epochs. This correlation between SFR and $M_{\star}$ has been observed out to $z \sim 6$ \citep[e.g.][]{Brinchmann+2004,Elbaz+2007,Daddi+2007,Gonzalez+2010,Speagle+2014,Schreiber+2015,Scoville+2017} and galaxies significantly above the main sequence are usually considered to be short-lived starbursts, whereas those significantly below the main sequence are typically quenched. The position of galaxies relative to the main sequence provides insights into their evolutionary state and can be used to infer the role of any environmentally-driven mechanisms enhancing or inhibiting star formation activity. 
Note that while ALESS~5.1 does not appear to reside in an overdense structure, the properties of the coeval \OIII+\Hb\ emitters in its vicinity are still of interest and we thus include them in this part of the analysis.

\begin{figure}
	\includegraphics[width=0.99\columnwidth]{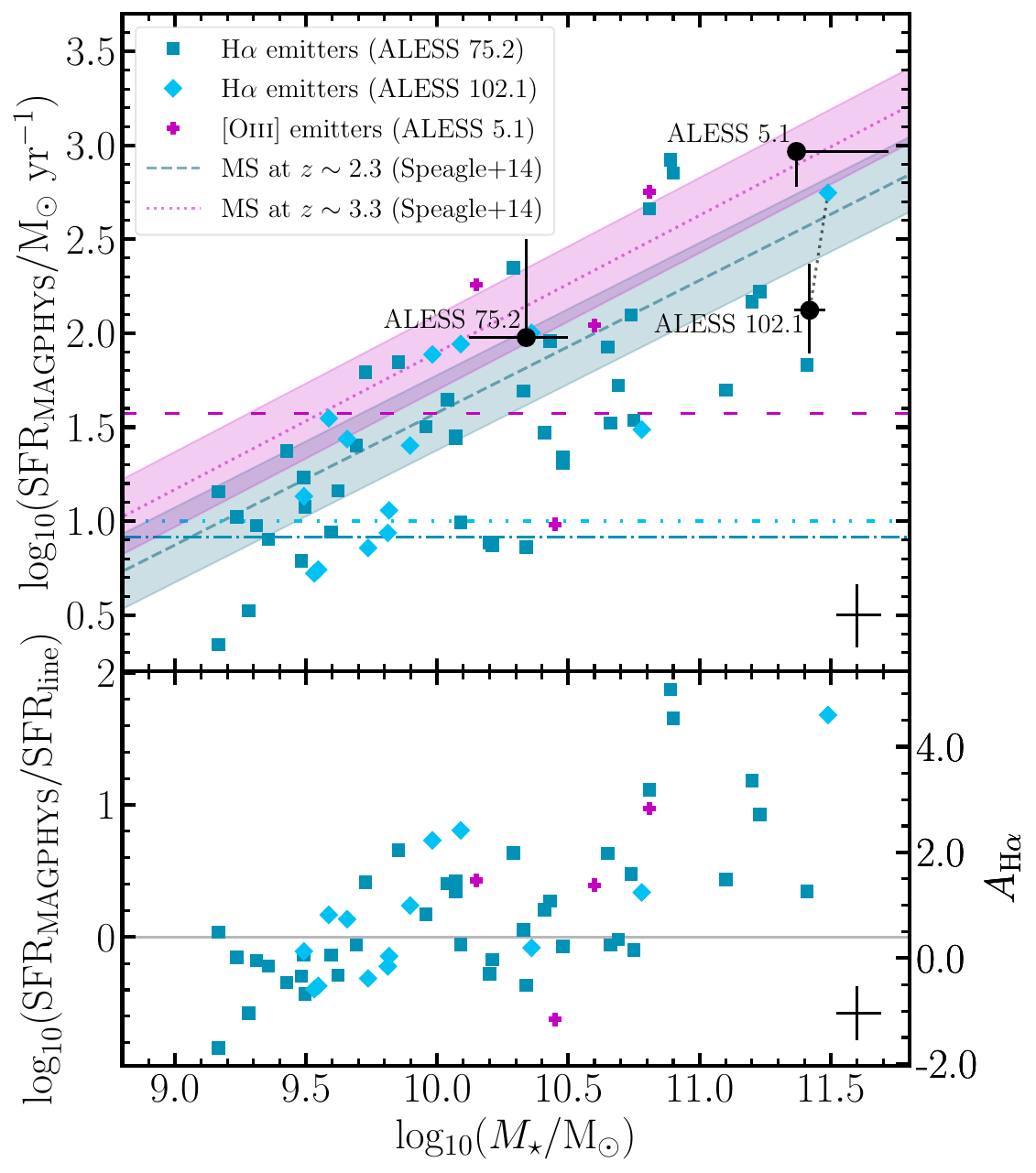}
	\caption{{\it (Top)} \magphys-derived SFR vs stellar mass for the \Ha\ and \OIII\ emitters identified in this study, compared with the $z {\sim} 2.3$ and $z {\sim} 3.3$ main sequence \citep[shaded regions represent 0.2~dex scatter;][]{Speagle+2014}. 
    The target SMGs are also shown, with masses and SFRs from \citet{Danielson+2017} and \citet{Birkin+2021} (black points; ALESS~102.1 is connected with a black dotted line to the counterpart \Ha\ emitter identified from our data; discussed in \S\ref{SSS:MS}). 
    Dashed, dot-dashed, and dot-dot-dashed horizontal lines correspond to the minimum SFR sensitivity of our survey in the environments of ALESS~5.1, 75.2, and 102.1, respectively, based on estimates using the line luminosities (Equation \ref{eq:SFR_Ha}). The galaxies generally follow the main sequence for their respective epochs, with some scatter in all three SMG environments.
    {\it (Bottom)} Ratio of \magphys-derived SFRs to the SFRs derived from line luminosities using fixed dust \Ha\ (\OIII) dust attenuations of 1.0 (1.35) mag, plotted as a function of stellar mass. The righthand axis shows the corresponding dust attenuation required to make the line luminosity-derived SFR match the \magphys-derived SFR for \Ha\ emitters, $A_{{\rm H}\alpha}$ (Equation \ref{eq:AHa}). (Analogous values for \OIII\ emitters, $A_{[{\rm O}\textsc{iii}]}$, can be obtained by adding 0.14.) The observed correlation suggests that assuming a constant dust attenuation for all \Ha/\OIII\ emitters results in underestimated SFRs at high stellar masses; such an approximation should therefore be used with caution. The black cross in the bottom-right of each panel shows the mean parameter uncertainties.
    }
	\label{fig:MS}
\end{figure}

We obtain stellar masses and SFRs for our galaxies by using the SED fitting code, \magphys\  \citep{daCunhaMAGPHYS+2008}, to fit SEDs to the same fixed-aperture photometry used to derive photometric redshifts in \S\ref{SS:contaminants}. Figure \ref{fig:MS} compares the relationship between SFR and stellar mass for the \Ha\ and \OIII+\Hb\ emitters that are SMG companions with the main-sequence at similar epochs using the prescription from \citet{Speagle+2014}. These galaxies generally scatter about the main sequence at their respective epochs, following a similar trend of increasing SFR with increasing stellar mass. 
We thus find no significant evidence of enhanced star formation in any of these SMG environments, despite the range of overdensities that they span; this is contrary to some previous studies in which enhanced SFRs have been found in overdense environments at $z \gtrsim 1$ \citep[e.g.][]{Elbaz+2007,Cooper+2008,Lemaux+2022}, however it is consistent with several other studies in which no environmentally-driven SFR enhancement is observed at high redshift \citep[e.g.][]{Scoville+2013,Darvish+2016,Zavala+2019}.

In addition to the \magphys-derived SFRs, we also calculate the dust-corrected star formation rates for the \Ha\ emitters using the scaling relation from \citet{Kennicutt1998}, modified for a \citet{Chabrier2003} IMF: 
\begin{equation}\label{eq:SFR_Ha}
\textrm{SFR} (\textrm{M}_{\odot}\ \textrm{yr}^{-1}) = 4.65 \times 10^{-42} L_{\textrm{\Ha}}\ (\textrm{erg s}^{-1})\,,
\end{equation}
\noindent where the \Ha\ line flux has been corrected for contamination by the nearby \NII\ doublet (see \S\ref{SSS:NII_correction}) and we assume a dust attenuation of $A_{\textrm{\Ha}} = 1$ mag (\S\ref{SSS:attenuation}). For the \OIII+\Hb\ emitters we assume an attenuation of $A_{\textrm{\OIII+\Hb}} = 1.35$\,mag following \citet{Khostovan+2015} and use the relation between SFR and $L_{\textrm{\OIII+\Hb}}$ from \citet{OsterbrockFerland2006}, similarly modified for a \citet{Chabrier2003} IMF:
\begin{equation}\label{eq:OIII_SFR} 
\textrm{SFR} (\textrm{M}_{\odot}\ \textrm{yr}^{-1}) = 4.32 \times 10^{-42} L_{\textrm{\OIII+\Hb}}\ (\textrm{erg s}^{-1})\,. 
\end{equation}
\noindent The bottom panel of Figure \ref{fig:MS} shows how the ratio of the {\sc magphys}-derived and line-derived SFR estimates varies with stellar mass. Also shown is the \Ha\ dust attenuation required for the SFR derived from Equation \ref{eq:SFR_Ha}, SFR$_{{\rm H}\alpha}$, to agree with the \magphys-derived value, SFR$_{\textsc{magphys}}$, as given by:
\begin{equation}\label{eq:AHa}
A_{{\rm H}\alpha} = 2.5\log_{10}\left(\frac{{\rm SFR}_{\textsc{magphys}}}{{\rm SFR}_{{\rm H}\alpha}}\right) + 0.4\,. 
\end{equation}
\noindent An analagous equation for $A_{\textrm{\OIII+\Hb}}$ can be obtained by adding 0.14 mag. It is evident that as one moves to higher stellar mass, the assumption that $A_{{\rm H}\alpha}~(A_{\textrm{\OIII+\Hb}}) = 1.0~(1.35)$ mag results in underestimated SFRs compared with the results from SED fitting. We therefore caution that while such an assumption may be suitable for galaxies with low-to-average stellar mass, it becomes less reliable for high-mass galaxies.

We also include the SMGs themselves in Figure~\ref{fig:MS}, with the SFRs and stellar masses for these calculated by \citet{Danielson+2017} and \citet{Birkin+2021} using \magphys. As expected for sources selected due to their infrared-brightness, the SMGs are among the most active galaxies in the observed fields. ALESS~102.1 is also one of the most massive galaxies in its environment, which suggests that if it is in a protocluster then it may be brightest cluster galaxy (BCG) progenitor, i.e. a proto-BCG. 
Similarly, ALESS~5.1 is massive relative to other galaxies in the surrounding region, but given the low density of this environment we deem it unlikely that this SMG is a proto-BCG. 
Conversely, ALESS~75.2 has a lower mass, which is not exceptional for its environment, and which points towards it being more likely to evolve into a normal cluster member. This is consistent with the spatial analysis of \Ha\ emitters and LAEs (\S\ref{SS:annuli}), which showed that ALESS~75.2 is offset from the densest regions of this field. 

Since ALESS~102.1 has a counterpart \Ha\ emitter in our sample, we also compare our \magphys-derived SFR and stellar mass with those derived by \citet{Danielson+2017}. Our stellar mass of $\log(M_{\star}/{\rm M}_{\odot}) = 11.49^{+0.18}_{-0.05}$ is in good agreement with their value of $\log(M_{\star}/{\rm M}_{\odot}) = 11.42^{+0.06}_{-0.06}$. Conversely, our SFR of $\log({\rm SFR}/{\rm M}_{\odot} {\rm yr}^{-1}) = 2.75^{+0.22}_{-0.25}$ is significantly higher than their estimate of $\log({\rm SFR}/{\rm M}_{\odot} {\rm yr}^{-1}) = 2.12^{+0.25}_{-0.23}$. This is likely due to the inclusion of FIR and radio photometry in their SED fitting which are absent from our own fit; the dust component (and thus the SFR) is better constrained in the \citet{Danielson+2017} SED fit. We therefore opt to use their values of SFR and stellar mass for this galaxy instead of our own.

\subsection{Stellar mass functions}\label{SSS:SMFs}

We next construct the stellar mass functions of the galaxies around each SMG and compare these with the blank field. The stellar mass functions are derived following a similar procedure as for the luminosity functions (see \S\ref{SS:LFs}), minus the corrections that are only relevant to luminosity functions (dust attenuation, line flux contamination, filter profile corrections). Completeness corrections were applied to each mass bin according to the completeness values estimated in \S\ref{SSS:completeness} based on the emission line fluxes. We then fit Schechter functions to the data:
\begin{equation}
\Phi(M_{\star})dM_{\star} = \ln(10) \Phi^{\ast} \left(\frac{M_{\star}}{M^{\ast}}\right)^{\alpha + 1} e^{-(M_{\star}/M^{\ast})} d\log M_{\star}\,,
\end{equation}
\noindent where $\Phi(M_{\star})$ is the number density at stellar mass $M_{\star}$, $\Phi^{\ast}$ is the normalisation of the stellar mass function, $M^{\ast}$ is the characteristic stellar mass, and $\alpha$ is the slope at the faint end of the stellar mass function. Mass bins that are less than 50 percent complete are excluded from the fitting procedure. 
As with the luminosity functions, we also fix the faint-end slope $\alpha$ to the  values derived for blank-field stellar mass functions by \citet{Sobral+2014} and \citet{Khostovan+2016} (i.e. $\alpha = -1.37$ and $\alpha = -1.3$ for the for the \Ha\ and \OIII+\Hb\ emitters, respectively). For \OIII+\Hb\ emitters we also fix the characteristic stellar mass, $M^{\ast}$, to the blank-field value of $\log(M^{\ast} / M_{\odot}) = 10.96$ \citep{Khostovan+2016}. 

\begin{figure*}
	\includegraphics[width=0.95\columnwidth]{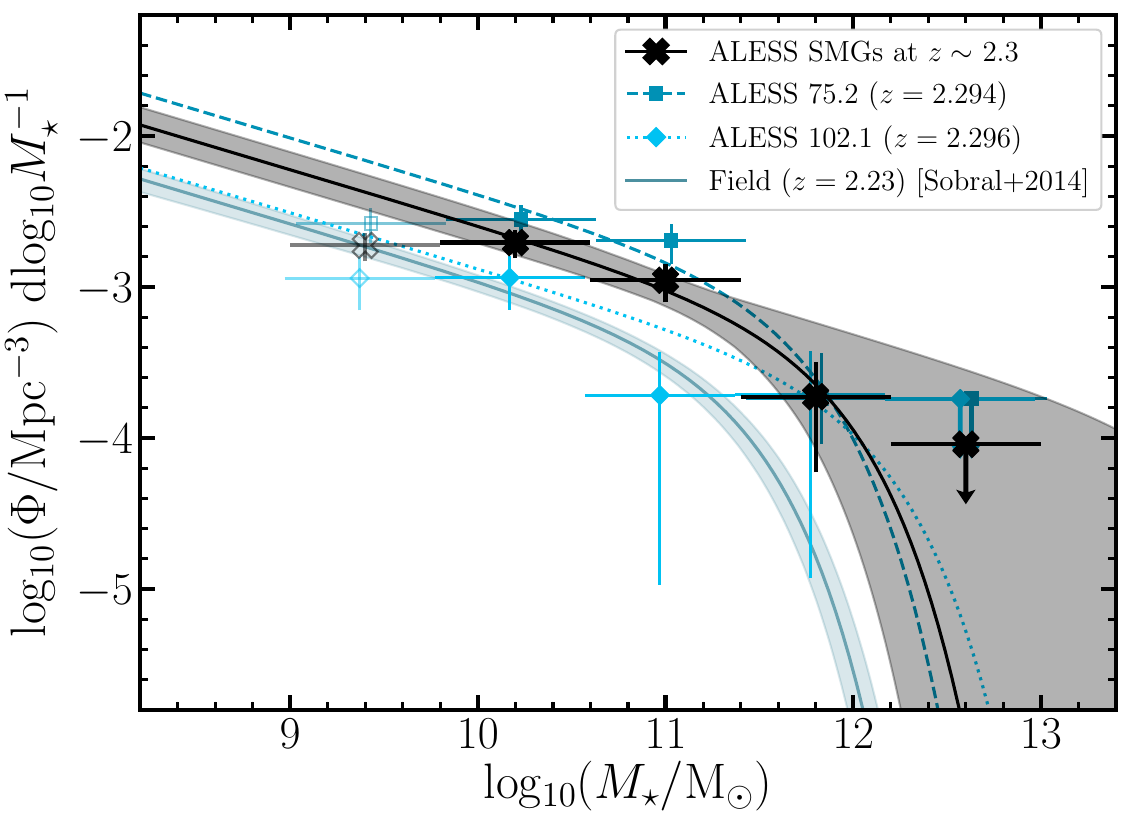}
	\includegraphics[width=0.95\columnwidth]{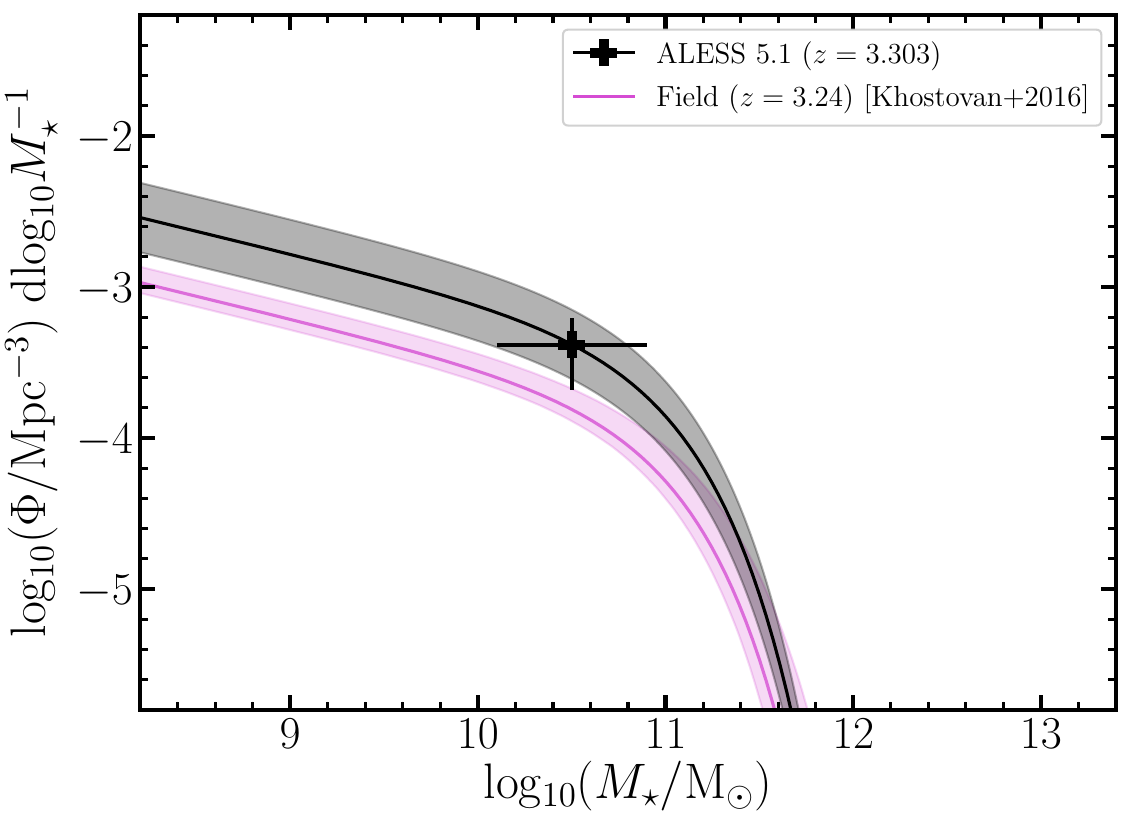}
	\caption{Stellar mass functions for the \Ha\ (\textit{left}) and \OIII\ (\textit{right}) emitters identified in this study. The data are compared with blank-field studies of emission line galaxies at similar redshifts \citep{Sobral+2014,Khostovan+2016} and shaded regions represent 1$\sigma$ uncertainties. For each of our stellar mass functions, we fix the faint-end slope to the value derived for the blank field at a similar redshift: $\alpha = {-}1.37$ \citep{Sobral+2014} and $\alpha = -1.3$ \citep{Khostovan+2016} for the \Ha\ and \OIII+\Hb\ stellar mass functions, respectively. For the \OIII+\Hb\ stellar mass function, we also fix the characteristic stellar mass to the blank-field value of $\log(M^{\ast}/M_{\odot}) = 10.96$ \citep{Khostovan+2016}. For the \Ha\ emitters, the upper limit of $M^{\ast}$ is poorly constrained (see also Figure~\ref{fig:SMF_contours}), which leads to large uncertainties at the high mass end. There are offsets between the blank field stellar mass functions and those around our SMGs in all targeted SMG regions; these are quantified in Figure~\ref{fig:SMF_contours} and Table~\ref{tab:SMF_summary}.
 }
	\label{fig:SMFs}
\end{figure*}

\begin{figure}
	\includegraphics[width=0.95\columnwidth]{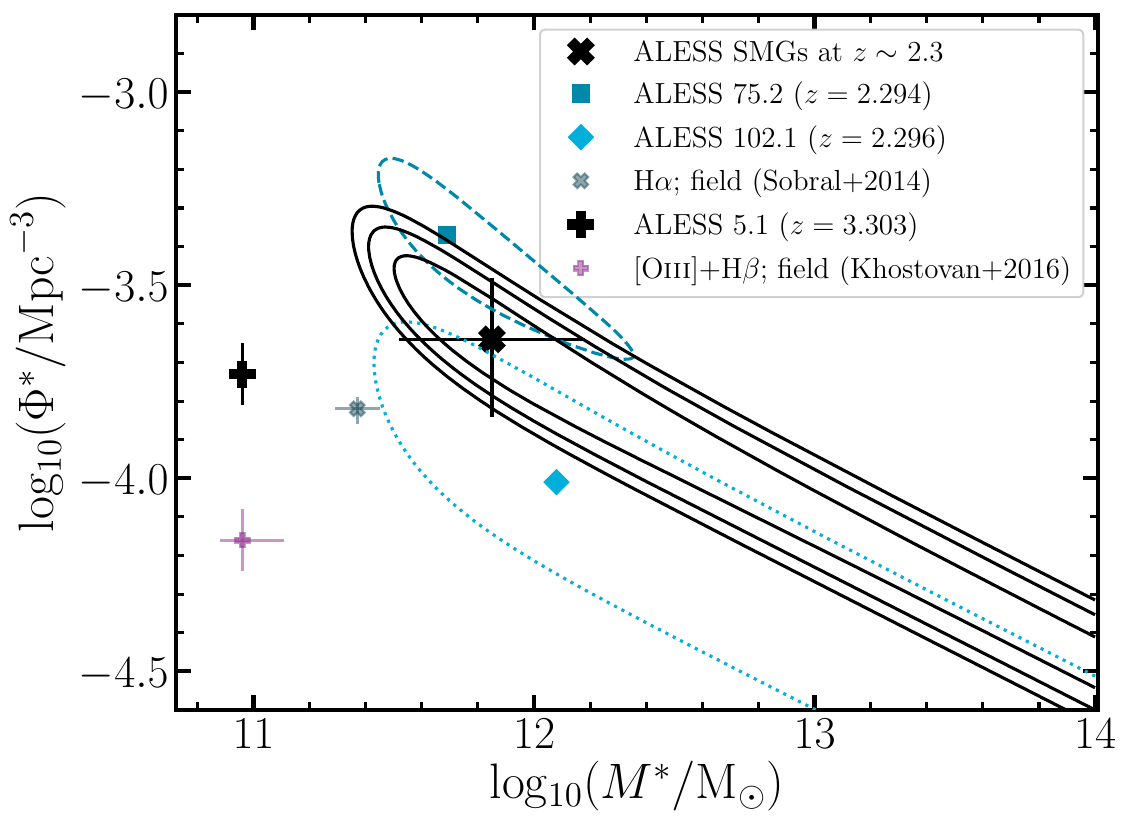}
	\caption{Contours showing the correlated uncertainties on the Schechter fit parameters for the stellar mass functions shown in figure \ref{fig:SMFs}. In all cases, the faint-end slope, $\alpha$, is fixed to match to the blank-field stellar mass functions \citep{Sobral+2014, Khostovan+2016}, and for ALESS~5.1 $\textrm{M}^{\ast}$ is also fixed.
 Symbols and contours have the same meaning as in Figure \ref{fig:contours}. For the \Ha\ emitters at $z\sim2.3$ the characteristic stellar mass, $M^{\ast}$, is effectively a lower limit due to the correlation with $\Phi^*$; this can also be seen in Figure~\ref{fig:SMFs}. 
 The offset between $M^*$ for field \Ha\ emitters \citep{Sobral+2014} and the galaxies around $z\sim2.3$ SMGs indicates that there is an excess of high-mass galaxies around the SMGs, and this is likely partially responsible for the overdensity around ALESS~75.2.} 
	\label{fig:SMF_contours}
\end{figure}

The stellar mass functions are presented in Figure~\ref{fig:SMFs}, with the parameters in  Table~\ref{tab:SMF_summary}. Uncertainties in each parameter are estimated following the same procedure as for those of the luminosity functions (see \S\ref{SSS:Schechter}) and the correlation between the parameters and their uncertainties are shown in  Figure~\ref{fig:SMF_contours}, which demonstrates that at the upper limit the characteristic stellar mass, $M^{\ast}$, is poorly constrained for all of our samples except the \Ha\ emitters around ALESS~75.2. 
However, the lower limit is sufficient to show that in the $z\sim2.3$ SMG regions the characteristic stellar mass is significantly higher than the $z\sim2.3$ field, which suggests that the stellar mass build-up in SMG companion galaxies is further advanced than the coeval field \citep[e.g.][]{Muzzin+2013}. Due to our selection of \Ha\ emitters the galaxies have non-negligible star-formation rates (though many are below the main sequence; \S\ref{SSS:MS}). Observations using a local galaxy density estimator suggest that local environment has minimal effect on the stellar mass function of star-forming or quiescent galaxies at $z=1.5$--2 \citep{Papovich+2018}. However, there is evidence of protocluster environments being skewed towards containing galaxies with higher masses than the field \citep[e.g.][]{Cooke+2014}, consistent with our results.

\begin{table}
	\centering
	\caption{Summary of the best-fit stellar mass function parameters for the companion galaxies in the environments of the target SMGs. In all cases, the faint-end slope of the stellar mass function, $\alpha$, is fixed to the value from the relevant blank-field stellar mass function \citep{Sobral+2014,Khostovan+2016}. Where indicated, the characteristic stellar mass, $M^{\ast}$, is also fixed to the blank-field value from \citet{Khostovan+2016}.}
	\label{tab:SMF_summary}
	\begin{tabular}{cccc}
		\hline
		SMG environment & $\log(M^{\ast} / M_{\odot})$ & $\log(\Phi^{\ast} / \textrm{Mpc}^{-3})$\\
		\hline
		\hline
		\vspace{0.5em}
		ALESS~75.2 & $11.69^{+0.41}_{-0.15}$ & $-3.37^{+0.09}_{-0.25}$\\
		\vspace{0.5em}
		ALESS~102.1 & $12.08^{+0.52}_{-0.59}$ & $-4.01^{+0.23}_{-0.41}$\\
		\hline
		SMGs at $z \sim 2.3$ & $11.85^{+0.30}_{-0.34}$ & $-3.64^{+0.16}_{-0.18}$\\
		\hline
		ALESS~5.1 & $10.96$ (fixed) & $-3.73^{+0.08}_{-0.08}$\\
		\hline
	\end{tabular}
\end{table}

\subsection{Dark matter halo masses and evolution}
\label{SSS:total_masses}
We next estimate the total halo masses of the SMG environments in order to place them within the context of existing protoclusters and trace their likely evolution, focusing primarily on the overdense environments of the two SMGs at $z \sim 2.3$. 
Since these overdensities are unvirialised and lack a detectable intracluster medium, the classic methods for weighing galaxy clusters cannot be used. 
Instead, we use two methods that have been used in protocluster studies, though the underlying assumptions required can lead to significant uncertainties, as discussed in the following subsections.
The first method is detailed in \S\ref{sec:masses_SHMR} and uses the stellar-to-halo mass relation (SHMR) to estimate the high-redshift mass of the clusters (hereafter SHMR method) and evolve it to the local Universe using the Millennium and Millennium-II simulations \citep{McBride+2009, Fakhouri+2010}. 
The second method follows \citet{Steidel+1998} and assumes the region of interest is a homogeneous sphere undergoing spherical collapse and uses the overdensity parameter to estimate the $z=0$ descendant mass, which we trace back to high-redshift using the Millennium and Millenium-II simulations. This is referred to as the spherical collapse model (SCM) method and detailed in \S\ref{sec:masses_SCM}.
In \S\ref{sec:masses_combined_evol} we discuss the evolution of the SMG environments compared with other systems and previous measurements.

\subsubsection{The SHMR method for deriving halo masses}
\label{sec:masses_SHMR}

The SHMR method for estimating protocluster masses involves identifying the most massive galaxy in the structure and converting its stellar mass to a halo mass, and taking this to be the halo mass of the whole structure. This method has been employed in recent protocluster studies \citep[e.g.][]{Long+2020,Calvi+2021,Sillassen+2022,Ito+2023} and implicitly assumes that all member galaxies occupy a single halo at the observed redshift of the structure, which may not be the case if some of the galaxies are still infalling. Nevertheless, we deem this assumption preferable to the commonly-used alternative of estimating the halo masses of each individual galaxy and summing them together \citep[e.g.][]{Long+2020,Calvi+2021}, which risks `double-counting' overlapping dark matter halos to produce an overestimate of the structure halo mass. 
Note that we perform this calculation even for ALESS~5.1 and its surrounding \OIII\ emitters despite our analyses revealing no signs of their environment being significantly overdense. This is because this method does not explicitly depend on the density of the surrounding environment, and the high stellar mass of ALESS~5.1 (see \S\ref{SSS:MS} and \ref{fig:MS}) suggests it may yet reside in a massive halo. We do however caution that the result obtained here likely represents an extreme upper limit on the mass of any possible structure around ALESS~5.1.

We estimate halo masses for each target SMG and their candidate companion galaxies using the SHMR from \citet{Behroozi+2013}. We use the relation as defined at $z = 2$ for galaxies in the environments of ALESS~75.2 and 102.1, and at $z = 3$ for galaxies in the environment of ALESS~5.1. 
For the SMGs themselves we use the stellar masses from the literature \citep[][see also \S\ref{SSS:MS}]{Danielson+2017,Birkin+2021}. 
Some of our galaxies have stellar masses which lie above the range at which the SHMR is defined and for these we use the stellar-to-halo mass ratio for the largest halo mass at which the relation is defined \citep[see Figure~7 of][]{Behroozi+2013} to convert the stellar mass to a halo mass. This affects only two \Ha\ emitters from P5\_75, along with one from P102 which we have identified as a counterpart to ALESS~102.1 (see \S\ref{SSS:MS} and Figure \ref{fig:MS}). None of the \OIII\ emitters have stellar masses above the range for which the SHMR is defined at $z \sim 3$, but ALESS~5.1 does lie above this range.  Uncertainties on individual galaxy halo masses are estimated based on the stellar mass uncertainties and the uncertainties in the SHMR derived by \citet{Behroozi+2013}.

We derive halo masses of $\log(M_{h}/{\rm M}_{\odot}) = 11.45\text{--}14.46$ for individual \Ha\ and \OIII\ emitters, with medians of $\log(M_{h}/{\rm M}_{\odot}) = 12.16^{+0.21}_{-0.16}, 11.94^{+0.46}_{-0.35}, 11.75^{+0.26}_{-0.12}$ for galaxies in the environments of ALESS~5.1, 75.2 and 102.1, respectively. The halo masses of the corresponding SMGs are $\log(M_{h}/{\rm M}_{\odot}) = 13.94^{+0.38}_{-0.21}, 12.02^{+0.08}_{-0.25}, 14.39^{+0.03}_{-0.37}$, derived using their stellar masses reported in the literature \citep{Danielson+2017,Birkin+2021}. 
ALESS~5.1 and 102.1 are both the most massive galaxies in their respective environments; we therefore adopt their halo masses as the total masses of the potential structures at the observed redshifts. ALESS~75.2 is not the dominant galaxy in its environment, being surpassed in stellar (and hence inferred halo) mass by ${\sim}40$ percent of its companion \Ha\ emitters. The most massive of these is a spectroscopically confirmed member \citep{Popesso+2009,Balestra+2010} located in the \Ha\ emitter density peak associated with the \Lya\ blob CDFS-LAB03 (see Figure \ref{fig:density_maps}), with a halo mass of $\log(M_{h}/{\rm M}_{\odot}) = 14.38^{+0.01}_{-0.40}$. We thus assume this is the total mass of the surrounding structure. Since high-redshift radio galaxies are commonly found in protocluster cores \citep[e.g.][]{Kurk+2000,Venemans+2002,Kuiper+2011a,Wylezalek+2013,Hayashi+2012,Cooke+2014}, we search for radio counterparts for this galaxy in the second data release from the Very Large Array 1.4 GHz survey of the ECDFS \citep{Miller+2013}, for which the typical sensitivity is 7.4 \ujy\ per $2\farcs8\times1\farcs6$ beam. We find no counterparts within $30\arcsec$ of this \Ha\ emitter and thus rule it out as being a high-redshift radio galaxy.

The total halo masses at the observed redshift obtained using the SHMR method are thus $\log(M_{h}/{\rm M}_{\odot}) = 13.94^{+0.38}_{-0.21}, 14.38^{+0.01}_{-0.40}, 14.39^{+0.03}_{-0.37}$ for the environments of ALESS~5.1, 75.2, and 102.1, respectively. 
We note that these masses may be affected by systematic uncertainties on the stellar masses (due to uncertainties on star-formation histories and resulting mass-to-light ratios, which is particularly relevant for young starbursts; e.g. \citealt{Wardlow+2011}) and on the SHMR for very high mass galaxies, which are present in the simulations from which the SHMR is derived \citep{Behroozi+2013}. Indeed, predictions from halo mass functions suggest that the halo masses inferred from this method should be sufficiently rare that finding three such structures in the $\sim0.25$\,deg$^2$ ECDFS is unlikely \citep[e.g.][]{PressSchechter1974,Tinker+2008}. 
Therefore, we consider these SHMR-derived halo masses to be upper limits and as such they are represented by the upper bounds on Figure~\ref{fig:PC_comparison} (the lower bounds are derived in \S\ref{sec:masses_SCM}), which compares the halo masses of the SMG environments with previously-studied galaxy clusters and protoclusters. 

To assess whether these SMG environments are true protoclusters, we evolve the masses derived from the SHMR method to the present day masses and compare with known galaxy clusters in the local Universe. This is done using the redshift-dependent formula for the mean mass accretion rate derived from the results of the Millennium and Millennium-II simulations \citep{McBride+2009,Fakhouri+2010}:
\begin{equation}
  \begin{aligned}
    \langle\dot {M} \rangle_{\textrm{mean}} = &\ 46.1 {\rm M}_{\odot} \textrm{yr}^{-1} \left(\frac{M_{z}}{10^{12}M_{\odot}}\right)(1 + 1.11z) \\ 
    & \times \sqrt{\Omega_{m,0}(1+z)^{3} + \Omega_{\Lambda,0}}\,,
  \end{aligned}
  \label{eq:dmdt}
\end{equation}
\noindent where $M_{z}$ is the halo mass of the structure at its observed redshift, and $\Omega_{m,0}$ and $\Omega_{\Lambda,0}$ are the present-day density parameters for matter and the cosmological constant according to our assumed cosmology \citep{Planck2018}. 

For each overdensity we begin with the total halo masses estimated using the SHMR method and apply Equation~\ref{eq:dmdt} to incrementally add mass in small time steps until the present day is reached. The present-day masses obtained with this method are $\log(M_{h,z=0}/{\rm M}_{\odot}) = 15.93^{+0.62}_{-0.33}, 15.81^{+0.01}_{-0.55}, 15.82^{+0.04}_{-0.50}$ for the overdensities containing ALESS~5.1, ALESS~75.2, and ALESS~102.1, respectively.
These masses suggest that these structures would all evolve into some of the most massive clusters in the Universe, rivalling that of the Coma cluster \citep[e.g.][]{Gavazzi+2009,Ho+2022} and other massive clusters at $z \lesssim 1$ such as those in the CLASH \citep[][]{Postman+2012,Merten+2015}. However, given the rarity of such massive structures seen in the local Universe, we posit that the identification of progenitor structures around all three of our target SMGs is due to the systematics in the calculations, rather than a real occurrence. 
In addition to the possible systematics in the stellar masses and SHMR, as previously described, we also note that the mean mass accretion rate given by Equation \ref{eq:dmdt} is poorly constrained for halos with masses of $\log(M_{h}/{\rm M}_{\odot}) \gtrsim 14$ beyond $z \sim 0.5$. 
Furthermore, Equation \ref{eq:dmdt} alone does not account for the diversity of evolutionary paths that real dark matter halos undergo, being the mean result for many halos in the Millennium simulation. 
Therefore, as for the high-redshift SHMR-derived halo masses, we also take these SHMR-derived $z=0$ and intermediate masses to be upper limits. 
This upper limit on the halo mass at the SMG redshift and the evolution to the present day is shown in Figure~\ref{fig:PC_comparison} as the upper edges of the shaded regions for ALESS~75.2 and 102.1, and as a single solid line for ALESS~5.1.

\subsubsection{The SCM method for deriving halo masses}
\label{sec:masses_SCM}

An alternative method for estimating the present-day mass of each SMG environment is obtained following \citet{Steidel+1998}, which approximates each SMG environment as a homogeneous sphere undergoing spherical collapse. In this case the total present-day mass is given by:
\begin{equation}
  M_{h,z=0} = \bar{\rho} V (1 + \delta_{m})\,,
\end{equation}
\noindent where $\bar{\rho}$ is the mean comoving matter density of the Universe, $\delta_{m}$ is the dark matter mass overdensity, and $V$ is the comoving volume of the structure. We refer to this method as the spherical collapse model (SCM) method. Since the assumption of spherical collapse is unphysical for environments that are not overdense, we only perform this calculation for the two SMG environments at $z \sim 2.3$.

To estimate the volume of each overdensity, we assume that the structures are spherical and use the spatial extent of the structure on the sky to infer the angular diameter of the sphere containing the member galaxies. 
As discussed in \S\ref{SS:annuli}, the overdensity around ALESS~75.2 extends beyond the confines of the HAWK-I pointing and therefore the size of the HAWK-I field-of-view can be used as a lower limit on the angular diameter of the structure. 
Therefore, for this environment we calculate the comoving volume for a spatial extent of $\sim 7.5\arcmin$ \citep{HAWKI2008}, which corresponds to a comoving volume of $V \sim 1000$~cMpc$^{3}$. This volume should be considered a lower limit, and thus the derived halo mass is also a lower limit. 
For the environment of ALESS~102.1, we assume that the structure is confined to the quadrant containing the SMG (\S\ref{SS:annuli}), such that the angular diameter of the sphere is then $\sim 220\arcsec$ \citep{HAWKI2008}, which gives a comoving volume of $V \sim 110$~cMpc$^{3}$.

The dark matter mass overdensity is linked to the observed galaxy overdensity, $\delta_{g}$, via 
\begin{equation}
    \delta_{m} = \delta_{g}/b\,,
\end{equation}
\noindent where $b$ is the bias parameter. To estimate the bias parameters for \Ha\ emitters at $z \sim 2.3$, we make use of the relationship between $b$ and $L_{\textrm{H}\alpha}$ derived by \citet{Cochrane+2017} at $z = 2.23$. Following a similar method to that of \citet{Stott+2020}, we derive a linear fit to this relation and estimate $b$ at the mean value of $L_{\textrm{H}\alpha}$ for each SMG environment. This gives $b_{\textrm{H}\alpha} = 2.9^{+0.2}_{-0.1}\ {\rm and}\ 2.8^{+0.2}_{-0.3}$ for the candidate companions of ALESS~75.2 and 102.1, respectively.

Using the above values of $b$ and $V$ along with the $\delta_{g}$ values calculated in \S\ref{SS:BF_compare}, we obtain present-day halo masses of 
${\rm log}(M_{h,z=0}/{\rm M}_{\odot}) = 13.90^{+0.09}_{-0.10}$ and $13.05^{+0.12}_{-0.14}$ 
for the environments of ALESS~75.2 and ALESS~102.1, respectively. Contrary to the previous present-day mass estimates, these masses suggest that descendant of the environment of ALESS~102.1 is more akin to a galaxy group than a galaxy cluster \citep[e.g.][]{Han+2015,Man+2019}, while the environment of ALESS~75.2 may evolve into a `Virgo-like' or `Fornax-like' cluster by $z = 0$ \citep[e.g.][]{Chiang+2013}.

\begin{figure}
  \includegraphics[width=0.95\columnwidth]{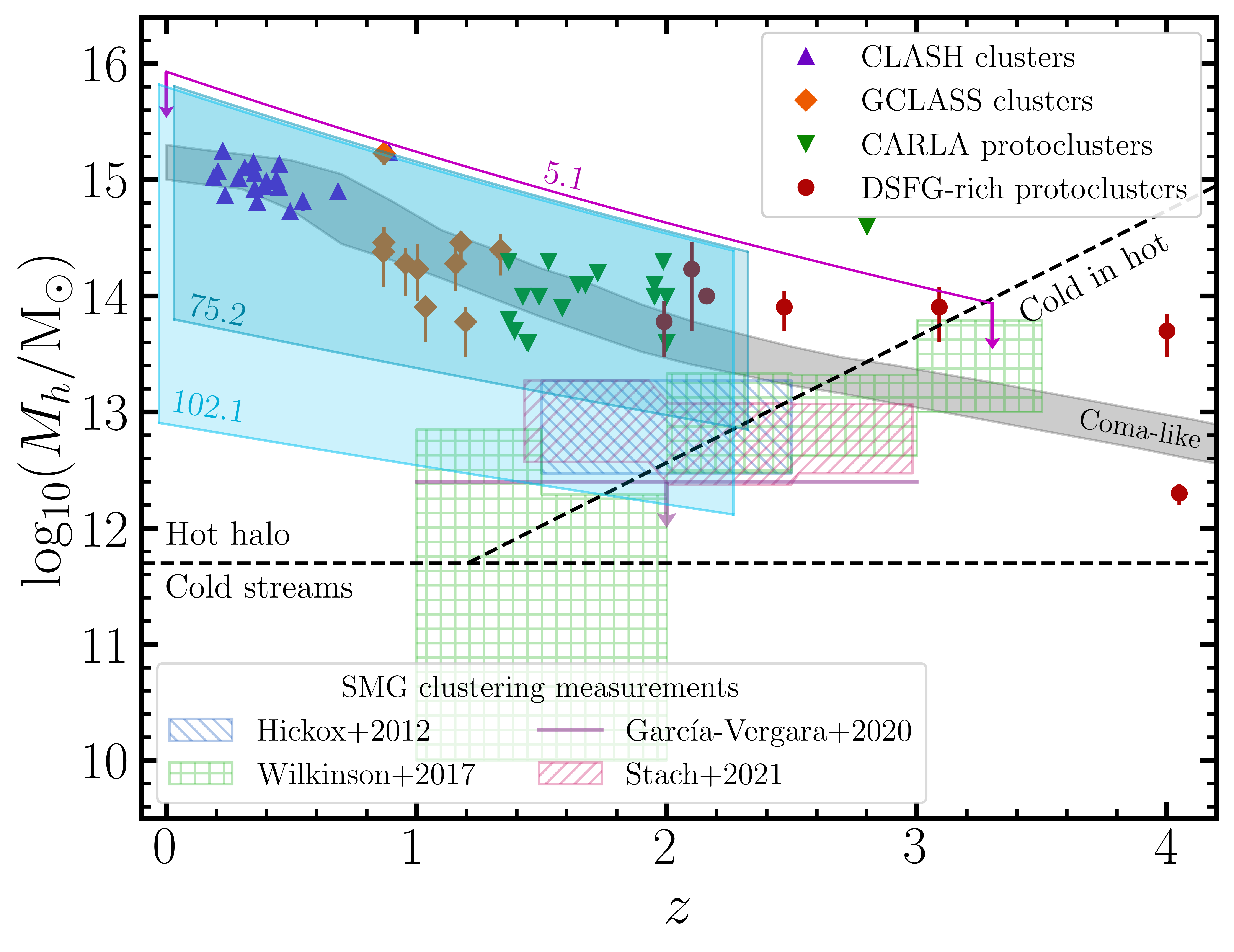}
  \caption{A comparison of protocluster halo masses across cosmic time. Two mass estimates are obtained for the environments of ALESS~75.2 and 102.1, at both the observed redshift of the potential structure and at $z = 0$, using the methods described in \S\ref{SSS:total_masses}. The evolutionary paths of these halos across cosmic time are estimated using the redshift-dependent mean mass accretion rate formula derived from the Millennium simulations \citep{McBride+2009,Fakhouri+2010}. 
 The coloured shaded regions show the possible mass ranges and evolution for each of our target SMG environments, and are labelled with the ALESS ID of the inhabiting SMG. The upper bounds of these mass ranges correspond to the masses estimated using the stellar-to-halo mass relations from \citet{Behroozi+2013}, while the lower bounds are derived by assuming a spherical collapse model (see text for details). For ALESS~5.1 only the former mass estimate is used, and is marked as an upper limit. 
 The grey shaded region shows the expected evolution of a Coma-like cluster \citep{Chiang+2013} and coloured symbols show samples of clusters and protocluster from CLASH, GCLASS and CARLA, and protoclusters targeted due to their richness in DSFGs, as detailed in \S\ref{sec:masses_combined_evol}. 
 We include regions showing measurements of SMG halo masses obtained from clustering studies \citep{Hickox+2012,Wilkinson+2017,Garcia-Vergara+2020,Stach+2021} and mark the borders between different gas regimes \citep{DekelBirnboim2006}. 
 The two $z \sim 2.3$ SMGs reside in environments consistent with protoclusters, although ALESS~102.1 may reside in a protogroup instead. The lower bounds of our mass estimates are broadly consistent with the masses obtained from SMG clustering studies, while the upper bounds imply these halos may evolve into Coma-like structures or larger. 
 }
  \label{fig:PC_comparison}
\end{figure}

The significant disparity between the present-day halo mass estimates from the SCM method compared with the SHMR method likely arises from the assumptions and uncertainties in the calculations, including the systematics discussed in \S\ref{sec:masses_SHMR} and the estimates of the structure volumes. 
Note that for ALESS~75.2 the derived $z=0$ SCM halo mass is a lower limit, due to the volume used being a lower limit. 
Since the masses derived from the SCM method are all lower than the those from the SHMR method and its uncertainties, we proceed with the SCM estimates and adopt the lower bound of the 1--$\sigma$ confidence intervals as lower limits for the present-day halo masses.
Thus, the present-day halo mass estimates from the two methods gives a range of plausible evolutionary pathways for the two $z \sim 2.3$ SMG environments as shown on Figure~\ref{fig:PC_comparison}.

We also use the present-day halo mass estimates from the SCM method in combination with  Equation~\ref{eq:dmdt} to trace the evolution of these SMG environments back in time to their observed redshifts, thereby obtaining a second estimate of the total mass at these redshifts. These masses are 
$\log(M_{h}/{\rm M}_{\odot}) = 12.93^{+0.07}_{-0.08}$ and $12.24^{+0.10}_{-0.11}$ 
for the potential structures around ALESS~75.2 and ALESS~102.1, respectively. 
This calculation provides an evolutionary track that connects the lower halo mass limit at the $z \sim 2.3$ to the corresponding value at $z = 0$, and this is what defines the bottom edge of each shaded region in Figure~\ref{fig:PC_comparison}.

\begin{table*}
	\centering
	\caption{Halo mass estimates for each SMG environment.}
	\label{tab:halo_masses}
	\begin{tabular}{ccccc}
		\hline
		SMG & $\log(M_{h}^{\rm SHMR}/{\rm M}_{\odot})^a$ & $\log(M_{h,z=0}^{\rm SHMR}/{\rm M}_{\odot})^b$ & $\log(M_{h}^{\rm SCM}/{\rm M}_{\odot})^c$ & $\log(M_{h,z=0}^{\rm SCM}/{\rm M}_{\odot})^d$\\
		\hline
		\hline
		\vspace{0.5em}
		ALESS 5.1 & ${13.94}^{+0.38}_{-0.21}$ & ${15.93}^{+0.63}_{-0.33}$ & -- & -- \\
		\vspace{0.5em}
		ALESS 75.2 & ${14.38}^{+0.01}_{-0.40}$ & ${15.81}^{+0.01}_{-0.55}$ & ${12.93}^{+0.07}_{-0.08}$ & ${13.90}^{+0.09}_{-0.10}$\\
		\vspace{0.5em}
		ALESS 102.1 & ${14.39}^{+0.03}_{-0.37}$ & ${15.82}^{+0.04}_{-0.50}$ & ${12.24}^{+0.10}_{-0.11}$ & ${13.05}^{+0.12}_{-0.14}$\\
		\hline
	\end{tabular}
	\begin{flushleft}
		$^a$ Halo mass derived using the SHMR method (\S\ref{sec:masses_SHMR}) at the redshift of the SMG (i.e.\ $z\sim2.3$ for ALESS~75.2 and ALESS~102.1, and $z\sim3.3$ for ALESS~5.1). As discussed in \S\ref{sec:masses_SHMR}, we consider these to be upper limits.\\
		$^b$ Halo mass derived using the SHMR method and evolved to $z=0$ using Equation~\ref{eq:dmdt}; these masses are considered to be upper limits (see \S\ref{sec:masses_SHMR}).\\
		$^c$ Halo mass derived using the SCM method and traced back to the SMG redshift using Equation~\ref{eq:dmdt} (\S\ref{sec:masses_SCM}).\\
		$^d$ Descendant halo mass at $z=0$ derived using the SCM method (\S\ref{sec:masses_SCM}).
	\end{flushleft}
\end{table*}

\subsubsection{Halo masses and evolution}
\label{sec:masses_combined_evol}

The halo mass estimates of each SMG environment are summarised in Table \ref{tab:halo_masses}, and Figure \ref{fig:PC_comparison} compares the SMG halo masses and their evolution with previously-studied galaxy clusters and protoclusters.  
As described in \S\ref{sec:masses_SHMR} and \S\ref{sec:masses_SCM}, the upper and lower bounds shown for the SMG halos are derived from the masses calculated using the SHMR method and the lower limits from the SCM method, respectively (with the exception of ALESS~5.1, for which only the SHMR method is used). Thus, this region encompasses the full range of possible halo masses and evolution for the SMGs. 
In Figure~\ref{fig:PC_comparison} these are compared with clusters from the Cluster Lensing And Supernova Survey with Hubble \citep[CLASH;][]{Postman+2012, Merten+2015}; clusters from the Gemini Cluster Astrophysics Spectroscopic Survey \citep[GCLASS;][]{Muzzin+2012,van_der_Burg+2014}; clusters and protoclusters from the Clusters Around Radio-Loud AGN program \citep[CARLA;][]{Wylezalek+2013,Mei+2022}. We also show the halo masses calculated by \citet{Casey2016} for overdense structures known to be rich in dusty star-forming galaxies (DSFGs), including: the GOODS-N protocluster at $z = 1.99$ \citep{Blain+2004,Chapman+2009}; the COSMOS protoclusters at $z = 2.10$ \citep{Yuan+2014} and $z = 2.47$ \citep{Casey+2015,Diener+2015,Chiang+2015}; the `Spiderweb' protocluster, MRC~1138-256, at $z = 2.16$ \citep{Kurk+2000,Kuiper+2011b}; the SSA~22 protocluster at $z = 3.09$ \citep{Steidel+1998,Hayashino+2004,Tamura+2009,Lehmer+2009,Umehata+2015}; the GN20 overdensity at $z = 4.05$ \citep{Daddi+2009,Hodge+2013b}. 
In this sample of DSFG-rich protoclusters we additionally include the halo mass of the DRC at $z = 4$ \citep{Long+2020}. 
Figure~\ref{fig:PC_comparison} shows that the potential structures surrounding ALESS~75.2 and 102.1 are consistent with being protoclusters at $z \sim 2.3$. The upper limit for the halo mass derived for ALESS~5.1 is also consistent with this environment being a protocluster, but we emphasise that our previous analyses suggest it is unlikely to reside in such a structure. 

Both the present-day  and  high-redshift masses obtained using the SHMR method (\S\ref{sec:masses_SHMR}) are significantly higher than those obtained using the SCM method (\S\ref{sec:masses_SCM}), and higher than expected for three structures all found in a survey of $\sim0.25$~deg$^2$ based on predictions from halo mass functions \citep[e.g.][]{PressSchechter1974,Tinker+2008}. This is consistent with the SHMR masses being affected by systematic effects that make them upper limits, as discussed in \S\ref{sec:masses_SHMR}. The SCM-derived masses are typically lower than the masses of protoclusters at similar redshifts (see Figure \ref{fig:PC_comparison}), but are consistent with the results of SMG clustering studies \citep[][see Figure \ref{fig:PC_comparison}]{Hickox+2012,Wilkinson+2017,Garcia-Vergara+2020,Stach+2021}, which generally agree that SMGs reside in halos with $\log(M_{h}/{\rm M}_{\odot}) \lesssim 13$ at $z = 1$--3. 
However, we note that the studies by \citet{Hickox+2012} and \citet{Wilkinson+2017} were both conducted using single-dish observations and are thus limited by false counterpart identification and source blending. Although the high-resolution interferometric studies by \citet{Garcia-Vergara+2020} and \citet{Stach+2021} are not afflicted by these limitations, they present conflicting results for the halo masses of the SMG population, likely stemming from the methodological differences described in \S\ref{S:intro}.

Figure~\ref{fig:PC_comparison} also includes the approximate boundaries separating different gas regimes, as proposed by \citet{DekelBirnboim2006}: in halos for which $\log(M_{h}/{\rm M}_{\odot}) \lesssim 12$, inflowing gas is predominantly cold and enables the growth of galaxies; in halos above this mass threshold, these gas inflows are shock-heated resulting in strangulation of the galaxy within. However, if these massive halos still fall below some other, redshift-dependent mass threshold (as marked by the `cold in hot' boundary in Figure \ref{fig:PC_comparison}), then penetrating cold gas may still be sustaining galaxy growth. At $z \sim 2.3$ this mass threshold is $\log(M_{h}/{\rm M}_{\odot}) \sim 12.9$, while at $z = 3.3$ it is $\log(M_{h}/{\rm M}_{\odot}) \sim 14.0$.

Based on our mass estimates, the halo of ALESS~5.1 is likely in the `cold in hot' category at its observed redshift, particularly when noting that the mass of this halo is possibly overestimated. It is therefore probable that ALESS~5.1, along with any other galaxy that may share its halo, is undergoing growth sustained by penetrating cold gas inflows. Conversely galaxies in the environment of ALESS~75.2 are more likely to be undergoing strangulation due to shock-heating in the halo at the observed redshift; the SHMR-derived halo mass lies significantly above the limit for `cold in hot' gas inflows, and the $1\sigma$ confidence interval for the SCM-derived mass only just crosses below the limit. 
We cannot conclude anything about the gas regime in the environment of ALESS~102.1 as the `cold in hot' boundary is straddled by the mass estimates for this structure.

\section{Conclusions}
We have conducted a wide-field narrowband survey of star-forming galaxies in the environments of three SMGs at $z \sim 2.3$ and $z \sim 3.3$ to determine whether these SMGs reside in protocluster-like environments. By studying individual SMGs selecting based only on their spectroscopic redshifts we have measured `typical' SMG environments. 
Our main conclusions are as follows:
\begin{itemize}
  \item Using HAWK-I \BrG\ and \Ks\ photometry, we identified a total of 147 candidate emission line galaxies in the two HAWK-I pointings containing the three target SMGs. After extracting photometry from archival UV-to-NIR broadband images, we performed SED fitting with \eazypy\ to obtain photometric redshifts for these galaxies and identified 44, 11, and 4 companion galaxies to the SMGs ALESS~75.2 ($z_{\textrm{spec}} = 2.294$), ALESS~102.1 ($z_{\textrm{spec}} = 2.296$), and ALESS~5.1 ($z_{\textrm{spec}} = 3.303$), respectively.
 
  \item By constructing luminosity functions for each SMG environment and comparing with blank-field luminosity functions from the literature at similar redshifts, we measure overdensity parameters of $\delta_{g} = 0.2^{+2.5}_{-0.7}$, $2.6^{+1.4}_{-1.2}$, and $0.2^{+0.6}_{-0.5}$ across the whole $\sim4$~Mpc HAWK-I field of view for ALESS~5.1, 75.2, and 102.1, respectively. 
    Whilst ALESS~102.1 is not overdense on these large scales, it does sit in a $\sim1$~Mpc region with $\delta_g=3.8^{+2.4}_{-1.8}$. Therefore 2/3 of the target SMGs reside in overdense environments.
 
  \item We considered the spatial distribution of companion \Ha\ emitters in the environments of the two SMGs at $z \sim 2.3$ (ALESS~75.2 and 102.1) by measuring their density in annuli around the SMGs and by constructing overdensity maps. For ALESS~75.2 the companion galaxies are spread out across the entire HAWK-I field of view, spanning a few Mpc. This is consistent with simulations, in which protoclusters are seen to extend over several Mpc at $z \sim 2$--3. The SMG resides near a possible density peak of \Ha\ emitters, although a greater peak is seen a few arcminutes eastward which appears to be associated with a previously discovered \Lya\ blob \citep{Yang+2010}. 
    The overdensity around ALESS~102.1 is smaller ($\sim1$~Mpc) and could instead evolve into a galaxy group locally. 
    
  \item Stellar masses and SFRs were obtained for the companion galaxies in each SMG environment by performing SED fitting with \magphys. The galaxies are generally scattered about the star-forming main sequence at their respective epochs, with no evidence of enhanced star formation activity in either environment.
 
  \item  Two methods were used to estimate the total halo mass of each of the two overdense SMG environments, which provided upper and lower bounds on the halo masses at the observed redshifts and evolving to the present day. 
  These reveal that ALESS~75.2 likely resides in a protocluster, while ALESS~102.1 resides in either a protocluster or a protogroup. 
  
	\item We therefore surmise that 2/3 of these SMGs are strong candidates for the progenitors of massive elliptical galaxies in clusters, although the possibility remains for them to end up in galaxy groups. If these targets are indeed representative of `typical' SMGs then this suggests that SMGs in general are likely to evolve into massive elliptical galaxies by the present day, as suggested by certain evolutionary models \citep[e.g.][]{Sanders+1988,Hopkins+2008}, but with significant variation in the surrounding environments.
\end{itemize}

With this study we have demonstrated the efficacy of narrowband surveys as a means of searching for galaxy overdensities around SMGs selected without bias towards particular environments. 
Future followup with larger samples of SMGs and/or spectroscopic confirmation of companion galaxies would confirm the nature of the overdensities that we have detected, and resolved analyses (e.g.\ with ALMA and/or JWST) will further reveal how the member galaxies evolve.

\bigskip

\section*{Acknowledgements}

We are extremely grateful to Ian Smail for numerous insightful discussions, which significantly improved this work. The research is based on observations collected at the European Southern Observatory under ESO programme 0103.A-0668(A).
TC received support from the Science and Technology Facilities Council (STFC; 2287406) and the Faculty of Science and Technology at Lancaster University. 
JW acknowledges an STFC Ernest Rutherford Fellowship (ST/P004784/2).
HD acknowledges support from the Agencia Estatal de Investigación del Ministerio de Ciencia, Innovaci{\'o}n y Universidades (MCIU/AEI) under grant (Construcci{\'o}n de cúmulos de galaxias en formaci{\'o}n a trav{\'e}s de la formaci{\'o}n estelar oscurecida por el polvo) and the European Regional Development Fund (ERDF) with reference (PID2022-143243NB-I00/10.13039/501100011033). 
The authors would also like to acknowledge the publicly available software {\sc{Topcat}} \citep{TOPCAT2005}, as well as the {\sc Python} packages {\sc{NumPy}} \citep{NUMPY2020}, {\sc{SciPy}} \citep{SCIPY2020}, {\sc{Astropy}} \citep{ASTROPYI,ASTROPYII}, {\sc{Matplotlib}} \citep{MATPLOTLIB2007}, and {\sc{Photutils}} \citep{Bradley_photutils+2022}.

\section*{Data Availability}

The data used in this publication can be accessed via the \href{http://archive.eso.org/eso/eso_archive_main.html}{ESO Science Archive} under Program ID: 0103-0668(A), PI: Wardlow.



\bibliographystyle{mnras}
\bibliography{References}

\begin{thebibliography}{}
\makeatletter
\relax
\def\mn@urlcharsother{\let\do\@makeother \do\$\do\&\do\#\do\^\do\_\do\%\do\~}
\def\mn@doi{\begingroup\mn@urlcharsother \@ifnextchar [ {\mn@doi@}
  {\mn@doi@[]}}
\def\mn@doi@[#1]#2{\def\@tempa{#1}\ifx\@tempa\@empty \href
  {http://dx.doi.org/#2} {doi:#2}\else \href {http://dx.doi.org/#2} {#1}\fi
  \endgroup}
\def\mn@eprint#1#2{\mn@eprint@#1:#2::\@nil}
\def\mn@eprint@arXiv#1{\href {http://arxiv.org/abs/#1} {{\tt arXiv:#1}}}
\def\mn@eprint@dblp#1{\href {http://dblp.uni-trier.de/rec/bibtex/#1.xml}
  {dblp:#1}}
\def\mn@eprint@#1:#2:#3:#4\@nil{\def\@tempa {#1}\def\@tempb {#2}\def\@tempc
  {#3}\ifx \@tempc \@empty \let \@tempc \@tempb \let \@tempb \@tempa \fi \ifx
  \@tempb \@empty \def\@tempb {arXiv}\fi \@ifundefined
  {mn@eprint@\@tempb}{\@tempb:\@tempc}{\expandafter \expandafter \csname
  mn@eprint@\@tempb\endcsname \expandafter{\@tempc}}}

\bibitem[\protect\citeauthoryear{{Angulo}, {Springel}, {White}, {Cole},
  {Jenkins}, {Baugh}  \& {Frenk}}{{Angulo} et~al.}{2012}]{Angulo+2012}
{Angulo} R.~E.,  {Springel} V.,  {White} S.~D.~M.,  {Cole} S.,  {Jenkins} A.,
  {Baugh} C.~M.,   {Frenk} C.~S.,  2012, \mn@doi [\mnras]
  {10.1111/j.1365-2966.2012.21783.x}, \href
  {https://ui.adsabs.harvard.edu/abs/2012MNRAS.425.2722A} {425, 2722}

\bibitem[\protect\citeauthoryear{{Arsenault} et~al.,}{{Arsenault}
  et~al.}{2008}]{HAWKI_GRAAL2008}
{Arsenault} R.,  et~al., 2008, in {Hubin} N.,  {Max} C.~E.,   {Wizinowich}
  P.~L.,  eds,  Society of Photo-Optical Instrumentation Engineers (SPIE)
  Conference Series Vol. 7015, Adaptive Optics Systems. p. 701524,
  \mn@doi{10.1117/12.790359}

\bibitem[\protect\citeauthoryear{{Astropy Collaboration} et~al.,}{{Astropy
  Collaboration} et~al.}{2013}]{ASTROPYI}
{Astropy Collaboration} et~al., 2013, \mn@doi [\aap]
  {10.1051/0004-6361/201322068}, \href
  {https://ui.adsabs.harvard.edu/abs/2013A&A...558A..33A} {558, A33}

\bibitem[\protect\citeauthoryear{{Astropy Collaboration} et~al.,}{{Astropy
  Collaboration} et~al.}{2018}]{ASTROPYII}
{Astropy Collaboration} et~al., 2018, \mn@doi [\aj] {10.3847/1538-3881/aabc4f},
  \href {https://ui.adsabs.harvard.edu/abs/2018AJ....156..123A} {156, 123}

\bibitem[\protect\citeauthoryear{{Balestra} et~al.,}{{Balestra}
  et~al.}{2010}]{Balestra+2010}
{Balestra} I.,  et~al., 2010, \mn@doi [\aap] {10.1051/0004-6361/200913626},
  \href {https://ui.adsabs.harvard.edu/abs/2010A&A...512A..12B} {512, A12}

\bibitem[\protect\citeauthoryear{{Barger}, {Cowie}, {Sanders}, {Fulton},
  {Taniguchi}, {Sato}, {Kawara}  \& {Okuda}}{{Barger}
  et~al.}{1998}]{Barger+1998}
{Barger} A.~J.,  {Cowie} L.~L.,  {Sanders} D.~B.,  {Fulton} E.,  {Taniguchi}
  Y.,  {Sato} Y.,  {Kawara} K.,   {Okuda} H.,  1998, \mn@doi [\nat]
  {10.1038/28338}, \href
  {https://ui.adsabs.harvard.edu/abs/1998Natur.394..248B} {394, 248}

\bibitem[\protect\citeauthoryear{{Barger}, {Wang}, {Cowie}, {Owen}, {Chen}  \&
  {Williams}}{{Barger} et~al.}{2012}]{Barger+2012}
{Barger} A.~J.,  {Wang} W.~H.,  {Cowie} L.~L.,  {Owen} F.~N.,  {Chen} C.~C.,
  {Williams} J.~P.,  2012, \mn@doi [\apj] {10.1088/0004-637X/761/2/89}, \href
  {https://ui.adsabs.harvard.edu/abs/2012ApJ...761...89B} {761, 89}

\bibitem[\protect\citeauthoryear{{Baugh}, {Cole}, {Frenk}  \& {Lacey}}{{Baugh}
  et~al.}{1998}]{Baugh+1998}
{Baugh} C.~M.,  {Cole} S.,  {Frenk} C.~S.,   {Lacey} C.~G.,  1998, \mn@doi
  [\apj] {10.1086/305563}, \href
  {https://ui.adsabs.harvard.edu/abs/1998ApJ...498..504B} {498, 504}

\bibitem[\protect\citeauthoryear{{Baugh}, {Lacey}, {Frenk}, {Granato}, {Silva},
  {Bressan}, {Benson}  \& {Cole}}{{Baugh} et~al.}{2005}]{Baugh+2005}
{Baugh} C.~M.,  {Lacey} C.~G.,  {Frenk} C.~S.,  {Granato} G.~L.,  {Silva} L.,
  {Bressan} A.,  {Benson} A.~J.,   {Cole} S.,  2005, \mn@doi [\mnras]
  {10.1111/j.1365-2966.2004.08553.x}, \href
  {https://ui.adsabs.harvard.edu/abs/2005MNRAS.356.1191B} {356, 1191}

\bibitem[\protect\citeauthoryear{{Behroozi}, {Wechsler}  \&
  {Conroy}}{{Behroozi} et~al.}{2013}]{Behroozi+2013}
{Behroozi} P.~S.,  {Wechsler} R.~H.,   {Conroy} C.,  2013, \mn@doi [\apj]
  {10.1088/0004-637X/770/1/57}, \href
  {https://ui.adsabs.harvard.edu/abs/2013ApJ...770...57B} {770, 57}

\bibitem[\protect\citeauthoryear{{Bertin}}{{Bertin}}{2006}]{SCAMP2006}
{Bertin} E.,  2006, in {Gabriel} C.,  {Arviset} C.,  {Ponz} D.,   {Enrique} S.,
   eds,  Astronomical Society of the Pacific Conference Series Vol. 351,
  Astronomical Data Analysis Software and Systems XV. p.~112

\bibitem[\protect\citeauthoryear{{Bertin}}{{Bertin}}{2010}]{SWarp2010}
{Bertin} E.,  2010, {SWarp: Resampling and Co-adding FITS Images Together}
  (\mn@eprint {ascl} {1010.068})

\bibitem[\protect\citeauthoryear{{Bertin} \& {Arnouts}}{{Bertin} \&
  {Arnouts}}{1996}]{SExtractor1996}
{Bertin} E.,  {Arnouts} S.,  1996, \mn@doi [\aaps] {10.1051/aas:1996164}, \href
  {https://ui.adsabs.harvard.edu/abs/1996A&AS..117..393B} {117, 393}

\bibitem[\protect\citeauthoryear{{B{\'e}thermin}, {Dole}, {Lagache}, {Le
  Borgne}  \& {Penin}}{{B{\'e}thermin} et~al.}{2011}]{Bethermin+2011}
{B{\'e}thermin} M.,  {Dole} H.,  {Lagache} G.,  {Le Borgne} D.,   {Penin} A.,
  2011, \mn@doi [\aap] {10.1051/0004-6361/201015841}, \href
  {https://ui.adsabs.harvard.edu/abs/2011A&A...529A...4B} {529, A4}

\bibitem[\protect\citeauthoryear{{Birkin} et~al.,}{{Birkin}
  et~al.}{2021}]{Birkin+2021}
{Birkin} J.~E.,  et~al., 2021, \mn@doi [\mnras] {10.1093/mnras/staa3862}, \href
  {https://ui.adsabs.harvard.edu/abs/2021MNRAS.501.3926B} {501, 3926}

\bibitem[\protect\citeauthoryear{{Blain}, {Smail}, {Ivison}, {Kneib}  \&
  {Frayer}}{{Blain} et~al.}{2002}]{Blain+2002}
{Blain} A.~W.,  {Smail} I.,  {Ivison} R.~J.,  {Kneib} J.~P.,   {Frayer} D.~T.,
  2002, \mn@doi [\physrep] {10.1016/S0370-1573(02)00134-5}, \href
  {https://ui.adsabs.harvard.edu/abs/2002PhR...369..111B} {369, 111}

\bibitem[\protect\citeauthoryear{{Blain}, {Chapman}, {Smail}  \&
  {Ivison}}{{Blain} et~al.}{2004}]{Blain+2004}
{Blain} A.~W.,  {Chapman} S.~C.,  {Smail} I.,   {Ivison} R.,  2004, \mn@doi
  [\apj] {10.1086/422353}, \href
  {https://ui.adsabs.harvard.edu/abs/2004ApJ...611..725B} {611, 725}

\bibitem[\protect\citeauthoryear{{Blakeslee} et~al.,}{{Blakeslee}
  et~al.}{2003}]{Blakeslee+2003}
{Blakeslee} J.~P.,  et~al., 2003, \mn@doi [\apjl] {10.1086/379234}, \href
  {https://ui.adsabs.harvard.edu/abs/2003ApJ...596L.143B} {596, L143}

\bibitem[\protect\citeauthoryear{{Bleem} et~al.,}{{Bleem}
  et~al.}{2015}]{Bleem+2015}
{Bleem} L.~E.,  et~al., 2015, \mn@doi [\apjs] {10.1088/0067-0049/216/2/27},
  \href {https://ui.adsabs.harvard.edu/abs/2015ApJS..216...27B} {216, 27}

\bibitem[\protect\citeauthoryear{{B{\"o}hringer} et~al.,}{{B{\"o}hringer}
  et~al.}{2001}]{Bohringer+2001}
{B{\"o}hringer} H.,  et~al., 2001, \mn@doi [\aap] {10.1051/0004-6361:20010240},
  \href {https://ui.adsabs.harvard.edu/abs/2001A&A...369..826B} {369, 826}

\bibitem[\protect\citeauthoryear{{Bond}, {Kofman}  \& {Pogosyan}}{{Bond}
  et~al.}{1996}]{Bond+1996}
{Bond} J.~R.,  {Kofman} L.,   {Pogosyan} D.,  1996, \mn@doi [\nat]
  {10.1038/380603a0}, \href
  {https://ui.adsabs.harvard.edu/abs/1996Natur.380..603B} {380, 603}

\bibitem[\protect\citeauthoryear{{Bothwell} et~al.,}{{Bothwell}
  et~al.}{2013}]{Bothwell+2013}
{Bothwell} M.~S.,  et~al., 2013, \mn@doi [\mnras] {10.1093/mnras/sts562}, \href
  {https://ui.adsabs.harvard.edu/abs/2013MNRAS.429.3047B} {429, 3047}

\bibitem[\protect\citeauthoryear{Bradley et~al.,}{Bradley
  et~al.}{2022}]{Bradley_photutils+2022}
Bradley L.,  et~al., 2022, astropy/photutils:, \mn@doi{10.5281/zenodo.6385735},
  \url {https://doi.org/10.5281/zenodo.6385735}

\bibitem[\protect\citeauthoryear{{Brammer}, {van Dokkum}  \& {Coppi}}{{Brammer}
  et~al.}{2008}]{Brammer+2008}
{Brammer} G.~B.,  {van Dokkum} P.~G.,   {Coppi} P.,  2008, \mn@doi [\apj]
  {10.1086/591786}, \href
  {https://ui.adsabs.harvard.edu/abs/2008ApJ...686.1503B} {686, 1503}

\bibitem[\protect\citeauthoryear{{Brinchmann}, {Charlot}, {White}, {Tremonti},
  {Kauffmann}, {Heckman}  \& {Brinkmann}}{{Brinchmann}
  et~al.}{2004}]{Brinchmann+2004}
{Brinchmann} J.,  {Charlot} S.,  {White} S.~D.~M.,  {Tremonti} C.,  {Kauffmann}
  G.,  {Heckman} T.,   {Brinkmann} J.,  2004, \mn@doi [\mnras]
  {10.1111/j.1365-2966.2004.07881.x}, \href
  {https://ui.adsabs.harvard.edu/abs/2004MNRAS.351.1151B} {351, 1151}

\bibitem[\protect\citeauthoryear{{Bunker}, {Warren}, {Hewett}  \&
  {Clements}}{{Bunker} et~al.}{1995}]{Bunker+1995}
{Bunker} A.~J.,  {Warren} S.~J.,  {Hewett} P.~C.,   {Clements} D.~L.,  1995,
  \mn@doi [\mnras] {10.1093/mnras/273.2.513}, \href
  {https://ui.adsabs.harvard.edu/abs/1995MNRAS.273..513B} {273, 513}

\bibitem[\protect\citeauthoryear{{Calhau}, {Sobral}, {Stroe}, {Best}, {Smail},
  {Lehmer}, {Harrison}  \& {Thomson}}{{Calhau} et~al.}{2017}]{Calhau+2017}
{Calhau} J.,  {Sobral} D.,  {Stroe} A.,  {Best} P.,  {Smail} I.,  {Lehmer} B.,
  {Harrison} C.,   {Thomson} A.,  2017, \mn@doi [\mnras]
  {10.1093/mnras/stw2295}, \href
  {https://ui.adsabs.harvard.edu/abs/2017MNRAS.464..303C} {464, 303}

\bibitem[\protect\citeauthoryear{{Calvi}, {Dannerbauer}, {Arrabal Haro},
  {Rodr{\'\i}guez Espinosa}, {Mu{\~n}oz-Tu{\~n}{\'o}n}, {P{\'e}rez
  Gonz{\'a}lez}  \& {Geier}}{{Calvi} et~al.}{2021}]{Calvi+2021}
{Calvi} R.,  {Dannerbauer} H.,  {Arrabal Haro} P.,  {Rodr{\'\i}guez Espinosa}
  J.~M.,  {Mu{\~n}oz-Tu{\~n}{\'o}n} C.,  {P{\'e}rez Gonz{\'a}lez} P.~G.,
  {Geier} S.,  2021, \mn@doi [\mnras] {10.1093/mnras/staa4037}, \href
  {https://ui.adsabs.harvard.edu/abs/2021MNRAS.502.4558C} {502, 4558}

\bibitem[\protect\citeauthoryear{{Calzetti}, {Armus}, {Bohlin}, {Kinney},
  {Koornneef}  \& {Storchi-Bergmann}}{{Calzetti} et~al.}{2000}]{Calzetti+2000}
{Calzetti} D.,  {Armus} L.,  {Bohlin} R.~C.,  {Kinney} A.~L.,  {Koornneef} J.,
   {Storchi-Bergmann} T.,  2000, \mn@doi [\apj] {10.1086/308692}, \href
  {https://ui.adsabs.harvard.edu/abs/2000ApJ...533..682C} {533, 682}

\bibitem[\protect\citeauthoryear{{Cardamone} et~al.,}{{Cardamone}
  et~al.}{2010}]{Cardamone+2010}
{Cardamone} C.~N.,  et~al., 2010, \mn@doi [\apjs]
  {10.1088/0067-0049/189/2/270}, \href
  {https://ui.adsabs.harvard.edu/abs/2010ApJS..189..270C} {189, 270}

\bibitem[\protect\citeauthoryear{{Casali} et~al.,}{{Casali}
  et~al.}{2006}]{HAWKI2006}
{Casali} M.,  et~al., 2006, in {McLean} I.~S.,  {Iye} M.,  eds,  Society of
  Photo-Optical Instrumentation Engineers (SPIE) Conference Series Vol. 6269,
  Society of Photo-Optical Instrumentation Engineers (SPIE) Conference Series.
  p. 62690W, \mn@doi{10.1117/12.670150}

\bibitem[\protect\citeauthoryear{{Casey}}{{Casey}}{2016}]{Casey2016}
{Casey} C.~M.,  2016, \mn@doi [\apj] {10.3847/0004-637X/824/1/36}, \href
  {https://ui.adsabs.harvard.edu/abs/2016ApJ...824...36C} {824, 36}

\bibitem[\protect\citeauthoryear{{Casey}, {Narayanan}  \& {Cooray}}{{Casey}
  et~al.}{2014}]{Casey2014}
{Casey} C.~M.,  {Narayanan} D.,   {Cooray} A.,  2014, \mn@doi [\physrep]
  {10.1016/j.physrep.2014.02.009}, \href
  {https://ui.adsabs.harvard.edu/abs/2014PhR...541...45C} {541, 45}

\bibitem[\protect\citeauthoryear{{Casey} et~al.,}{{Casey}
  et~al.}{2015}]{Casey+2015}
{Casey} C.~M.,  et~al., 2015, \mn@doi [\apjl] {10.1088/2041-8205/808/2/L33},
  \href {https://ui.adsabs.harvard.edu/abs/2015ApJ...808L..33C} {808, L33}

\bibitem[\protect\citeauthoryear{{Chabrier}}{{Chabrier}}{2003}]{Chabrier2003}
{Chabrier} G.,  2003, \mn@doi [\pasp] {10.1086/376392}, \href
  {https://ui.adsabs.harvard.edu/abs/2003PASP..115..763C} {115, 763}

\bibitem[\protect\citeauthoryear{{Chapman} et~al.}{{Chapman}
  et~al.}{2005}]{Chapman+2005}
{Chapman} S.~C.,  et~al., 2005, \mn@doi [\apj] {10.1086/428082}, \href
  {https://ui.adsabs.harvard.edu/abs/2005ApJ...622..772C} {622, 772}

\bibitem[\protect\citeauthoryear{{Chapman}, {Blain}, {Ibata}, {Ivison}, {Smail}
   \& {Morrison}}{{Chapman} et~al.}{2009}]{Chapman+2009}
{Chapman} S.~C.,  {Blain} A.,  {Ibata} R.,  {Ivison} R.~J.,  {Smail} I.,
  {Morrison} G.,  2009, \mn@doi [\apj] {10.1088/0004-637X/691/1/560}, \href
  {https://ui.adsabs.harvard.edu/abs/2009ApJ...691..560C} {691, 560}

\bibitem[\protect\citeauthoryear{{Cheng} et~al.,}{{Cheng}
  et~al.}{2019}]{Cheng+2019}
{Cheng} T.,  et~al., 2019, \mn@doi [\mnras] {10.1093/mnras/stz2640}, \href
  {https://ui.adsabs.harvard.edu/abs/2019MNRAS.490.3840C} {490, 3840}

\bibitem[\protect\citeauthoryear{{Chiang}, {Overzier}  \& {Gebhardt}}{{Chiang}
  et~al.}{2013}]{Chiang+2013}
{Chiang} Y.-K.,  {Overzier} R.,   {Gebhardt} K.,  2013, \mn@doi [\apj]
  {10.1088/0004-637X/779/2/127}, \href
  {https://ui.adsabs.harvard.edu/abs/2013ApJ...779..127C} {779, 127}

\bibitem[\protect\citeauthoryear{{Chiang} et~al.,}{{Chiang}
  et~al.}{2015}]{Chiang+2015}
{Chiang} Y.-K.,  et~al., 2015, \mn@doi [\apj] {10.1088/0004-637X/808/1/37},
  \href {https://ui.adsabs.harvard.edu/abs/2015ApJ...808...37C} {808, 37}

\bibitem[\protect\citeauthoryear{{Chiang} et~al.}{{Chiang}
  et~al.}{2017}]{Chiang+2017}
{Chiang} Y.-K.,  et~al., 2017, \mn@doi [\apjl] {10.3847/2041-8213/aa7e7b},
  \href {https://ui.adsabs.harvard.edu/abs/2017ApJ...844L..23C} {844, L23}

\bibitem[\protect\citeauthoryear{{Clements}, {Dunne}  \& {Eales}}{{Clements}
  et~al.}{2010}]{Clements+2010}
{Clements} D.~L.,  {Dunne} L.,   {Eales} S.,  2010, \mn@doi [\mnras]
  {10.1111/j.1365-2966.2009.16064.x}, \href
  {https://ui.adsabs.harvard.edu/abs/2010MNRAS.403..274C} {403, 274}

\bibitem[\protect\citeauthoryear{{Cochrane}, {Best}, {Sobral}, {Smail}, {Wake},
  {Stott}  \& {Geach}}{{Cochrane} et~al.}{2017}]{Cochrane+2017}
{Cochrane} R.~K.,  {Best} P.~N.,  {Sobral} D.,  {Smail} I.,  {Wake} D.~A.,
  {Stott} J.~P.,   {Geach} J.~E.,  2017, \mn@doi [\mnras]
  {10.1093/mnras/stx957}, \href
  {https://ui.adsabs.harvard.edu/abs/2017MNRAS.469.2913C} {469, 2913}

\bibitem[\protect\citeauthoryear{{Cooke}, {Hatch}, {Muldrew}, {Rigby}  \&
  {Kurk}}{{Cooke} et~al.}{2014}]{Cooke+2014}
{Cooke} E.~A.,  {Hatch} N.~A.,  {Muldrew} S.~I.,  {Rigby} E.~E.,   {Kurk}
  J.~D.,  2014, \mn@doi [\mnras] {10.1093/mnras/stu522}, \href
  {https://ui.adsabs.harvard.edu/abs/2014MNRAS.440.3262C} {440, 3262}

\bibitem[\protect\citeauthoryear{{Cooper} et~al.,}{{Cooper}
  et~al.}{2008}]{Cooper+2008}
{Cooper} M.~C.,  et~al., 2008, \mn@doi [\mnras]
  {10.1111/j.1365-2966.2007.12613.x}, \href
  {https://ui.adsabs.harvard.edu/abs/2008MNRAS.383.1058C} {383, 1058}

\bibitem[\protect\citeauthoryear{{Coppin} et~al.,}{{Coppin}
  et~al.}{2006}]{Coppin+2006}
{Coppin} K.,  et~al., 2006, \mn@doi [\mnras]
  {10.1111/j.1365-2966.2006.10961.x}, \href
  {https://ui.adsabs.harvard.edu/abs/2006MNRAS.372.1621C} {372, 1621}

\bibitem[\protect\citeauthoryear{{Cucciati} et~al.,}{{Cucciati}
  et~al.}{2014}]{Cucchiati+2014}
{Cucciati} O.,  et~al., 2014, \mn@doi [\aap] {10.1051/0004-6361/201423811},
  \href {https://ui.adsabs.harvard.edu/abs/2014A&A...570A..16C} {570, A16}

\bibitem[\protect\citeauthoryear{{Daddi} et~al.,}{{Daddi}
  et~al.}{2007}]{Daddi+2007}
{Daddi} E.,  et~al., 2007, \mn@doi [\apj] {10.1086/521818}, \href
  {https://ui.adsabs.harvard.edu/abs/2007ApJ...670..156D} {670, 156}

\bibitem[\protect\citeauthoryear{{Daddi} et~al.,}{{Daddi}
  et~al.}{2009}]{Daddi+2009}
{Daddi} E.,  et~al., 2009, \mn@doi [\apj] {10.1088/0004-637X/694/2/1517}, \href
  {https://ui.adsabs.harvard.edu/abs/2009ApJ...694.1517D} {694, 1517}

\bibitem[\protect\citeauthoryear{{Damen} et~al.,}{{Damen}
  et~al.}{2011}]{SIMPLE2011}
{Damen} M.,  et~al., 2011, \mn@doi [\apj] {10.1088/0004-637X/727/1/1}, \href
  {https://ui.adsabs.harvard.edu/abs/2011ApJ...727....1D} {727, 1}

\bibitem[\protect\citeauthoryear{{Danielson} et~al.,}{{Danielson}
  et~al.}{2017}]{Danielson+2017}
{Danielson} A.~L.~R.,  et~al., 2017, \mn@doi [\apj] {10.3847/1538-4357/aa6caf},
  \href {https://ui.adsabs.harvard.edu/abs/2017ApJ...840...78D} {840, 78}

\bibitem[\protect\citeauthoryear{{Darvish}, {Mobasher}, {Sobral}, {Rettura},
  {Scoville}, {Faisst}  \& {Capak}}{{Darvish} et~al.}{2016}]{Darvish+2016}
{Darvish} B.,  {Mobasher} B.,  {Sobral} D.,  {Rettura} A.,  {Scoville} N.,
  {Faisst} A.,   {Capak} P.,  2016, \mn@doi [\apj]
  {10.3847/0004-637X/825/2/113}, \href
  {https://ui.adsabs.harvard.edu/abs/2016ApJ...825..113D} {825, 113}

\bibitem[\protect\citeauthoryear{{Dav{\'e}}, {Finlator}, {Oppenheimer},
  {Fardal}, {Katz}, {Kere{\v{s}}}  \& {Weinberg}}{{Dav{\'e}}
  et~al.}{2010}]{Dave+2010}
{Dav{\'e}} R.,  {Finlator} K.,  {Oppenheimer} B.~D.,  {Fardal} M.,  {Katz} N.,
  {Kere{\v{s}}} D.,   {Weinberg} D.~H.,  2010, \mn@doi [\mnras]
  {10.1111/j.1365-2966.2010.16395.x}, \href
  {https://ui.adsabs.harvard.edu/abs/2010MNRAS.404.1355D} {404, 1355}

\bibitem[\protect\citeauthoryear{{Davies}, {Bremer}, {Stanway}, {Husband},
  {Lehnert}  \& {Mannering}}{{Davies} et~al.}{2014}]{Davies+2014}
{Davies} L.~J.~M.,  {Bremer} M.~N.,  {Stanway} E.~R.,  {Husband} K.,  {Lehnert}
  M.~D.,   {Mannering} E.~J.~A.,  2014, \mn@doi [\mnras]
  {10.1093/mnras/stt2306}, \href
  {https://ui.adsabs.harvard.edu/abs/2014MNRAS.438.2732D} {438, 2732}

\bibitem[\protect\citeauthoryear{{De Lucia} \& {Blaizot}}{{De Lucia} \&
  {Blaizot}}{2007}]{DeLucia+2007b}
{De Lucia} G.,  {Blaizot} J.,  2007, \mn@doi [\mnras]
  {10.1111/j.1365-2966.2006.11287.x}, \href
  {https://ui.adsabs.harvard.edu/abs/2007MNRAS.375....2D} {375, 2}

\bibitem[\protect\citeauthoryear{{Dekel} \& {Birnboim}}{{Dekel} \&
  {Birnboim}}{2006}]{DekelBirnboim2006}
{Dekel} A.,  {Birnboim} Y.,  2006, \mn@doi [\mnras]
  {10.1111/j.1365-2966.2006.10145.x}, \href
  {https://ui.adsabs.harvard.edu/abs/2006MNRAS.368....2D} {368, 2}

\bibitem[\protect\citeauthoryear{{Diener} et~al.,}{{Diener}
  et~al.}{2013}]{Diener+2013}
{Diener} C.,  et~al., 2013, \mn@doi [\apj] {10.1088/0004-637X/765/2/109}, \href
  {https://ui.adsabs.harvard.edu/abs/2013ApJ...765..109D} {765, 109}

\bibitem[\protect\citeauthoryear{{Diener} et~al.,}{{Diener}
  et~al.}{2015}]{Diener+2015}
{Diener} C.,  et~al., 2015, \mn@doi [\apj] {10.1088/0004-637X/802/1/31}, \href
  {https://ui.adsabs.harvard.edu/abs/2015ApJ...802...31D} {802, 31}

\bibitem[\protect\citeauthoryear{{Dressler}}{{Dressler}}{1980}]{Dressler1980}
{Dressler} A.,  1980, \mn@doi [\apj] {10.1086/157753}, \href
  {https://ui.adsabs.harvard.edu/#abs/1980ApJ...236..351D} {236, 351}

\bibitem[\protect\citeauthoryear{{Dudzevi{\v{c}}i{\={u}}t{\.{e}}}
  et~al.,}{{Dudzevi{\v{c}}i{\={u}}t{\.{e}}} et~al.}{2020}]{Dudzeviciute+2020a}
{Dudzevi{\v{c}}i{\={u}}t{\.{e}}} U.,  et~al., 2020, \mn@doi [\mnras]
  {10.1093/mnras/staa769}, \href
  {https://ui.adsabs.harvard.edu/abs/2020MNRAS.494.3828D} {494, 3828}

\bibitem[\protect\citeauthoryear{{Eales}, {Lilly}, {Gear}, {Dunne}, {Bond},
  {Hammer}, {Le F{\`e}vre}  \& {Crampton}}{{Eales} et~al.}{1999}]{Eales+1999}
{Eales} S.,  {Lilly} S.,  {Gear} W.,  {Dunne} L.,  {Bond} J.~R.,  {Hammer} F.,
  {Le F{\`e}vre} O.,   {Crampton} D.,  1999, \mn@doi [\apj] {10.1086/307069},
  \href {https://ui.adsabs.harvard.edu/abs/1999ApJ...515..518E} {515, 518}

\bibitem[\protect\citeauthoryear{{Eddington}}{{Eddington}}{1913}]{Eddington1913}
{Eddington} A.~S.,  1913, \mn@doi [\mnras] {10.1093/mnras/73.5.359}, \href
  {https://ui.adsabs.harvard.edu/abs/1913MNRAS..73..359E} {73, 359}

\bibitem[\protect\citeauthoryear{{Elbaz} et~al.,}{{Elbaz}
  et~al.}{2007}]{Elbaz+2007}
{Elbaz} D.,  et~al., 2007, \mn@doi [\aap] {10.1051/0004-6361:20077525}, \href
  {https://ui.adsabs.harvard.edu/abs/2007A&A...468...33E} {468, 33}

\bibitem[\protect\citeauthoryear{{Ellis}, {Smail}, {Dressler}, {Couch},
  {Oemler}, {Butcher}  \& {Sharples}}{{Ellis} et~al.}{1997}]{Ellis+1997}
{Ellis} R.~S.,  {Smail} I.,  {Dressler} A.,  {Couch} W.~J.,  {Oemler} Augustus
  J.,  {Butcher} H.,   {Sharples} R.~M.,  1997, \mn@doi [\apj]
  {10.1086/304261}, \href
  {https://ui.adsabs.harvard.edu/abs/1997ApJ...483..582E} {483, 582}

\bibitem[\protect\citeauthoryear{{Fakhouri}, {Ma}  \&
  {Boylan-Kolchin}}{{Fakhouri} et~al.}{2010}]{Fakhouri+2010}
{Fakhouri} O.,  {Ma} C.-P.,   {Boylan-Kolchin} M.,  2010, \mn@doi [\mnras]
  {10.1111/j.1365-2966.2010.16859.x}, \href
  {https://ui.adsabs.harvard.edu/abs/2010MNRAS.406.2267F} {406, 2267}

\bibitem[\protect\citeauthoryear{{Finkelstein} et~al.,}{{Finkelstein}
  et~al.}{2022}]{Finkelstein+2022}
{Finkelstein} S.~L.,  et~al., 2022, \mn@doi [\apj] {10.3847/1538-4357/ac3aed},
  \href {https://ui.adsabs.harvard.edu/abs/2022ApJ...928...52F} {928, 52}

\bibitem[\protect\citeauthoryear{{Garc{\'\i}a-Vergara}, {Hodge}, {Hennawi},
  {Weiss}, {Wardlow}, {Myers}  \& {Hickox}}{{Garc{\'\i}a-Vergara}
  et~al.}{2020}]{Garcia-Vergara+2020}
{Garc{\'\i}a-Vergara} C.,  {Hodge} J.,  {Hennawi} J.~F.,  {Weiss} A.,
  {Wardlow} J.,  {Myers} A.~D.,   {Hickox} R.,  2020, \mn@doi [\apj]
  {10.3847/1538-4357/abbdfe}, \href
  {https://ui.adsabs.harvard.edu/abs/2020ApJ...904....2G} {904, 2}

\bibitem[\protect\citeauthoryear{{Garn} et~al.,}{{Garn}
  et~al.}{2010}]{Garn+2010}
{Garn} T.,  et~al., 2010, \mn@doi [\mnras] {10.1111/j.1365-2966.2009.16042.x},
  \href {https://ui.adsabs.harvard.edu/abs/2010MNRAS.402.2017G} {402, 2017}

\bibitem[\protect\citeauthoryear{{Gavazzi}, {Adami}, {Durret}, {Cuillandre},
  {Ilbert}, {Mazure}, {Pell{\'o}}  \& {Ulmer}}{{Gavazzi}
  et~al.}{2009}]{Gavazzi+2009}
{Gavazzi} R.,  {Adami} C.,  {Durret} F.,  {Cuillandre} J.~C.,  {Ilbert} O.,
  {Mazure} A.,  {Pell{\'o}} R.,   {Ulmer} M.~P.,  2009, \mn@doi [\aap]
  {10.1051/0004-6361/200911841}, \href
  {https://ui.adsabs.harvard.edu/abs/2009A&A...498L..33G} {498, L33}

\bibitem[\protect\citeauthoryear{{Gawiser} et~al.,}{{Gawiser}
  et~al.}{2006a}]{MUSYC2006a}
{Gawiser} E.,  et~al., 2006a, \mn@doi [\apjs] {10.1086/497644}, \href
  {https://ui.adsabs.harvard.edu/abs/2006ApJS..162....1G} {162, 1}

\bibitem[\protect\citeauthoryear{{Gawiser} et~al.,}{{Gawiser}
  et~al.}{2006b}]{MUSYC2006b}
{Gawiser} E.,  et~al., 2006b, \mn@doi [\apjl] {10.1086/504467}, \href
  {https://ui.adsabs.harvard.edu/abs/2006ApJ...642L..13G} {642, L13}

\bibitem[\protect\citeauthoryear{{Geach} et~al.,}{{Geach}
  et~al.}{2005}]{Geach+2005}
{Geach} J.~E.,  et~al., 2005, \mn@doi [\mnras]
  {10.1111/j.1365-2966.2005.09538.x}, \href
  {https://ui.adsabs.harvard.edu/abs/2005MNRAS.363.1398G} {363, 1398}

\bibitem[\protect\citeauthoryear{{Geach}, {Smail}, {Best}, {Kurk}, {Casali},
  {Ivison}  \& {Coppin}}{{Geach} et~al.}{2008}]{Geach+2008}
{Geach} J.~E.,  {Smail} I.,  {Best} P.~N.,  {Kurk} J.,  {Casali} M.,  {Ivison}
  R.~J.,   {Coppin} K.,  2008, \mn@doi [\mnras]
  {10.1111/j.1365-2966.2008.13481.x}, \href
  {https://ui.adsabs.harvard.edu/abs/2008MNRAS.388.1473G} {388, 1473}

\bibitem[\protect\citeauthoryear{{Geach} et~al.,}{{Geach}
  et~al.}{2016}]{Geach+2016}
{Geach} J.~E.,  et~al., 2016, \mn@doi [\apj] {10.3847/0004-637X/832/1/37},
  \href {https://ui.adsabs.harvard.edu/abs/2016ApJ...832...37G} {832, 37}

\bibitem[\protect\citeauthoryear{{Gilbank}, {Gladders}, {Yee}  \&
  {Hsieh}}{{Gilbank} et~al.}{2011}]{Gilbank+2011}
{Gilbank} D.~G.,  {Gladders} M.~D.,  {Yee} H.~K.~C.,   {Hsieh} B.~C.,  2011,
  \mn@doi [\aj] {10.1088/0004-6256/141/3/94}, \href
  {https://ui.adsabs.harvard.edu/abs/2011AJ....141...94G} {141, 94}

\bibitem[\protect\citeauthoryear{{Gladders} \& {Yee}}{{Gladders} \&
  {Yee}}{2000}]{GladdersYee2000}
{Gladders} M.~D.,  {Yee} H.~K.~C.,  2000, \mn@doi [\aj] {10.1086/301557}, \href
  {https://ui.adsabs.harvard.edu/abs/2000AJ....120.2148G} {120, 2148}

\bibitem[\protect\citeauthoryear{{Gladders} \& {Yee}}{{Gladders} \&
  {Yee}}{2005}]{GladdersYee2005}
{Gladders} M.~D.,  {Yee} H.~K.~C.,  2005, \mn@doi [\apjs] {10.1086/427327},
  \href {https://ui.adsabs.harvard.edu/abs/2005ApJS..157....1G} {157, 1}

\bibitem[\protect\citeauthoryear{{Gonz{\'a}lez}, {Labb{\'e}}, {Bouwens},
  {Illingworth}, {Franx}, {Kriek}  \& {Brammer}}{{Gonz{\'a}lez}
  et~al.}{2010}]{Gonzalez+2010}
{Gonz{\'a}lez} V.,  {Labb{\'e}} I.,  {Bouwens} R.~J.,  {Illingworth} G.,
  {Franx} M.,  {Kriek} M.,   {Brammer} G.~B.,  2010, \mn@doi [\apj]
  {10.1088/0004-637X/713/1/115}, \href
  {https://ui.adsabs.harvard.edu/abs/2010ApJ...713..115G} {713, 115}

\bibitem[\protect\citeauthoryear{{Greenslade}, {Clements}, {Petitpas},
  {Asboth}, {Conley}, {P{\'e}rez-Fournon}  \& {Riechers}}{{Greenslade}
  et~al.}{2020}]{Greenslade+2020}
{Greenslade} J.,  {Clements} D.~L.,  {Petitpas} G.,  {Asboth} V.,  {Conley} A.,
   {P{\'e}rez-Fournon} I.,   {Riechers} D.,  2020, \mn@doi [\mnras]
  {10.1093/mnras/staa1637}, \href
  {https://ui.adsabs.harvard.edu/abs/2020MNRAS.496.2315G} {496, 2315}

\bibitem[\protect\citeauthoryear{{Greve} et~al.,}{{Greve}
  et~al.}{2005}]{Greve+2005}
{Greve} T.~R.,  et~al., 2005, \mn@doi [\mnras]
  {10.1111/j.1365-2966.2005.08979.x}, \href
  {https://ui.adsabs.harvard.edu/abs/2005MNRAS.359.1165G} {359, 1165}

\bibitem[\protect\citeauthoryear{{Gruppioni} et~al.,}{{Gruppioni}
  et~al.}{2013}]{Gruppioni+2013}
{Gruppioni} C.,  et~al., 2013, \mn@doi [\mnras] {10.1093/mnras/stt308}, \href
  {https://ui.adsabs.harvard.edu/abs/2013MNRAS.432...23G} {432, 23}

\bibitem[\protect\citeauthoryear{{Gullberg} et~al.,}{{Gullberg}
  et~al.}{2019}]{Gullberg+2019}
{Gullberg} B.,  et~al., 2019, \mn@doi [\mnras] {10.1093/mnras/stz2835}, \href
  {https://ui.adsabs.harvard.edu/abs/2019MNRAS.490.4956G} {490, 4956}

\bibitem[\protect\citeauthoryear{{Hainline}, {Blain}, {Smail}, {Alexander},
  {Armus}, {Chapman}  \& {Ivison}}{{Hainline} et~al.}{2011}]{Hainline+2011}
{Hainline} L.~J.,  {Blain} A.~W.,  {Smail} I.,  {Alexander} D.~M.,  {Armus} L.,
   {Chapman} S.~C.,   {Ivison} R.~J.,  2011, \mn@doi [\apj]
  {10.1088/0004-637X/740/2/96}, \href
  {https://ui.adsabs.harvard.edu/abs/2011ApJ...740...96H} {740, 96}

\bibitem[\protect\citeauthoryear{{Han} et~al.,}{{Han} et~al.}{2015}]{Han+2015}
{Han} J.,  et~al., 2015, \mn@doi [\mnras] {10.1093/mnras/stu2178}, \href
  {https://ui.adsabs.harvard.edu/abs/2015MNRAS.446.1356H} {446, 1356}

\bibitem[\protect\citeauthoryear{Harris et~al.,}{Harris
  et~al.}{2020}]{NUMPY2020}
Harris C.~R.,  et~al., 2020, \mn@doi [Nature] {10.1038/s41586-020-2649-2}, 585,
  357

\bibitem[\protect\citeauthoryear{{Hasselfield} et~al.,}{{Hasselfield}
  et~al.}{2013}]{Hasselfield+2013}
{Hasselfield} M.,  et~al., 2013, \mn@doi [\jcap]
  {10.1088/1475-7516/2013/07/008}, \href
  {https://ui.adsabs.harvard.edu/abs/2013JCAP...07..008H} {2013, 008}

\bibitem[\protect\citeauthoryear{{Hatch}, {Kurk}, {Pentericci}, {Venemans},
  {Kuiper}, {Miley}  \& {R{\"o}ttgering}}{{Hatch} et~al.}{2011}]{Hatch+2011}
{Hatch} N.~A.,  {Kurk} J.~D.,  {Pentericci} L.,  {Venemans} B.~P.,  {Kuiper}
  E.,  {Miley} G.~K.,   {R{\"o}ttgering} H.~J.~A.,  2011, \mn@doi [\mnras]
  {10.1111/j.1365-2966.2011.18735.x}, \href
  {https://ui.adsabs.harvard.edu/abs/2011MNRAS.415.2993H} {415, 2993}

\bibitem[\protect\citeauthoryear{{Hayashi}, {Kodama}, {Tadaki}, {Koyama}  \&
  {Tanaka}}{{Hayashi} et~al.}{2012}]{Hayashi+2012}
{Hayashi} M.,  {Kodama} T.,  {Tadaki} K.-i.,  {Koyama} Y.,   {Tanaka} I.,
  2012, \mn@doi [\apj] {10.1088/0004-637X/757/1/15}, \href
  {https://ui.adsabs.harvard.edu/abs/2012ApJ...757...15H} {757, 15}

\bibitem[\protect\citeauthoryear{{Hayashino} et~al.,}{{Hayashino}
  et~al.}{2004}]{Hayashino+2004}
{Hayashino} T.,  et~al., 2004, \mn@doi [\aj] {10.1086/424935}, \href
  {https://ui.adsabs.harvard.edu/abs/2004AJ....128.2073H} {128, 2073}

\bibitem[\protect\citeauthoryear{{Hayward} et~al.,}{{Hayward}
  et~al.}{2021}]{Hayward+2021}
{Hayward} C.~C.,  et~al., 2021, \mn@doi [\mnras] {10.1093/mnras/stab246}, \href
  {https://ui.adsabs.harvard.edu/abs/2021MNRAS.502.2922H} {502, 2922}

\bibitem[\protect\citeauthoryear{{Henry}, {Mullis}, {Voges}, {B{\"o}hringer},
  {Briel}, {Gioia}  \& {Huchra}}{{Henry} et~al.}{2006}]{Henry+2006}
{Henry} J.~P.,  {Mullis} C.~R.,  {Voges} W.,  {B{\"o}hringer} H.,  {Briel}
  U.~G.,  {Gioia} I.~M.,   {Huchra} J.~P.,  2006, \mn@doi [\apjs]
  {10.1086/498749}, \href
  {https://ui.adsabs.harvard.edu/abs/2006ApJS..162..304H} {162, 304}

\bibitem[\protect\citeauthoryear{{Hickox} et~al.,}{{Hickox}
  et~al.}{2012}]{Hickox+2012}
{Hickox} R.~C.,  et~al., 2012, \mn@doi [\mnras]
  {10.1111/j.1365-2966.2011.20303.x}, \href
  {https://ui.adsabs.harvard.edu/abs/2012MNRAS.421..284H} {421, 284}

\bibitem[\protect\citeauthoryear{{Hildebrandt} et~al.,}{{Hildebrandt}
  et~al.}{2006}]{Hildebrandt+2006}
{Hildebrandt} H.,  et~al., 2006, \mn@doi [\aap] {10.1051/0004-6361:20054278},
  \href {https://ui.adsabs.harvard.edu/abs/2006A&A...452.1121H} {452, 1121}

\bibitem[\protect\citeauthoryear{{Ho}, {Ntampaka}, {Rau}, {Chen}, {Lansberry},
  {Ruehle}  \& {Trac}}{{Ho} et~al.}{2022}]{Ho+2022}
{Ho} M.,  {Ntampaka} M.,  {Rau} M.~M.,  {Chen} M.,  {Lansberry} A.,  {Ruehle}
  F.,   {Trac} H.,  2022, \mn@doi [Nature Astronomy]
  {10.1038/s41550-022-01711-1}, \href
  {https://ui.adsabs.harvard.edu/abs/2022NatAs...6..936H} {6, 936}

\bibitem[\protect\citeauthoryear{{Hodge} \& {da Cunha}}{{Hodge} \& {da
  Cunha}}{2020}]{Hodge&daCunha2020}
{Hodge} J.~A.,  {da Cunha} E.,  2020, \mn@doi [Royal Society Open Science]
  {10.1098/rsos.200556}, \href
  {https://ui.adsabs.harvard.edu/abs/2020RSOS....700556H} {7, 200556}

\bibitem[\protect\citeauthoryear{{Hodge} et~al.,}{{Hodge}
  et~al.}{2013a}]{Hodge+2013a}
{Hodge} J.~A.,  et~al., 2013a, \mn@doi [\apj] {10.1088/0004-637X/768/1/91},
  \href {https://ui.adsabs.harvard.edu/abs/2013ApJ...768...91H} {768, 91}

\bibitem[\protect\citeauthoryear{{Hodge}, {Carilli}, {Walter}, {Daddi}  \&
  {Riechers}}{{Hodge} et~al.}{2013b}]{Hodge+2013b}
{Hodge} J.~A.,  {Carilli} C.~L.,  {Walter} F.,  {Daddi} E.,   {Riechers} D.,
  2013b, \mn@doi [\apj] {10.1088/0004-637X/776/1/22}, \href
  {https://ui.adsabs.harvard.edu/abs/2013ApJ...776...22H} {776, 22}

\bibitem[\protect\citeauthoryear{{Hodge} et~al.,}{{Hodge}
  et~al.}{2016}]{Hodge+2016}
{Hodge} J.~A.,  et~al., 2016, \mn@doi [\apj] {10.3847/1538-4357/833/1/103},
  \href {https://ui.adsabs.harvard.edu/abs/2016ApJ...833..103H} {833, 103}

\bibitem[\protect\citeauthoryear{{Hopkins}, {Hernquist}, {Cox}  \&
  {Kere{\v{s}}}}{{Hopkins} et~al.}{2008}]{Hopkins+2008}
{Hopkins} P.~F.,  {Hernquist} L.,  {Cox} T.~J.,   {Kere{\v{s}}} D.,  2008,
  \mn@doi [\apjs] {10.1086/524362}, \href
  {https://ui.adsabs.harvard.edu/abs/2008ApJS..175..356H} {175, 356}

\bibitem[\protect\citeauthoryear{{Hsieh}, {Wang}, {Hsieh}, {Lin}, {Yan}, {Lim}
  \& {Ho}}{{Hsieh} et~al.}{2012}]{TENIS2012}
{Hsieh} B.-C.,  {Wang} W.-H.,  {Hsieh} C.-C.,  {Lin} L.,  {Yan} H.,  {Lim} J.,
   {Ho} P. T.~P.,  2012, \mn@doi [\apjs] {10.1088/0067-0049/203/2/23}, \href
  {https://ui.adsabs.harvard.edu/abs/2012ApJS..203...23H} {203, 23}

\bibitem[\protect\citeauthoryear{{Hughes} et~al.,}{{Hughes}
  et~al.}{1998}]{Hughes+1998}
{Hughes} D.~H.,  et~al., 1998, \mn@doi [\nat] {10.1038/28328}, \href
  {https://ui.adsabs.harvard.edu/abs/1998Natur.394..241H} {394, 241}

\bibitem[\protect\citeauthoryear{Hunter}{Hunter}{2007}]{MATPLOTLIB2007}
Hunter J.~D.,  2007, \mn@doi [Computing in Science \& Engineering]
  {10.1109/MCSE.2007.55}, 9, 90

\bibitem[\protect\citeauthoryear{{Ikarashi} et~al.,}{{Ikarashi}
  et~al.}{2015}]{Ikarashi+2015}
{Ikarashi} S.,  et~al., 2015, \mn@doi [\apj] {10.1088/0004-637X/810/2/133},
  \href {https://ui.adsabs.harvard.edu/abs/2015ApJ...810..133I} {810, 133}

\bibitem[\protect\citeauthoryear{{Ito} et~al.,}{{Ito} et~al.}{2023}]{Ito+2023}
{Ito} K.,  et~al., 2023, \mn@doi [arXiv e-prints] {10.48550/arXiv.2301.08845},
  \href {https://ui.adsabs.harvard.edu/abs/2023arXiv230108845I} {p.
  arXiv:2301.08845}

\bibitem[\protect\citeauthoryear{{Ivison} et~al.}{{Ivison}
  et~al.}{2000}]{Ivison+2000}
{Ivison} R.~J.,  et~al., 2000, \mn@doi [\apj] {10.1086/309536}, \href
  {https://ui.adsabs.harvard.edu/abs/2000ApJ...542...27I} {542, 27}

\bibitem[\protect\citeauthoryear{{Ivison} et~al.,}{{Ivison}
  et~al.}{2013}]{Ivison+2013}
{Ivison} R.~J.,  et~al., 2013, \mn@doi [\apj] {10.1088/0004-637X/772/2/137},
  \href {https://ui.adsabs.harvard.edu/abs/2013ApJ...772..137I} {772, 137}

\bibitem[\protect\citeauthoryear{{Kennicutt}}{{Kennicutt}}{1998}]{Kennicutt1998}
{Kennicutt} Robert~C. J.,  1998, \mn@doi [\araa]
  {10.1146/annurev.astro.36.1.189}, \href
  {https://ui.adsabs.harvard.edu/abs/1998ARA&A..36..189K} {36, 189}

\bibitem[\protect\citeauthoryear{{Khostovan} et~al.}{{Khostovan}
  et~al.}{2015}]{Khostovan+2015}
{Khostovan} A.~A.,  et~al., 2015, \mn@doi [\mnras] {10.1093/mnras/stv1474},
  \href {https://ui.adsabs.harvard.edu/abs/2015MNRAS.452.3948K} {452, 3948}

\bibitem[\protect\citeauthoryear{{Khostovan}, {Sobral}, {Mobasher}, {Smail},
  {Darvish}, {Nayyeri}, {Hemmati}  \& {Stott}}{{Khostovan}
  et~al.}{2016}]{Khostovan+2016}
{Khostovan} A.~A.,  {Sobral} D.,  {Mobasher} B.,  {Smail} I.,  {Darvish} B.,
  {Nayyeri} H.,  {Hemmati} S.,   {Stott} J.~P.,  2016, \mn@doi [\mnras]
  {10.1093/mnras/stw2174}, \href
  {https://ui.adsabs.harvard.edu/abs/2016MNRAS.463.2363K} {463, 2363}

\bibitem[\protect\citeauthoryear{{Kissler-Patig} et~al.,}{{Kissler-Patig}
  et~al.}{2008}]{HAWKI2008}
{Kissler-Patig} M.,  et~al., 2008, \mn@doi [\aap]
  {10.1051/0004-6361:200809910}, \href
  {https://ui.adsabs.harvard.edu/abs/2008A&A...491..941K} {491, 941}

\bibitem[\protect\citeauthoryear{{Koprowski}, {Dunlop}, {Micha{\l}owski},
  {Cirasuolo}  \& {Bowler}}{{Koprowski} et~al.}{2014}]{Koprowski+2014}
{Koprowski} M.~P.,  {Dunlop} J.~S.,  {Micha{\l}owski} M.~J.,  {Cirasuolo} M.,
  {Bowler} R.~A.~A.,  2014, \mn@doi [\mnras] {10.1093/mnras/stu1402}, \href
  {https://ui.adsabs.harvard.edu/abs/2014MNRAS.444..117K} {444, 117}

\bibitem[\protect\citeauthoryear{{Koyama}, {Kodama}, {Tadaki}, {Hayashi},
  {Tanaka}, {Smail}, {Tanaka}  \& {Kurk}}{{Koyama} et~al.}{2013}]{Koyama+2013b}
{Koyama} Y.,  {Kodama} T.,  {Tadaki} K.-i.,  {Hayashi} M.,  {Tanaka} M.,
  {Smail} I.,  {Tanaka} I.,   {Kurk} J.,  2013, \mn@doi [\mnras]
  {10.1093/mnras/sts133}, \href
  {https://ui.adsabs.harvard.edu/abs/2013MNRAS.428.1551K} {428, 1551}

\bibitem[\protect\citeauthoryear{{Kriek} \& {Conroy}}{{Kriek} \&
  {Conroy}}{2013}]{KriekConroy2013}
{Kriek} M.,  {Conroy} C.,  2013, \mn@doi [\apjl] {10.1088/2041-8205/775/1/L16},
  \href {https://ui.adsabs.harvard.edu/abs/2013ApJ...775L..16K} {775, L16}

\bibitem[\protect\citeauthoryear{{Kron}}{{Kron}}{1980}]{Kron1980}
{Kron} R.~G.,  1980, \mn@doi [\apjs] {10.1086/190669}, \href
  {https://ui.adsabs.harvard.edu/abs/1980ApJS...43..305K} {43, 305}

\bibitem[\protect\citeauthoryear{{Kuiper} et~al.,}{{Kuiper}
  et~al.}{2011a}]{Kuiper+2011b}
{Kuiper} E.,  et~al., 2011a, \mn@doi [\mnras]
  {10.1111/j.1365-2966.2011.18852.x}, \href
  {https://ui.adsabs.harvard.edu/abs/2011MNRAS.415.2245K} {415, 2245}

\bibitem[\protect\citeauthoryear{{Kuiper} et~al.,}{{Kuiper}
  et~al.}{2011b}]{Kuiper+2011a}
{Kuiper} E.,  et~al., 2011b, \mn@doi [\mnras]
  {10.1111/j.1365-2966.2011.19324.x}, \href
  {https://ui.adsabs.harvard.edu/abs/2011MNRAS.417.1088K} {417, 1088}

\bibitem[\protect\citeauthoryear{{Kurk} et~al.,}{{Kurk}
  et~al.}{2000}]{Kurk+2000}
{Kurk} J.~D.,  et~al., 2000, \aap, \href
  {https://ui.adsabs.harvard.edu/abs/2000A&A...358L...1K} {358, L1}

\bibitem[\protect\citeauthoryear{{Lacey}, {Baugh}, {Frenk}, {Silva}, {Granato}
  \& {Bressan}}{{Lacey} et~al.}{2008}]{Lacey+2008}
{Lacey} C.~G.,  {Baugh} C.~M.,  {Frenk} C.~S.,  {Silva} L.,  {Granato} G.~L.,
  {Bressan} A.,  2008, \mn@doi [\mnras] {10.1111/j.1365-2966.2008.12949.x},
  \href {https://ui.adsabs.harvard.edu/abs/2008MNRAS.385.1155L} {385, 1155}

\bibitem[\protect\citeauthoryear{{Lacey}, {Baugh}, {Frenk}, {Benson}, {Orsi},
  {Silva}, {Granato}  \& {Bressan}}{{Lacey} et~al.}{2010}]{Lacey+2010}
{Lacey} C.~G.,  {Baugh} C.~M.,  {Frenk} C.~S.,  {Benson} A.~J.,  {Orsi} A.,
  {Silva} L.,  {Granato} G.~L.,   {Bressan} A.,  2010, \mn@doi [\mnras]
  {10.1111/j.1365-2966.2010.16463.x}, \href
  {https://ui.adsabs.harvard.edu/abs/2010MNRAS.405....2L} {405, 2}

\bibitem[\protect\citeauthoryear{{Le F{\`e}vre} et~al.,}{{Le F{\`e}vre}
  et~al.}{2005}]{LeFevre+2005}
{Le F{\`e}vre} O.,  et~al., 2005, \mn@doi [\aap] {10.1051/0004-6361:20041960},
  \href {https://ui.adsabs.harvard.edu/abs/2005A&A...439..845L} {439, 845}

\bibitem[\protect\citeauthoryear{{Lehmer} et~al.,}{{Lehmer}
  et~al.}{2005}]{Lehmer+2005}
{Lehmer} B.~D.,  et~al., 2005, \mn@doi [\apjs] {10.1086/444590}, \href
  {https://ui.adsabs.harvard.edu/abs/2005ApJS..161...21L} {161, 21}

\bibitem[\protect\citeauthoryear{{Lehmer} et~al.,}{{Lehmer}
  et~al.}{2009}]{Lehmer+2009}
{Lehmer} B.~D.,  et~al., 2009, \mn@doi [\mnras]
  {10.1111/j.1365-2966.2009.15449.x}, \href
  {https://ui.adsabs.harvard.edu/abs/2009MNRAS.400..299L} {400, 299}

\bibitem[\protect\citeauthoryear{{Lemaux} et~al.,}{{Lemaux}
  et~al.}{2014}]{Lemaux+2014}
{Lemaux} B.~C.,  et~al., 2014, \mn@doi [\aap] {10.1051/0004-6361/201423828},
  \href {https://ui.adsabs.harvard.edu/abs/2014A&A...572A..41L} {572, A41}

\bibitem[\protect\citeauthoryear{{Lemaux} et~al.,}{{Lemaux}
  et~al.}{2022}]{Lemaux+2022}
{Lemaux} B.~C.,  et~al., 2022, \mn@doi [\aap] {10.1051/0004-6361/202039346},
  \href {https://ui.adsabs.harvard.edu/abs/2022A&A...662A..33L} {662, A33}

\bibitem[\protect\citeauthoryear{{Long} et~al.,}{{Long}
  et~al.}{2020}]{Long+2020}
{Long} A.~S.,  et~al., 2020, \mn@doi [\apj] {10.3847/1538-4357/ab9d1f}, \href
  {https://ui.adsabs.harvard.edu/abs/2020ApJ...898..133L} {898, 133}

\bibitem[\protect\citeauthoryear{{Lovell}, {Geach}, {Dav{\'e}}, {Narayanan}  \&
  {Li}}{{Lovell} et~al.}{2021}]{Lovell+2021}
{Lovell} C.~C.,  {Geach} J.~E.,  {Dav{\'e}} R.,  {Narayanan} D.,   {Li} Q.,
  2021, \mn@doi [\mnras] {10.1093/mnras/staa4043}, \href
  {https://ui.adsabs.harvard.edu/abs/2021MNRAS.502..772L} {502, 772}

\bibitem[\protect\citeauthoryear{{MacKenzie} et~al.,}{{MacKenzie}
  et~al.}{2017}]{Mackenzie+2017}
{MacKenzie} T.~P.,  et~al., 2017, \mn@doi [\mnras] {10.1093/mnras/stx512},
  \href {https://ui.adsabs.harvard.edu/abs/2017MNRAS.468.4006M} {468, 4006}

\bibitem[\protect\citeauthoryear{{Madau} \& {Dickinson}}{{Madau} \&
  {Dickinson}}{2014}]{Madau+Dickinson2014}
{Madau} P.,  {Dickinson} M.,  2014, \mn@doi [\araa]
  {10.1146/annurev-astro-081811-125615}, \href
  {https://ui.adsabs.harvard.edu/abs/2014ARA&A..52..415M} {52, 415}

\bibitem[\protect\citeauthoryear{{Magnelli} et~al.,}{{Magnelli}
  et~al.}{2012}]{Magnelli+2012}
{Magnelli} B.,  et~al., 2012, \mn@doi [\aap] {10.1051/0004-6361/201118312},
  \href {https://ui.adsabs.harvard.edu/abs/2012A&A...539A.155M} {539, A155}

\bibitem[\protect\citeauthoryear{{Man}, {Peng}, {Shi}, {Kong}, {Zhang}, {Dou}
  \& {Guo}}{{Man} et~al.}{2019}]{Man+2019}
{Man} Z.-Y.,  {Peng} Y.-J.,  {Shi} J.-J.,  {Kong} X.,  {Zhang} C.-P.,  {Dou}
  J.,   {Guo} K.-X.,  2019, \mn@doi [\apj] {10.3847/1538-4357/ab2ece}, \href
  {https://ui.adsabs.harvard.edu/abs/2019ApJ...881...74M} {881, 74}

\bibitem[\protect\citeauthoryear{{Matsuda} et~al.,}{{Matsuda}
  et~al.}{2004}]{Matsuda+2004}
{Matsuda} Y.,  et~al., 2004, \mn@doi [\aj] {10.1086/422020}, \href
  {https://ui.adsabs.harvard.edu/abs/2004AJ....128..569M} {128, 569}

\bibitem[\protect\citeauthoryear{{Matsuda} et~al.,}{{Matsuda}
  et~al.}{2011}]{Matsuda+2011}
{Matsuda} Y.,  et~al., 2011, \mn@doi [\mnras]
  {10.1111/j.1365-2966.2011.19179.x}, \href
  {https://ui.adsabs.harvard.edu/abs/2011MNRAS.416.2041M} {416, 2041}

\bibitem[\protect\citeauthoryear{{McBride}, {Fakhouri}  \& {Ma}}{{McBride}
  et~al.}{2009}]{McBride+2009}
{McBride} J.,  {Fakhouri} O.,   {Ma} C.-P.,  2009, \mn@doi [\mnras]
  {10.1111/j.1365-2966.2009.15329.x}, \href
  {https://ui.adsabs.harvard.edu/abs/2009MNRAS.398.1858M} {398, 1858}

\bibitem[\protect\citeauthoryear{{Mei} et~al.,}{{Mei} et~al.}{2022}]{Mei+2022}
{Mei} S.,  et~al., 2022, arXiv e-prints, \href
  {https://ui.adsabs.harvard.edu/abs/2022arXiv220902078M} {p. arXiv:2209.02078}

\bibitem[\protect\citeauthoryear{{Merten} et~al.,}{{Merten}
  et~al.}{2015}]{Merten+2015}
{Merten} J.,  et~al., 2015, \mn@doi [\apj] {10.1088/0004-637X/806/1/4}, \href
  {https://ui.adsabs.harvard.edu/abs/2015ApJ...806....4M} {806, 4}

\bibitem[\protect\citeauthoryear{{Micha{\l}owski}, {Dunlop}, {Cirasuolo},
  {Hjorth}, {Hayward}  \& {Watson}}{{Micha{\l}owski}
  et~al.}{2012}]{Michalowski+2012}
{Micha{\l}owski} M.~J.,  {Dunlop} J.~S.,  {Cirasuolo} M.,  {Hjorth} J.,
  {Hayward} C.~C.,   {Watson} D.,  2012, \mn@doi [\aap]
  {10.1051/0004-6361/201016308}, \href
  {https://ui.adsabs.harvard.edu/abs/2012A&A...541A..85M} {541, A85}

\bibitem[\protect\citeauthoryear{{Micha{\l}owski} et~al.,}{{Micha{\l}owski}
  et~al.}{2017}]{Michalowski+2017}
{Micha{\l}owski} M.~J.,  et~al., 2017, \mn@doi [\mnras] {10.1093/mnras/stx861},
  \href {https://ui.adsabs.harvard.edu/abs/2017MNRAS.469..492M} {469, 492}

\bibitem[\protect\citeauthoryear{{Miettinen} et~al.,}{{Miettinen}
  et~al.}{2017}]{Miettinen+2017}
{Miettinen} O.,  et~al., 2017, \mn@doi [\aap] {10.1051/0004-6361/201730762},
  \href {https://ui.adsabs.harvard.edu/abs/2017A&A...606A..17M} {606, A17}

\bibitem[\protect\citeauthoryear{{Miller} et~al.,}{{Miller}
  et~al.}{2013}]{Miller+2013}
{Miller} N.~A.,  et~al., 2013, \mn@doi [\apjs] {10.1088/0067-0049/205/2/13},
  \href {https://ui.adsabs.harvard.edu/abs/2013ApJS..205...13M} {205, 13}

\bibitem[\protect\citeauthoryear{{Moorwood}, {van der Werf}, {Cuby}  \&
  {Oliva}}{{Moorwood} et~al.}{2000}]{Moorwood+2000}
{Moorwood} A.~F.~M.,  {van der Werf} P.~P.,  {Cuby} J.~G.,   {Oliva} E.,  2000,
  \aap, \href {https://ui.adsabs.harvard.edu/abs/2000A&A...362....9M} {362, 9}

\bibitem[\protect\citeauthoryear{{Muldrew}, {Hatch}  \& {Cooke}}{{Muldrew}
  et~al.}{2015}]{Muldrew+2015}
{Muldrew} S.~I.,  {Hatch} N.~A.,   {Cooke} E.~A.,  2015, \mn@doi [\mnras]
  {10.1093/mnras/stv1449}, \href
  {https://ui.adsabs.harvard.edu/abs/2015MNRAS.452.2528M} {452, 2528}

\bibitem[\protect\citeauthoryear{{Muzzin} et~al.,}{{Muzzin}
  et~al.}{2009}]{Muzzin+2009}
{Muzzin} A.,  et~al., 2009, \mn@doi [\apj] {10.1088/0004-637X/698/2/1934},
  \href {https://ui.adsabs.harvard.edu/abs/2009ApJ...698.1934M} {698, 1934}

\bibitem[\protect\citeauthoryear{{Muzzin} et~al.,}{{Muzzin}
  et~al.}{2012}]{Muzzin+2012}
{Muzzin} A.,  et~al., 2012, \mn@doi [\apj] {10.1088/0004-637X/746/2/188}, \href
  {https://ui.adsabs.harvard.edu/abs/2012ApJ...746..188M} {746, 188}

\bibitem[\protect\citeauthoryear{{Muzzin} et~al.,}{{Muzzin}
  et~al.}{2013}]{Muzzin+2013}
{Muzzin} A.,  et~al., 2013, \mn@doi [\apj] {10.1088/0004-637X/777/1/18}, \href
  {https://ui.adsabs.harvard.edu/abs/2013ApJ...777...18M} {777, 18}

\bibitem[\protect\citeauthoryear{{Narayanan}, {Hayward}, {Cox}, {Hernquist},
  {Jonsson}, {Younger}  \& {Groves}}{{Narayanan} et~al.}{2010}]{Narayanan+2010}
{Narayanan} D.,  {Hayward} C.~C.,  {Cox} T.~J.,  {Hernquist} L.,  {Jonsson} P.,
   {Younger} J.~D.,   {Groves} B.,  2010, \mn@doi [\mnras]
  {10.1111/j.1365-2966.2009.15790.x}, \href
  {https://ui.adsabs.harvard.edu/abs/2010MNRAS.401.1613N} {401, 1613}

\bibitem[\protect\citeauthoryear{{Narayanan} et~al.,}{{Narayanan}
  et~al.}{2015}]{Narayanan+2015}
{Narayanan} D.,  et~al., 2015, \mn@doi [\nat] {10.1038/nature15383}, \href
  {https://ui.adsabs.harvard.edu/abs/2015Natur.525..496N} {525, 496}

\bibitem[\protect\citeauthoryear{{Niemi}, {Somerville}, {Ferguson}, {Huang},
  {Lotz}  \& {Koekemoer}}{{Niemi} et~al.}{2012}]{Niemi+2012}
{Niemi} S.-M.,  {Somerville} R.~S.,  {Ferguson} H.~C.,  {Huang} K.-H.,  {Lotz}
  J.,   {Koekemoer} A.~M.,  2012, \mn@doi [\mnras]
  {10.1111/j.1365-2966.2012.20425.x}, \href
  {https://ui.adsabs.harvard.edu/abs/2012MNRAS.421.1539N} {421, 1539}

\bibitem[\protect\citeauthoryear{{Oke} \& {Gunn}}{{Oke} \&
  {Gunn}}{1983}]{Oke+Gunn1983}
{Oke} J.~B.,  {Gunn} J.~E.,  1983, \mn@doi [\apj] {10.1086/160817}, \href
  {https://ui.adsabs.harvard.edu/abs/1983ApJ...266..713O} {266, 713}

\bibitem[\protect\citeauthoryear{{Osterbrock} \& {Ferland}}{{Osterbrock} \&
  {Ferland}}{2006}]{OsterbrockFerland2006}
{Osterbrock} D.~E.,  {Ferland} G.~J.,  2006, {Astrophysics of gaseous nebulae
  and active galactic nuclei}.
CA: University Science Books

\bibitem[\protect\citeauthoryear{{Overzier}}{{Overzier}}{2016}]{Overzier2016}
{Overzier} R.~A.,  2016, \mn@doi [\aapr] {10.1007/s00159-016-0100-3}, \href
  {https://ui.adsabs.harvard.edu/abs/2016A&ARv..24...14O} {24, 14}

\bibitem[\protect\citeauthoryear{{Overzier}, {Guo}, {Kauffmann}, {De Lucia},
  {Bouwens}  \& {Lemson}}{{Overzier} et~al.}{2009}]{Overzier+2009}
{Overzier} R.~A.,  {Guo} Q.,  {Kauffmann} G.,  {De Lucia} G.,  {Bouwens} R.,
  {Lemson} G.,  2009, \mn@doi [\mnras] {10.1111/j.1365-2966.2008.14264.x},
  \href {https://ui.adsabs.harvard.edu/abs/2009MNRAS.394..577O} {394, 577}

\bibitem[\protect\citeauthoryear{{Pacaud} et~al.,}{{Pacaud}
  et~al.}{2016}]{Pacaud+2016}
{Pacaud} F.,  et~al., 2016, \mn@doi [\aap] {10.1051/0004-6361/201526891}, \href
  {https://ui.adsabs.harvard.edu/abs/2016A&A...592A...2P} {592, A2}

\bibitem[\protect\citeauthoryear{{Palunas}, {Teplitz}, {Francis}, {Williger}
  \& {Woodgate}}{{Palunas} et~al.}{2004}]{Palunas+2004}
{Palunas} P.,  {Teplitz} H.~I.,  {Francis} P.~J.,  {Williger} G.~M.,
  {Woodgate} B.~E.,  2004, \mn@doi [\apj] {10.1086/381145}, \href
  {https://ui.adsabs.harvard.edu/abs/2004ApJ...602..545P} {602, 545}

\bibitem[\protect\citeauthoryear{{Pantoni} et~al.,}{{Pantoni}
  et~al.}{2021}]{Pantoni+2021}
{Pantoni} L.,  et~al., 2021, \mn@doi [\mnras] {10.1093/mnras/stab674}, \href
  {https://ui.adsabs.harvard.edu/abs/2021MNRAS.504..928P} {504, 928}

\bibitem[\protect\citeauthoryear{{Papovich} et~al.,}{{Papovich}
  et~al.}{2018}]{Papovich+2018}
{Papovich} C.,  et~al., 2018, \mn@doi [\apj] {10.3847/1538-4357/aaa766}, \href
  {https://ui.adsabs.harvard.edu/abs/2018ApJ...854...30P} {854, 30}

\bibitem[\protect\citeauthoryear{{Paufique} et~al.,}{{Paufique}
  et~al.}{2010}]{HAWKI_GRAAL2010}
{Paufique} J.,  et~al., 2010, in {Ellerbroek} B.~L.,  {Hart} M.,  {Hubin} N.,
  {Wizinowich} P.~L.,  eds,  Society of Photo-Optical Instrumentation Engineers
  (SPIE) Conference Series Vol. 7736, Adaptive Optics Systems II. p. 77361P,
  \mn@doi{10.1117/12.858261}

\bibitem[\protect\citeauthoryear{{Pirard} et~al.,}{{Pirard}
  et~al.}{2004}]{HAWKI2004}
{Pirard} J.-F.,  et~al., 2004, in {Moorwood} A. F.~M.,  {Iye} M.,  eds,
  Society of Photo-Optical Instrumentation Engineers (SPIE) Conference Series
  Vol. 5492, Ground-based Instrumentation for Astronomy. pp 1763--1772,
  \mn@doi{10.1117/12.578293}

\bibitem[\protect\citeauthoryear{{Planck Collaboration}}{{Planck
  Collaboration}}{2020}]{Planck2018}
{Planck Collaboration} 2020, \mn@doi [\aap] {10.1051/0004-6361/201833910},
  \href {https://ui.adsabs.harvard.edu/abs/2020A&A...641A...6P} {641, A6}

\bibitem[\protect\citeauthoryear{{Planck Collaboration} et~al.,}{{Planck
  Collaboration} et~al.}{2016}]{Planck2016}
{Planck Collaboration} et~al., 2016, \mn@doi [\aap]
  {10.1051/0004-6361/201525823}, \href
  {https://ui.adsabs.harvard.edu/abs/2016A&A...594A..27P} {594, A27}

\bibitem[\protect\citeauthoryear{{Pope} et~al.,}{{Pope}
  et~al.}{2006}]{Pope+2006}
{Pope} A.,  et~al., 2006, \mn@doi [\mnras] {10.1111/j.1365-2966.2006.10575.x},
  \href {https://ui.adsabs.harvard.edu/abs/2006MNRAS.370.1185P} {370, 1185}

\bibitem[\protect\citeauthoryear{{Popesso} et~al.,}{{Popesso}
  et~al.}{2009}]{Popesso+2009}
{Popesso} P.,  et~al., 2009, \mn@doi [\aap] {10.1051/0004-6361:200809617},
  \href {https://ui.adsabs.harvard.edu/abs/2009A&A...494..443P} {494, 443}

\bibitem[\protect\citeauthoryear{{Postman} et~al.,}{{Postman}
  et~al.}{2012}]{Postman+2012}
{Postman} M.,  et~al., 2012, \mn@doi [\apjs] {10.1088/0067-0049/199/2/25},
  \href {https://ui.adsabs.harvard.edu/abs/2012ApJS..199...25P} {199, 25}

\bibitem[\protect\citeauthoryear{{Press} \& {Schechter}}{{Press} \&
  {Schechter}}{1974}]{PressSchechter1974}
{Press} W.~H.,  {Schechter} P.,  1974, \mn@doi [\apj] {10.1086/152650}, \href
  {https://ui.adsabs.harvard.edu/abs/1974ApJ...187..425P} {187, 425}

\bibitem[\protect\citeauthoryear{{Ramsay}, {Mountain}  \& {Geballe}}{{Ramsay}
  et~al.}{1992}]{Ramsay+1992}
{Ramsay} S.~K.,  {Mountain} C.~M.,   {Geballe} T.~R.,  1992, \mn@doi [\mnras]
  {10.1093/mnras/259.4.751}, \href
  {https://ui.adsabs.harvard.edu/abs/1992MNRAS.259..751R} {259, 751}

\bibitem[\protect\citeauthoryear{{Rowan-Robinson} et~al.,}{{Rowan-Robinson}
  et~al.}{2018}]{RowanRobinson+2018}
{Rowan-Robinson} M.,  et~al., 2018, \mn@doi [\aap]
  {10.1051/0004-6361/201832671}, \href
  {https://ui.adsabs.harvard.edu/abs/2018A&A...619A.169R} {619, A169}

\bibitem[\protect\citeauthoryear{{Sanders}, {Soifer}, {Elias}, {Madore},
  {Matthews}, {Neugebauer}  \& {Scoville}}{{Sanders}
  et~al.}{1988}]{Sanders+1988}
{Sanders} D.~B.,  {Soifer} B.~T.,  {Elias} J.~H.,  {Madore} B.~F.,  {Matthews}
  K.,  {Neugebauer} G.,   {Scoville} N.~Z.,  1988, \mn@doi [\apj]
  {10.1086/165983}, \href
  {https://ui.adsabs.harvard.edu/abs/1988ApJ...325...74S} {325, 74}

\bibitem[\protect\citeauthoryear{{Schechter}}{{Schechter}}{1976}]{Schechter1976}
{Schechter} P.,  1976, \mn@doi [\apj] {10.1086/154079}, \href
  {https://ui.adsabs.harvard.edu/abs/1976ApJ...203..297S} {203, 297}

\bibitem[\protect\citeauthoryear{{Schreiber} et~al.,}{{Schreiber}
  et~al.}{2015}]{Schreiber+2015}
{Schreiber} C.,  et~al., 2015, \mn@doi [\aap] {10.1051/0004-6361/201425017},
  \href {https://ui.adsabs.harvard.edu/abs/2015A&A...575A..74S} {575, A74}

\bibitem[\protect\citeauthoryear{{Scoville} et~al.,}{{Scoville}
  et~al.}{2013}]{Scoville+2013}
{Scoville} N.,  et~al., 2013, \mn@doi [\apjs] {10.1088/0067-0049/206/1/3},
  \href {https://ui.adsabs.harvard.edu/abs/2013ApJS..206....3S} {206, 3}

\bibitem[\protect\citeauthoryear{{Scoville} et~al.,}{{Scoville}
  et~al.}{2017}]{Scoville+2017}
{Scoville} N.,  et~al., 2017, \mn@doi [\apj] {10.3847/1538-4357/aa61a0}, \href
  {https://ui.adsabs.harvard.edu/abs/2017ApJ...837..150S} {837, 150}

\bibitem[\protect\citeauthoryear{{Shimasaku} et~al.,}{{Shimasaku}
  et~al.}{2003}]{Shimasaku+2003}
{Shimasaku} K.,  et~al., 2003, \mn@doi [\apjl] {10.1086/374880}, \href
  {https://ui.adsabs.harvard.edu/abs/2003ApJ...586L.111S} {586, L111}

\bibitem[\protect\citeauthoryear{{Siebenmorgen}, {Carraro}, {Valenti},
  {Petr-Gotzens}, {Brammer}, {Garcia}  \& {Casali}}{{Siebenmorgen}
  et~al.}{2011}]{HAWKI2011}
{Siebenmorgen} R.,  {Carraro} G.,  {Valenti} E.,  {Petr-Gotzens} M.,  {Brammer}
  G.,  {Garcia} E.,   {Casali} M.,  2011, The Messenger, \href
  {https://ui.adsabs.harvard.edu/abs/2011Msngr.144....9S} {144, 9}

\bibitem[\protect\citeauthoryear{{Sillassen} et~al.,}{{Sillassen}
  et~al.}{2022}]{Sillassen+2022}
{Sillassen} N.~B.,  et~al., 2022, \mn@doi [\aap] {10.1051/0004-6361/202244661},
  \href {https://ui.adsabs.harvard.edu/abs/2022A&A...665L...7S} {665, L7}

\bibitem[\protect\citeauthoryear{{Silverman} et~al.,}{{Silverman}
  et~al.}{2010}]{Silverman+2010}
{Silverman} J.~D.,  et~al., 2010, \mn@doi [\apjs]
  {10.1088/0067-0049/191/1/124}, \href
  {https://ui.adsabs.harvard.edu/abs/2010ApJS..191..124S} {191, 124}

\bibitem[\protect\citeauthoryear{{Simpson} et~al.,}{{Simpson}
  et~al.}{2014}]{Simpson+2014}
{Simpson} J.~M.,  et~al., 2014, \mn@doi [\apj] {10.1088/0004-637X/788/2/125},
  \href {https://ui.adsabs.harvard.edu/abs/2014ApJ...788..125S} {788, 125}

\bibitem[\protect\citeauthoryear{{Smail}, {Ivison}  \& {Blain}}{{Smail}
  et~al.}{1997}]{Smail+1997}
{Smail} I.,  {Ivison} R.~J.,   {Blain} A.~W.,  1997, \mn@doi [\apjl]
  {10.1086/311017}, \href
  {https://ui.adsabs.harvard.edu/abs/1997ApJ...490L...5S} {490, L5}

\bibitem[\protect\citeauthoryear{{Smail}, {Ivison}, {Gilbank}, {Dunlop},
  {Keel}, {Motohara}  \& {Stevens}}{{Smail} et~al.}{2003}]{Smail+2003a}
{Smail} I.,  {Ivison} R.~J.,  {Gilbank} D.~G.,  {Dunlop} J.~S.,  {Keel} W.~C.,
  {Motohara} K.,   {Stevens} J.~A.,  2003, \mn@doi [\apj] {10.1086/345474},
  \href {https://ui.adsabs.harvard.edu/abs/2003ApJ...583..551S} {583, 551}

\bibitem[\protect\citeauthoryear{{Smith}, {Hayward}, {Jarvis}  \&
  {Simpson}}{{Smith} et~al.}{2017}]{Smith+2017}
{Smith} D.~J.~B.,  {Hayward} C.~C.,  {Jarvis} M.~J.,   {Simpson} C.,  2017,
  \mn@doi [\mnras] {10.1093/mnras/stx1689}, \href
  {https://ui.adsabs.harvard.edu/abs/2017MNRAS.471.2453S} {471, 2453}

\bibitem[\protect\citeauthoryear{{Smol{\v{c}}i{\'c}}
  et~al.,}{{Smol{\v{c}}i{\'c}} et~al.}{2017}]{Smolcic+2017a}
{Smol{\v{c}}i{\'c}} V.,  et~al., 2017, \mn@doi [\aap]
  {10.1051/0004-6361/201526989}, \href
  {https://ui.adsabs.harvard.edu/abs/2017A&A...597A...4S} {597, A4}

\bibitem[\protect\citeauthoryear{{Sobral} et~al.,}{{Sobral}
  et~al.}{2009}]{Sobral+2009}
{Sobral} D.,  et~al., 2009, \mn@doi [\mnras]
  {10.1111/j.1365-2966.2009.15129.x}, \href
  {https://ui.adsabs.harvard.edu/abs/2009MNRAS.398...75S} {398, 75}

\bibitem[\protect\citeauthoryear{{Sobral}, {Best}, {Matsuda}, {Smail}, {Geach}
  \& {Cirasuolo}}{{Sobral} et~al.}{2012}]{Sobral+2012}
{Sobral} D.,  {Best} P.~N.,  {Matsuda} Y.,  {Smail} I.,  {Geach} J.~E.,
  {Cirasuolo} M.,  2012, \mn@doi [\mnras] {10.1111/j.1365-2966.2011.19977.x},
  \href {https://ui.adsabs.harvard.edu/abs/2012MNRAS.420.1926S} {420, 1926}

\bibitem[\protect\citeauthoryear{{Sobral}, {Smail}, {Best}, {Geach}, {Matsuda},
  {Stott}, {Cirasuolo}  \& {Kurk}}{{Sobral} et~al.}{2013}]{Sobral+2013}
{Sobral} D.,  {Smail} I.,  {Best} P.~N.,  {Geach} J.~E.,  {Matsuda} Y.,
  {Stott} J.~P.,  {Cirasuolo} M.,   {Kurk} J.,  2013, \mn@doi [\mnras]
  {10.1093/mnras/sts096}, \href
  {https://ui.adsabs.harvard.edu/abs/2013MNRAS.428.1128S} {428, 1128}

\bibitem[\protect\citeauthoryear{{Sobral}, {Best}, {Smail}, {Mobasher}, {Stott}
   \& {Nisbet}}{{Sobral} et~al.}{2014}]{Sobral+2014}
{Sobral} D.,  {Best} P.~N.,  {Smail} I.,  {Mobasher} B.,  {Stott} J.,
  {Nisbet} D.,  2014, \mn@doi [\mnras] {10.1093/mnras/stt2159}, \href
  {https://ui.adsabs.harvard.edu/abs/2014MNRAS.437.3516S} {437, 3516}

\bibitem[\protect\citeauthoryear{{Sobral} et~al.,}{{Sobral}
  et~al.}{2015}]{Sobral+2015}
{Sobral} D.,  et~al., 2015, \mn@doi [\mnras] {10.1093/mnras/stv1076}, \href
  {https://ui.adsabs.harvard.edu/abs/2015MNRAS.451.2303S} {451, 2303}

\bibitem[\protect\citeauthoryear{{Speagle}, {Steinhardt}, {Capak}  \&
  {Silverman}}{{Speagle} et~al.}{2014}]{Speagle+2014}
{Speagle} J.~S.,  {Steinhardt} C.~L.,  {Capak} P.~L.,   {Silverman} J.~D.,
  2014, \mn@doi [\apjs] {10.1088/0067-0049/214/2/15}, \href
  {https://ui.adsabs.harvard.edu/abs/2014ApJS..214...15S} {214, 15}

\bibitem[\protect\citeauthoryear{{Stach} et~al.,}{{Stach}
  et~al.}{2019}]{Stach+2019}
{Stach} S.~M.,  et~al., 2019, \mn@doi [\mnras] {10.1093/mnras/stz1536}, \href
  {https://ui.adsabs.harvard.edu/abs/2019MNRAS.487.4648S} {487, 4648}

\bibitem[\protect\citeauthoryear{{Stach} et~al.,}{{Stach}
  et~al.}{2021}]{Stach+2021}
{Stach} S.~M.,  et~al., 2021, \mn@doi [\mnras] {10.1093/mnras/stab714}, \href
  {https://ui.adsabs.harvard.edu/abs/2021MNRAS.504..172S} {504, 172}

\bibitem[\protect\citeauthoryear{{Staniszewski} et~al.,}{{Staniszewski}
  et~al.}{2009}]{Staniszewski+2009}
{Staniszewski} Z.,  et~al., 2009, \mn@doi [\apj] {10.1088/0004-637X/701/1/32},
  \href {https://ui.adsabs.harvard.edu/abs/2009ApJ...701...32S} {701, 32}

\bibitem[\protect\citeauthoryear{{Steidel}, {Adelberger}, {Dickinson},
  {Giavalisco}, {Pettini}  \& {Kellogg}}{{Steidel} et~al.}{1998}]{Steidel+1998}
{Steidel} C.~C.,  {Adelberger} K.~L.,  {Dickinson} M.,  {Giavalisco} M.,
  {Pettini} M.,   {Kellogg} M.,  1998, \mn@doi [\apj] {10.1086/305073}, \href
  {https://ui.adsabs.harvard.edu/abs/1998ApJ...492..428S} {492, 428}

\bibitem[\protect\citeauthoryear{{Steidel}, {Adelberger}, {Shapley}, {Pettini},
  {Dickinson}  \& {Giavalisco}}{{Steidel} et~al.}{2000}]{Steidel+2000}
{Steidel} C.~C.,  {Adelberger} K.~L.,  {Shapley} A.~E.,  {Pettini} M.,
  {Dickinson} M.,   {Giavalisco} M.,  2000, \mn@doi [\apj] {10.1086/308568},
  \href {https://ui.adsabs.harvard.edu/abs/2000ApJ...532..170S} {532, 170}

\bibitem[\protect\citeauthoryear{{Steidel}, {Adelberger}, {Shapley}, {Erb},
  {Reddy}  \& {Pettini}}{{Steidel} et~al.}{2005}]{Steidel+2005}
{Steidel} C.~C.,  {Adelberger} K.~L.,  {Shapley} A.~E.,  {Erb} D.~K.,  {Reddy}
  N.~A.,   {Pettini} M.,  2005, \mn@doi [\apj] {10.1086/429989}, \href
  {https://ui.adsabs.harvard.edu/abs/2005ApJ...626...44S} {626, 44}

\bibitem[\protect\citeauthoryear{{Stevans} et~al.,}{{Stevans}
  et~al.}{2021}]{Stevans+2021}
{Stevans} M.~L.,  et~al., 2021, \mn@doi [\apj] {10.3847/1538-4357/ac0cf6},
  \href {https://ui.adsabs.harvard.edu/abs/2021ApJ...921...58S} {921, 58}

\bibitem[\protect\citeauthoryear{{Stott} et~al.,}{{Stott}
  et~al.}{2020}]{Stott+2020}
{Stott} J.~P.,  et~al., 2020, \mn@doi [\mnras] {10.1093/mnras/staa2096}, \href
  {https://ui.adsabs.harvard.edu/abs/2020MNRAS.497.3083S} {497, 3083}

\bibitem[\protect\citeauthoryear{{Sunyaev} \& {Zeldovich}}{{Sunyaev} \&
  {Zeldovich}}{1972}]{SunyaevZeldovich1972}
{Sunyaev} R.~A.,  {Zeldovich} Y.~B.,  1972, Comments on Astrophysics and Space
  Physics, \href {https://ui.adsabs.harvard.edu/abs/1972CoASP...4..173S} {4,
  173}

\bibitem[\protect\citeauthoryear{{Swinbank} et~al.,}{{Swinbank}
  et~al.}{2014}]{Swinbank+2014}
{Swinbank} A.~M.,  et~al., 2014, \mn@doi [\mnras] {10.1093/mnras/stt2273},
  \href {https://ui.adsabs.harvard.edu/abs/2014MNRAS.438.1267S} {438, 1267}

\bibitem[\protect\citeauthoryear{{Tacconi} et~al.,}{{Tacconi}
  et~al.}{2006}]{Tacconi+2006}
{Tacconi} L.~J.,  et~al., 2006, \mn@doi [\apj] {10.1086/499933}, \href
  {https://ui.adsabs.harvard.edu/abs/2006ApJ...640..228T} {640, 228}

\bibitem[\protect\citeauthoryear{{Tamura} et~al.,}{{Tamura}
  et~al.}{2009}]{Tamura+2009}
{Tamura} Y.,  et~al., 2009, \mn@doi [\nat] {10.1038/nature07947}, \href
  {https://ui.adsabs.harvard.edu/abs/2009Natur.459...61T} {459, 61}

\bibitem[\protect\citeauthoryear{{Tanaka} et~al.,}{{Tanaka}
  et~al.}{2011}]{Tanaka+2011}
{Tanaka} I.,  et~al., 2011, \mn@doi [\pasj] {10.1093/pasj/63.sp2.S415}, \href
  {https://ui.adsabs.harvard.edu/abs/2011PASJ...63S.415T} {63, 415}

\bibitem[\protect\citeauthoryear{{Taylor}}{{Taylor}}{2005}]{TOPCAT2005}
{Taylor} M.~B.,  2005, in {Shopbell} P.,  {Britton} M.,   {Ebert} R.,  eds,
  Astronomical Society of the Pacific Conference Series Vol. 347, Astronomical
  Data Analysis Software and Systems XIV. p.~29

\bibitem[\protect\citeauthoryear{{Taylor} et~al.}{{Taylor}
  et~al.}{2009}]{MUSYC2009}
{Taylor} E.~N.,  et~al., 2009, \mn@doi [\apjs] {10.1088/0067-0049/183/2/295},
  \href {https://ui.adsabs.harvard.edu/abs/2009ApJS..183..295T} {183, 295}

\bibitem[\protect\citeauthoryear{{Tinker}, {Kravtsov}, {Klypin}, {Abazajian},
  {Warren}, {Yepes}, {Gottl{\"o}ber}  \& {Holz}}{{Tinker}
  et~al.}{2008}]{Tinker+2008}
{Tinker} J.,  {Kravtsov} A.~V.,  {Klypin} A.,  {Abazajian} K.,  {Warren} M.,
  {Yepes} G.,  {Gottl{\"o}ber} S.,   {Holz} D.~E.,  2008, \mn@doi [\apj]
  {10.1086/591439}, \href
  {https://ui.adsabs.harvard.edu/abs/2008ApJ...688..709T} {688, 709}

\bibitem[\protect\citeauthoryear{{Toft} et~al.,}{{Toft}
  et~al.}{2014}]{Toft+2014}
{Toft} S.,  et~al., 2014, \mn@doi [\apj] {10.1088/0004-637X/782/2/68}, \href
  {https://ui.adsabs.harvard.edu/abs/2014ApJ...782...68T} {782, 68}

\bibitem[\protect\citeauthoryear{{Treister} et~al.,}{{Treister}
  et~al.}{2009}]{Treister+2009}
{Treister} E.,  et~al., 2009, \mn@doi [\apj] {10.1088/0004-637X/693/2/1713},
  \href {https://ui.adsabs.harvard.edu/abs/2009ApJ...693.1713T} {693, 1713}

\bibitem[\protect\citeauthoryear{{Tr\"umper}}{{Tr\"umper}}{1993}]{Truemper1993}
{Tr\"umper} J.,  1993, \mn@doi [Science] {10.1126/science.260.5115.1769}, \href
  {https://ui.adsabs.harvard.edu/abs/1993Sci...260.1769T} {260, 1769}

\bibitem[\protect\citeauthoryear{{Umehata} et~al.,}{{Umehata}
  et~al.}{2015}]{Umehata+2015}
{Umehata} H.,  et~al., 2015, \mn@doi [\apjl] {10.1088/2041-8205/815/1/L8},
  \href {https://ui.adsabs.harvard.edu/abs/2015ApJ...815L...8U} {815, L8}

\bibitem[\protect\citeauthoryear{{Umehata} et~al.,}{{Umehata}
  et~al.}{2019}]{Umehata+2019}
{Umehata} H.,  et~al., 2019, \mn@doi [Science] {10.1126/science.aaw5949}, \href
  {https://ui.adsabs.harvard.edu/abs/2019Sci...366...97U} {366, 97}

\bibitem[\protect\citeauthoryear{{Venemans} et~al.,}{{Venemans}
  et~al.}{2002}]{Venemans+2002}
{Venemans} B.~P.,  et~al., 2002, \mn@doi [\apjl] {10.1086/340563}, \href
  {https://ui.adsabs.harvard.edu/abs/2002ApJ...569L..11V} {569, L11}

\bibitem[\protect\citeauthoryear{{Venemans} et~al.,}{{Venemans}
  et~al.}{2005}]{Venemans+2005}
{Venemans} B.~P.,  et~al., 2005, \mn@doi [\aap] {10.1051/0004-6361:20042038},
  \href {https://ui.adsabs.harvard.edu/abs/2005A&A...431..793V} {431, 793}

\bibitem[\protect\citeauthoryear{Virtanen et~al.,}{Virtanen
  et~al.}{2020}]{SCIPY2020}
Virtanen P.,  et~al., 2020, \mn@doi [Nature Methods]
  {10.1038/s41592-019-0686-2}, \href {https://rdcu.be/b08Wh} {17, 261}

\bibitem[\protect\citeauthoryear{{Wardlow} et~al.}{{Wardlow}
  et~al.}{2011}]{Wardlow+2011}
{Wardlow} J.~L.,  et~al., 2011, \mn@doi [\mnras]
  {10.1111/j.1365-2966.2011.18795.x}, \href
  {https://ui.adsabs.harvard.edu/abs/2011MNRAS.415.1479W} {415, 1479}

\bibitem[\protect\citeauthoryear{{Wei{\ss}} et~al.,}{{Wei{\ss}}
  et~al.}{2009}]{Weiss+2009}
{Wei{\ss}} A.,  et~al., 2009, \mn@doi [\apj] {10.1088/0004-637X/707/2/1201},
  \href {https://ui.adsabs.harvard.edu/abs/2009ApJ...707.1201W} {707, 1201}

\bibitem[\protect\citeauthoryear{{Wilkinson} et~al.,}{{Wilkinson}
  et~al.}{2017}]{Wilkinson+2017}
{Wilkinson} A.,  et~al., 2017, \mn@doi [\mnras] {10.1093/mnras/stw2405}, \href
  {https://ui.adsabs.harvard.edu/abs/2017MNRAS.464.1380W} {464, 1380}

\bibitem[\protect\citeauthoryear{{Williamson} et~al.,}{{Williamson}
  et~al.}{2011}]{Williamson+2011}
{Williamson} R.,  et~al., 2011, \mn@doi [\apj] {10.1088/0004-637X/738/2/139},
  \href {https://ui.adsabs.harvard.edu/abs/2011ApJ...738..139W} {738, 139}

\bibitem[\protect\citeauthoryear{{Wilson} et~al.,}{{Wilson}
  et~al.}{2009}]{Wilson+2009}
{Wilson} G.,  et~al., 2009, \mn@doi [\apj] {10.1088/0004-637X/698/2/1943},
  \href {https://ui.adsabs.harvard.edu/abs/2009ApJ...698.1943W} {698, 1943}

\bibitem[\protect\citeauthoryear{{Wylezalek} et~al.,}{{Wylezalek}
  et~al.}{2013}]{Wylezalek+2013}
{Wylezalek} D.,  et~al., 2013, \mn@doi [\apj] {10.1088/0004-637X/769/1/79},
  \href {https://ui.adsabs.harvard.edu/abs/2013ApJ...769...79W} {769, 79}

\bibitem[\protect\citeauthoryear{{Yajima} et~al.,}{{Yajima}
  et~al.}{2022}]{Yajima+2022}
{Yajima} H.,  et~al., 2022, \mn@doi [\mnras] {10.1093/mnras/stab3092}, \href
  {https://ui.adsabs.harvard.edu/abs/2022MNRAS.509.4037Y} {509, 4037}

\bibitem[\protect\citeauthoryear{{Yang}, {Zabludoff}, {Eisenstein}  \&
  {Dav{\'e}}}{{Yang} et~al.}{2010}]{Yang+2010}
{Yang} Y.,  {Zabludoff} A.,  {Eisenstein} D.,   {Dav{\'e}} R.,  2010, \mn@doi
  [\apj] {10.1088/0004-637X/719/2/1654}, \href
  {https://ui.adsabs.harvard.edu/abs/2010ApJ...719.1654Y} {719, 1654}

\bibitem[\protect\citeauthoryear{{Yuan} et~al.,}{{Yuan}
  et~al.}{2014}]{Yuan+2014}
{Yuan} T.,  et~al., 2014, \mn@doi [\apjl] {10.1088/2041-8205/795/1/L20}, \href
  {https://ui.adsabs.harvard.edu/abs/2014ApJ...795L..20Y} {795, L20}

\bibitem[\protect\citeauthoryear{{Zavala} et~al.,}{{Zavala}
  et~al.}{2019}]{Zavala+2019}
{Zavala} J.~A.,  et~al., 2019, \mn@doi [\apj] {10.3847/1538-4357/ab5302}, \href
  {https://ui.adsabs.harvard.edu/abs/2019ApJ...887..183Z} {887, 183}

\bibitem[\protect\citeauthoryear{{Zheng}, {Cai}, {An}, {Fan}  \& {Shi}}{{Zheng}
  et~al.}{2021}]{Zheng+2021}
{Zheng} X.~Z.,  {Cai} Z.,  {An} F.~X.,  {Fan} X.,   {Shi} D.~D.,  2021, \mn@doi
  [\mnras] {10.1093/mnras/staa2882}, \href
  {https://ui.adsabs.harvard.edu/abs/2021MNRAS.500.4354Z} {500, 4354}

\bibitem[\protect\citeauthoryear{{da Cunha}, {Charlot}  \& {Elbaz}}{{da Cunha}
  et~al.}{2008}]{daCunhaMAGPHYS+2008}
{da Cunha} E.,  {Charlot} S.,   {Elbaz} D.,  2008, \mn@doi [\mnras]
  {10.1111/j.1365-2966.2008.13535.x}, \href
  {https://ui.adsabs.harvard.edu/abs/2008MNRAS.388.1595D} {388, 1595}

\bibitem[\protect\citeauthoryear{{da Cunha} et~al.,}{{da Cunha}
  et~al.}{2015}]{daCunha+2015}
{da Cunha} E.,  et~al., 2015, \mn@doi [\apj] {10.1088/0004-637X/806/1/110},
  \href {https://ui.adsabs.harvard.edu/abs/2015ApJ...806..110D} {806, 110}

\bibitem[\protect\citeauthoryear{{da Cunha} et~al.,}{{da Cunha}
  et~al.}{2021}]{daCunha+2021}
{da Cunha} E.,  et~al., 2021, \mn@doi [\apj] {10.3847/1538-4357/ac0ae0}, \href
  {https://ui.adsabs.harvard.edu/abs/2021ApJ...919...30D} {919, 30}

\bibitem[\protect\citeauthoryear{{van der Burg}, {Muzzin}, {Hoekstra},
  {Wilson}, {Lidman}  \& {Yee}}{{van der Burg}
  et~al.}{2014}]{van_der_Burg+2014}
{van der Burg} R.~F.~J.,  {Muzzin} A.,  {Hoekstra} H.,  {Wilson} G.,  {Lidman}
  C.,   {Yee} H.~K.~C.,  2014, \mn@doi [\aap] {10.1051/0004-6361/201322771},
  \href {https://ui.adsabs.harvard.edu/abs/2014A&A...561A..79V} {561, A79}

\makeatother
\end{thebibliography}




\bsp	
\label{lastpage}
\end{document}